\documentclass[utf8,loar,contribinloar,hyperref]{jydiss_phys}

\usepackage{textcomp}
\usepackage{upgreek}
\usepackage{tabularx}

\usepackage[LGR, T1]{fontenc}

\usepackage{physics}
\usepackage{graphicx}
\usepackage{amsmath,amssymb}
\usepackage{mathtools}
\usepackage{slashed}

\usepackage{siunitx}

\usepackage{xcolor}
\definecolor{lcolor}{rgb}{0.5,0,0}
\definecolor{citcolor}{rgb}{0,0.3,0.0}

\usepackage[breaklinks,colorlinks,urlcolor=blue,citecolor=citcolor,linkcolor=lcolor]{hyperref}

\usepackage{csquotes}
\usepackage[numbers,sort&compress]{natbib}
\usepackage{subcaption}

\usepackage[percent]{overpic}

\usepackage{pdfpages}

\newcommand{\Pt}{{\mathbf{P}}}
\newcommand{\rt}{{\mathbf{r}}}
\newcommand{\xt}{{\mathbf{x}}}

\newcommand{\Kt}{{\mathbf{K}}}
\newcommand{\bt}{{\mathbf{b}}}
\newcommand{\yt}{{\mathbf{y}}}
\newcommand{\zt}{{\mathbf{z}}}
\newcommand{\pt}{{\mathbf{p}}}
\newcommand{\qt}{{\mathbf{q}}}

\newcommand{\kt}{{\mathbf{k}}}

\newcommand{\Deltat}{{\boldsymbol{\Delta}}}
\newcommand{\ov}{\overline}

\newcommand{\W}{\mathcal{W}}

\newcommand{\cf}{C_\mathrm{F}}

\newcommand{\nc}{N_\mathrm{c}}

\newcommand{\qs}{Q_\mathrm{s}}

\newcommand{\lqcd}{\Lambda_{\mathrm{QCD}}}
\newcommand{\as}{\alpha_{\mathrm{s}}}
\newcommand{\aem}{\alpha_{\mathrm{em}}}

\newcommand{\jpsi}{\mathrm{J}/\Psi}

\newcommand{\so}{\text{SO}}

\newcommand{\Ydip}{Y_\text{dip}}

\newcommand{\msbar}{\overline{\text{MS}}}

\newcommand{\xpom}{{x_\mathbb{P}}}
\newcommand{\Ypom}{{Y_\mathbb{P}}}
\newcommand{\Ygap}{{Y_\text{gap}}}
\newcommand{\Yshow}{{Y_X}}

\newcommand{\ps}{\textnormal{PS}}

\newcommand{\D}{\textnormal{D}}

\newcommand{\iin}{{\text{in}}}
\newcommand{\oout}{{\text{out}}}

\newcommand{\dipole}{ \hat{S} }
\newcommand{\tripole}{ \hat{S} }

\newcommand{\Ybkzero}{ Y_{0,\text{BK}} }
\newcommand{\Yif}{ Y_{0,\text{if}} }
\newcommand{\Qszero}{ Q^2_{s,0} }

\newcommand{\azero}{ \hyperref[fig:A0]{\text{A0}} }
\newcommand{\aone}{  \hyperref[fig:A1]{\text{A1}} }
\newcommand{\atwo}{  \hyperref[fig:A2]{\text{A2}} }
\newcommand{\bzero}{ \hyperref[fig:B0]{\text{B0}} }
\newcommand{\bone}{  \hyperref[fig:B1]{\text{B1}} }
\newcommand{\btwo}{  \hyperref[fig:B2]{\text{B2}} }
\newcommand{\czero}{ \hyperref[fig:C0]{\text{C0}} }

\newcommand{\ctwo}{  \hyperref[fig:C2]{\text{C2}} }

\newcommand{\done}{  \hyperref[fig:D1]{\text{D1}} }

\newcommand{\eone}{  \hyperref[fig:E1]{\text{E1}} }

\newcommand{\adiag}{ {\hyperref[fig:A1]{\text{A}}} }
\newcommand{\bdiag}{ {\hyperref[fig:B1]{\text{B}}} }
\newcommand{\cdiag}{ {\hyperref[fig:C0]{\text{C}}} }

\newcommand{\aamp}{ (a) }
\newcommand{\bamp}{ (b) }
\newcommand{\camp}{ (c) }
\newcommand{\damp}{ (d) }

\newcommand{\gs}{ g_\text{s} }

\newcommand{\F}{ F }

\newcommand{\cl}{ \text{cl} }

\newcommand{\Qs}{ Q_\text{s} }

\newcommand{\btvar}{ \bt' }

\newcommand{\jimwlk}{ \mathcal{H}_\text{JIMWLK} }

\newcommand{\zmin}{ z_\text{min} }

\series{JYU Dissertations}
\issn{2489-9003}
\serialnumber{xxx}
\isbn[PDF]{xxx-xxx-xx-xxxx-x}

\title{Diffractive Processes at Next-to-Leading Order in the Dipole Picture}
\entitle{foo}
\setauthor{Jani}{Penttala}
\abstract{
Diffractive processes are very sensitive to the target's gluon distribution in the high-energy limit, making them a good candidate for probing the target in the nonlinear region of quantum chromodynamics. The nonlinear effects are expected to eventually lead to gluon saturation which is naturally described in the color-glass condensate (CGC) effective field theory.
While there are strong hints of gluon saturation in the currently available data, no unambiguous signal has been observed.
It is then important to improve the theoretical understanding of processes sensitive to saturation to find 
a clear difference between predictions from the linear and nonlinear regions of QCD.
This includes calculating diffractive processes beyond the leading order in perturbation theory.

In this thesis, we calculate diffractive processes at next-to-leading order (NLO) in the high-energy limit, with an emphasis on exclusive vector meson production and inclusive diffraction in deep inelastic scattering (DIS).
Calculations in the high-energy limit can be done using the dipole picture, the basics of which are briefly reviewed.
This includes using the CGC effective field theory to describe the nonperturbative dipole-target scattering amplitude which appears in practically all calculations in the dipole picture.
The universality of the dipole-target scattering amplitude at NLO is shown numerically, in the sense that the same dipole-target scattering amplitude can be used to describe the data in both massless and massive quark production in inclusive DIS, and also in diffractive processes where exclusive vector meson production is considered.
The analytical NLO calculations of exclusive vector meson production and inclusive diffraction in DIS  are also explained. Exclusive vector meson production is calculated in the nonrelativistic limit for heavy mesons and the limit of large photon virtuality for light mesons.
Also, the importance of including relativistic corrections to the heavy vector meson wave function in exclusive vector meson production is considered.
For inclusive diffraction in DIS, we focus on the NLO corrections to the final state and show how the divergences cancel.

}

\fintitle{Diffraktiiviset prosessit dipolikuvassa alinta seuraavassa kertaluvussa}
\finabstract{
Diffraktiiviset prosessit ovat korkean energian rajalla sensitiivisiä kohdehiukkasen gluonijakaumalle, mistä johtuen niiden avulla voidaan tutkia kohdetta kvanttiväridynamiikan epälineaarisessa alueessa.
Näiden epälineaaristen ilmiöiden odotetaan johtavan gluonisaturaatioon, jota voidaan kuvata luontevasti värilasikondensaatiksi kutsutun efektiivisen kenttäteorian avulla.
Vaikka saatavilla olevassa kokeellisessa datassa onkin vahvoja viitteitä gluonisaturaatiosta, yksiselitteistä merkkiä saturaatiosta ei ole havaittu.
Tämän vuoksi on tärkeää parantaa saturaatiolle sensitiivisten prosessien teoreettista ymmärrystä, jotta pystytään löytämään selkeitä eroja 
kvanttiväridynamiikan lineaarisen ja epälineaarisen alueen ennusteiden väillä.
Diffraktiivisten prosessien laskeminen korkeammille kertaluvuille häiriöteoriassa on osa tätä kehitystä.

Tässä väitöskirjassa lasketaan diffraktiivisia prosesseja korkean energian rajalla alinta seuraavassa kertaluvussa, ja näistä tarkastellaan erityisesti eksklusiivista vektorimesonituottoa sekä inklusiivista diffraktiota syvässä epäelastisessa sironnassa.
Korkean energian rajalla laskuissa voidaan käyttää niin sanottua dipolikuvaa, jonka perusteet käydään lyhyesti läpi.
Tähän kuuluu värilasikondensaattiteorian käyttäminen epäperturbatiivisen dipoliamplitudin kuvaamiseen, joka esiintyy oleellisesti kaikissa dipolikuvassa tehdyissä laskuissa.
Dipoliamplitudin universaalius alinta seuraavassa kertaluvussa näytetään numeerisesti siinä mielessä, että samaa dipoliamplitudia voidaan käyttää sekä massattomien ja massallisten kvarkkien tuoton kuvaamiseen inklusiivisessa syvässä epäelastisessa sironnassa että diffraktiivisissa prosesseissa, joista tarkastellaan eksklusiivista vektorimesonituottoa.
Eksklusiivisen vektorimesonituoton ja inklusiivisen diffraktion analyyttinen lasku alinta seuraavassa kertaluvussa käydään myös läpi.
Näistä eksklusiivinen vektorimesonituotto lasketaan epärelativisella rajalla raskaiden mesonien tapauksessa ja suuren fotonin virtualiteetin rajalla kevyiden mesonien tapauksessa. 
Tämän lisäksi tarkastellaan relativististen korjausten tärkeyttä raskaiden mesonien aaltofunktioon tässä prosessissa.
Inklusiivisen diffraktion tapauksessa keskitytään erityisesti alinta seuraavan kertaluvun korjauksiin lopputilassa sekä osoitetaan divergenssien kumoutuminen.
}

\people{
\item[Author]
    Jani Penttala\\
    Department of Physics\\
    University of Jyväskylä\\
    Finland

\item[Supervisors]
    Dr. Heikki Mäntysaari\\
    Department of Physics\\
    University of Jyväskylä\\
    Finland

    Prof. Tuomas Lappi\\
    Department of Physics\\
    University of Jyväskylä\\
    Finland

\item[Reviewers]
    Prof. Zhongbo Kang\\
    Department of Physics and Astronomy\\
    University of California, Los Angeles (UCLA)\\
    USA

    Dr. Renaud Boussarie\\
    CPHT, CNRS, École polytechnique \\
    Institut polytechnique de Paris \\
    France

\item[Opponent]
    Prof. Lech Szymanowski\\
    National Centre for Nuclear Research (NCBJ)\\
    Poland
}

\begin{document}

\preface

\begin{quote}
    \textit{There is no magic. \\
    There is only knowledge, \\
    more or less hidden.
    }
    \end{quote}

    \hfill
    --- Gene Wolfe, \textit{The Book of the New Sun}
\\ 

\noindent
When I started studying physics everything seemed like magic, incomprehensible to the untrained eye. 
Even after many years of studying in the field a lot of things still do seem like that, but now I feel like I can see a glimpse of meaning behind it all. 
This does not mean that the magic has faded away, but instead 
everything starts to seem even more impressive.
Of course, physics is still far from finished and there are a lot of questions unanswered.
While working on this thesis it has become clear that there is still much work to be done.

As is common in modern research, this thesis has been done in collaboration with many people. 
First of all, I would like to thank my supervisors for their guidance along the journey. 
I have had the joy to work with Dr. Heikki Mäntysaari on projects even before this thesis, and needless to say this work could not have been done without him. 
His help has made the whole thesis project feel possible, even more so than it probably should be.
My thanks go also to Prof. Tuomas Lappi for plentiful insights and for helping us find a way forward when stuck, often giving us that small piece of information that was needed to solve everything.  
My collaborators, Dr. Risto Paatelainen, Dr. Henri Hänninen, and  Dr. Guillaume Beuf, also have my gratitude for finding the time for the many meetings we have had, and for many fruitful discussions which have deepened my understanding.
I would like to thank the reviewers Prof. Zhongbo Kang and Dr. Renaud Boussarie for their kind comments on this thesis.
Prof. Lech Szymanowski has my thanks for agreeing to be my opponent for the public review of the thesis, and I look forward to our discussion at the actual event.
I would also like to thank the Finnish Cultural Foundation and the Research Council of Finland (project 321840 and the Centre of Excellence in Quark Matter, project  346324) for financial support while working on this thesis.

Throughout my studies, I have been inspired and helped by many people.
I am grateful to Prof. Kari J. Eskola and Dr. Hannu Paukkunen for their engaging lectures which have taught me the basic tools needed in particle physics. I have found myself returning to their lecture notes time after time, and they have turned the field from undecipherable hieroglyphs to a delightful puzzle.
Dr. DongJo Kim has my thanks for giving me the introduction to experimental particle physics.
I would also like to thank Prof. Christophe Royon for inviting me to Kansas and for his kind hospitality there, along with many discussions that have helped me get a better understanding of the experimental point of view.
Finally, I would like to thank Dr. Farid Salazar for numerous discussions and also for giving me essential advice for surviving my next destination on this journey.

The long hours of work have been made much more pleasant by the whole community in the physics department, which I am very grateful for.
Special thanks go to my office mates in Holvi:
Lotta, Henri, Topi, Oskari, Mikko, Miha, Mikko, Sami and Carlisle.
Your presence has truly made the office a lively place without a dull moment, and I will certainly miss the days spent there. 
The Holvi alumni have my thanks for showing that there is also life outside academia.
My thanks go also outside the office to the whole Holvi extended universe{\texttrademark}
that shows up to drag me to eat at the grand canonical lunchtime. 
This lunch collaboration has had an exponential growth in recent years, and it has become so big that it is impossible to name everyone here.
At this pace the cafeteria will soon be too small for us.

Of course, my days have not been spent purely writing this thesis.
Jyväskylä has become a really special place for me thanks to all of the wonderful people here.
I should especially thank Harri and Henry with whom I have spent countless days and nights studying and not-studying. 
Working out the exercises with you has been a lot of fun! 
I should also thank the rest of our group of friends, which has somewhat changed over the years: Sami, Kasperi, Jouni, Joona and Tatu. Many wonderful memories have been made, which I will carry on for the rest of my life.

Finally, I would like to express my thanks to my family. You have supported me in everything throughout my life and have allowed me to keep on following this path even further.

Kiitos.

\hfill In Jyväskylä, June 2023

\hfill \textit{Jani Penttala}

\begin{contribution}

The author performed all analytical and numerical calculations in Articles~\cite{relativistic,heavy_long,light,heavy_trans}.
For Article~\cite{structure}, the author implemented the massive structure functions into the existing code for the massless case.
The author wrote the original drafts of the manuscripts for Articles~\cite{light,heavy_trans}, and participated in writing Articles~\cite{relativistic,heavy_long} and editing Article~\cite{structure}.
Chapter~\ref{sec:DDIS} contains discussion about an unpublished work with Guillaume Beuf, Tuomas Lappi, Heikki Mäntysaari and Risto Paatelainen, where the author has performed all analytical calculations.

\end{contribution}

\hypersetup{linkcolor=black}
\mainmatter
\hypersetup{linkcolor=lcolor}

\chapter{Introduction}

The recent century has brought us a vast amount of insight into the fine details of the smallest particles in the world.
Especially interesting are the particles that form the atomic nuclei, protons and neutrons, as it turns out that their internal structure is much more complicated than what was originally thought.
The first observations of this were found in the 1960s when it was realized that they are not actually elementary particles but contain a sub-structure of smaller constituents.
These were later understood to be new particles called quarks, with protons and neutrons both comprised of three quarks each.

When the internal structure of nucleons was studied at higher energies, it was found that this simple quark model is not enough: a vast amount of gluons and virtual quarks also populate the nucleons. 
These constituents of nucleons are collectively called partons.
At even higher energies,
the inner structure is dominated by gluons~\cite{ZEUS:2002xjx}.
This can be understood by
the partons emitting gluons, 
with gluon emission becoming more likely with increasing energy.
This makes the cross sections rise very steeply in energy.
However, this steep increase cannot go on indefinitely.
It was realized that this would eventually break the unitarity of quantum chromodynamics (QCD), and subsequently gluon absorption also has to become important at some point. Indeed, this was found to be a theoretical prediction of QCD, 
which gave rise to saturation models~\cite{Mueller:1994gb}.

While these saturation effects are theoretically well-motivated,
current experiments have not been able to distinguish between saturation and non-saturation models within the uncertainties of the theory and the experiment.
However, there have been hints of saturation in heavy-ion collisions at the Relativistic Heavy-Ion Collider (RHIC)~\cite{Braidot:2011zj,PHENIX:2011puq,STAR:2021fgw} and the Large Hadron Collider (LHC)~\cite{ALICE:2012mj,ALICE:2012yye,CMS:2016itn,LHCb:2021vww}, 
and it is hoped that even clearer signals will be found in electron-nucleus collisions at the future Electron-Ion Collider (EIC)~\cite{AbdulKhalek:2021gbh}.
These experimental endeavors for precise data make studying saturation effects very topical, and
there has been a very active collaboration in the high-energy physics community
to improve the theoretical understanding of saturation physics.
This includes promoting calculations to the next-to-leading order (NLO), and
this thesis is a part of that process focusing on a subset of diffractive processes.
Diffractive processes are especially sensitive to the high-energy gluon distribution of the target, making them a good candidate for trying to find the ``smoking gun'' of saturation.

Diffraction in high-energy physics is briefly considered in Ch.~\ref{sec:dipole_picture}, along with the dipole picture which forms the basis of calculations in the high-energy limit.
Saturation physics is explained in Ch.~\ref{sec:dipole-target_scattering_amplitude} where the interaction of the probe with a highly energetic target is considered.
We also discuss Article~\cite{structure} which highlights the importance of including massive quarks for constraining the nonperturbative interaction with the target.
Chapter~\ref{sec:vector_meson_production} discusses higher-order corrections to exclusive vector meson production. 
We consider first relativistic effects to the heavy vector meson wave function, which is the topic of Article~\cite{relativistic},
and then the next-to-leading order corrections to exclusive vector meson production are discussed in detail.
This discussion is based on the work in Articles~\cite{heavy_long,light,heavy_trans} where the NLO correction to the production of heavy vector mesons is considered in the nonrelativistic limit and the production of light vector mesons in the limit of large photon virtuality.
Finally, inclusive diffraction is the topic of Sec.~\ref{sec:inclusive_DIS} where an unpublished NLO calculation of diffractive structure functions is briefly discussed.

\chapter{Diffractive processes in the dipole picture}
\label{sec:dipole_picture}

\section{High-energy diffraction}

Interest in diffractive processes in high-energy physics began to rise in the 1990s when the results from the HERA collider in DESY started to become available. 
At HERA, it was found that in about $10\%$ of Deep Inelastic Scattering (DIS) events there is a large rapidity gap present in the distribution of the final-state particles~\cite{ZEUS:1993vio,H1:1994ahk}. 
The ratio of such events also remains roughly constant as a function of energy~\cite{H1:1997bdi,ZEUS:1997fox}.
This finding was largely surprising as one would expect the target to break up into a shower of particles filling the rapidity gap, and while this was true for the majority of events, events with a large rapidity gap were expected to get more suppressed with higher energies. In fact, this is true if one expects an exchange of color between the virtual photon and the target~\cite{Bjorken:1992er}.
The large rapidity gap can then be understood as a signature of \textit{diffractive} events where the interaction between the photon and the target is color neutral\footnote{This definition of diffraction has an analog in optics where 
diffraction refers to light meeting an obstacle that has a size comparable to the wavelength of the light. While this analogy between high-energy physics and optics is far from perfect, diffractive processes in the two fields have some properties in common. For example, diffractive cross sections tend to decrease rapidly as a function of the momentum transfer, expressing also \textit{diffractive dips}.}.
Especially, this means that the target and the projectile remain in a color-singlet state which is essential for the large rapidity gap:
as a result of confinement, color-octet final states would start to radiate gluons as their separation grows, and this would produce a plethora of soft particles that would fill the whole rapidity spectrum. 

\begin{figure}[t]
    \centering
    \begin{overpic}[width=0.5\textwidth]{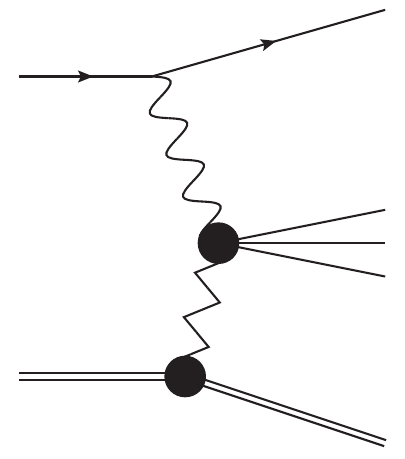}
     \put (5,88) {$e^-$}
     \put (65,95) {$e^{-}$}
     \put (5,22) {$A$}
     \put (70,10) {$A'$}
     \put (67,57) {$X$}
     \put (45,65) {$\gamma^*$}
     \put (83,21){\vector(0,1){15}}
     \put (83,21){\vector(0,-1){15}}
     \put (85,21) {$\Ygap$}
    \end{overpic}
    \caption{
        Diffractive DIS depicted as an exchange of a color-neutral quasiparticle.
        The large rapidity gap $\Ygap$ is an experimental sign of a diffractive process.
    }
    \label{fig:diffractive_process}
\end{figure}
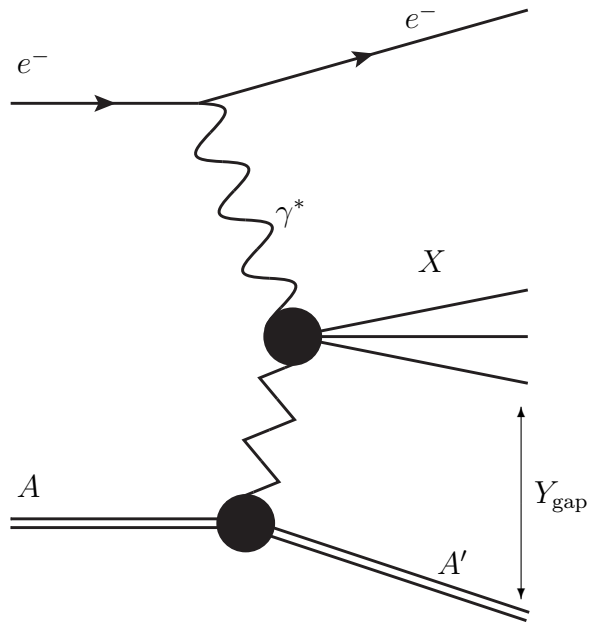

The theory of high-energy diffraction dates back to 1960s when the Regge theory~\cite{Regge:1959mz,Chew:1961ev,Gribov:1961fr} was developed to give a qualitative explanation of a diffractive process as an exchange of the so-called \textit{pomeron}\footnote{A pomeron exchange corresponds to the dominating $C$-parity even interaction. The $C$-parity odd interactions corresponding to the so-called \textit{odderon} exchange are more suppressed.}, shown in Fig.~\ref{fig:diffractive_process}.
A pomeron is a theoretical quasiparticle that has the quantum numbers of a vacuum, and thus this process satisfies the theoretical definition of diffraction as a color-neutral interaction.
Even without understanding the whole structure of the pomeron, the Regge theory was quite successful in describing the general energy and momentum transfer dependence of the cross section~\cite{Barone:2002cv}.
While this explanation via a pomeron exchange is these days mostly of historical interest, the vocabulary dating to the Regge theory is still largely used also in modern literature.

\begin{figure}[t]
    \centering
    \begin{overpic}[width=0.7\textwidth]{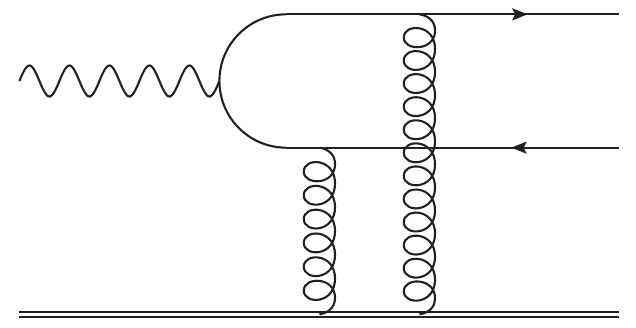}
        \put (10,5) {$A$}
        \put (90,5) {$A'$}
        \put (10,45) {$\gamma^*$}
    \end{overpic}
    \caption{
        The leading-order picture for a color-neutral interaction with the target.
    }
    \label{fig:pomeron_LO}
\end{figure}

After the discovery of QCD it was possible to give the pomeron a more rigorous definition starting from the first principles.
It was understood that at high energies the gluon distribution starts to dominate in the nucleus, and thus the lowest-order color-neutral interaction is an exchange of two gluons between the target and a quark-antiquark pair~\cite{Low:1975sv,Nussinov:1975qb}, shown in Fig.~\ref{fig:pomeron_LO}.
This makes diffractive processes highly sensitive to the gluon distribution of the target.
The enhancement of gluon emission at high energies also means that such a two-gluon exchange is not enough to fully describe the process, but instead one has to resum these gluon exchanges to all orders.
This was first done by the Balitsky--Fadin--Kuraev--Lipatov (BFKL) equation~\cite{Lipatov:1976zz,Kuraev:1977fs,Balitsky:1978ic}.
The resummation of gluon exchanges in the high-energy limit is quite general and applies also to inclusive processes.

These theoretical endeavors to understand the interaction with the target run into problems at even higher energies.
They predict a rapid rise of the gluon distribution which cannot go on indefinitely, as the power-like growth of the cross section predicted by the BFKL equation would eventually break the so-called Froissart--Martin bound~\cite{Froissart:1961ux,Martin:1962rt} for the energy dependence of the cross section. 
This is in violation of the unitarity of the $S$-matrix and thus the probability conservation in particle scattering.
The enhancement of gluon radiation at high energies should eventually be compensated by nonlinear gluon recombination effects such that the energy dependence of the cross section is tamed from power-like to logarithmic, which also results in a slower growth of the target's gluon distribution.
This phenomenon is called {gluon saturation}.
Saturation can be taken into account quite naturally by considering the interaction with the target in terms of nonperturbative \textit{Wilson lines} instead of gluon exchanges, which leads to the celebrated Balitsky--Kovchegov (BK)~\cite{Balitsky:1995ub,Kovchegov:1999yj} and Jalilian-Marian--Iancu--McLerran--Weigert--Leonidov--Kovner (JIMWLK)~\cite{Iancu:2000hn,Iancu:2001ad,Iancu:2001md,Ferreiro:2001qy,Jalilian-Marian:1996mkd,Jalilian-Marian:1997jhx,Jalilian-Marian:1997qno} evolution equations for the interaction with the target.
This is also the framework used in this thesis for describing the energy dependence of the nonperturbative \textit{dipole-target scattering amplitude}.
It turns out that this nonperturbative part can, in general, be treated separately from the rest of the process in the so-called \textit{dipole picture}~\cite{Nikolaev:1990ja,Nikolaev:1991et,Mueller:1993rr,Mueller:1994gb,Mueller:1994jq} which allows for perturbative calculations of processes in the high-energy limit.
The rest of this chapter is devoted to explaining the basics of this approach.

\section{Factorization in the high-energy limit}

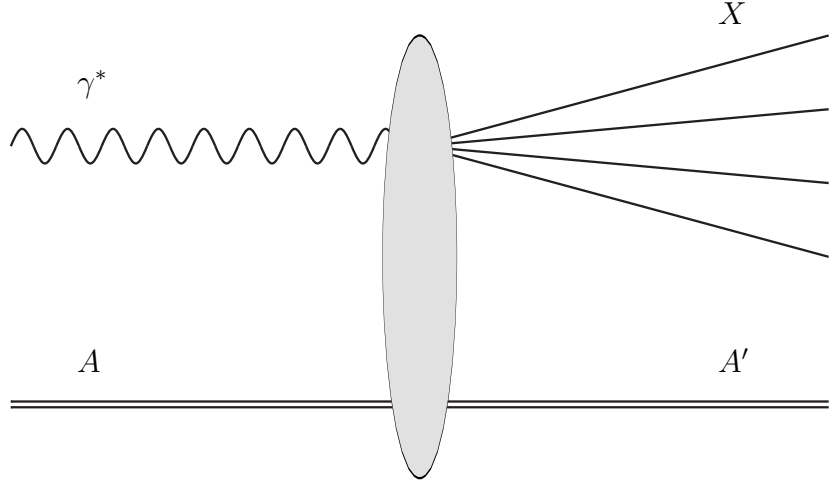
\begin{figure}[t]
    \centering
    \begin{overpic}[width=0.75\textwidth]{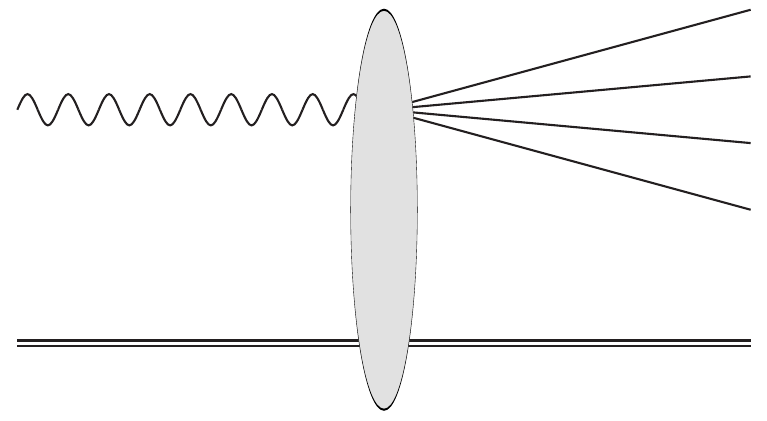}
    \put (10,47) {$\gamma^*$}
    \put (10,14) {$A$}
    \put (85,55) {$X$}
    \put (85,14) {$A'$}
   \end{overpic}
    
    \caption{Diffractive interaction of a virtual photon with a target nucleus.
    }
    \label{fig:photon-target-interaction}
\end{figure}

The class of processes we are interested in involves a photon interacting with a target nucleus as shown in Fig.~\ref{fig:photon-target-interaction}. This photon can be real, as in ultra-peripheral collisions, or virtual, as in DIS. 
In general, the interaction with the target is highly nonperturbative, but in the high-energy limit it can be written in a simplified form. This can be seen by considering the interaction in the light-cone coordinates
\begin{align}
    p^+ &= \frac{1}{\sqrt{2}} ( p^0 + p^3 ) &
    p^- &= \frac{1}{\sqrt{2}} ( p^0 - p^3 ) &
    \pt &= (p^1, p^2).
\end{align}
It is also convenient to use \textit{light-cone quantization} in the light-cone coordinates, which essentially means that the role of the Hamiltonian is played by the minus component of the momentum operator, $\hat{P}^-$, and the time is given by the light-cone time $x^+$. This also means that the components $(p^+, \pt)$ of the momenta are conserved during the interactions but the minus component $p^-$ is not. Instead, it is determined by the on-shell condition $p^- = \frac{m^2 + \pt^2}{2p^+}$.

We will also choose a frame where the photon and the target are moving along the $x^3$-axis, with the photon going in the positive direction.
For heavy nuclei, it is customary to consider the average momentum of a nucleon instead of the nucleus, which corresponds to dividing the momentum of the nucleus by its mass number. This makes it easier to study nuclear effects that modify the simple assumption that the nucleus consists of a number of free nucleons.
The momenta of the photon and the average nucleon are then given by
\begin{equation}
\begin{split}
    q &= (q^+, q^- , \qt) = \left(q^+, -\frac{Q^2}{2 q^+}, \mathbf{0} \right) \\
    P_n &= (P_n^+, P_n^- , \Pt_n) =  \left( \frac{m_n^2}{2P_n^-}, P^-_n, \mathbf{0} \right)
\end{split}
\end{equation}
where $Q^2 = -q^2 \geq 0 $ is the photon virtuality
and $m_n$ is the mass of the nucleon.  
In the high-energy limit the center-of-mass energy is large, $W^2 \gg Q^2, m_n^2$, and thus the energy  is given by
\begin{equation}
 W^2 = ( q+P_n )^2 = 2q^+ P_n^- - \frac{Q^2 m_n^2}{2 q^+ P_n^-} - Q^2 + m_n^2 \approx 2q^+ P_n^-.
\end{equation}

The photon-target interaction is dominated by strong interactions.
As the photon is color neutral, it has to first fluctuate into a quark-antiquark pair which acts as a color dipole, and it can be shown that in the high-energy limit this fluctuation has to happen before the interaction. The reason for this is that the target gets Lorentz-contracted so that the duration of the interaction in the light-cone time is
\begin{equation}
    x^+_\text{interaction} \sim \frac{1}{P^-_n}
\end{equation}
which is much smaller the lifetime of the virtual photon\footnote{By the Heisenberg uncertainty principle deviations from the on-shell condition for the minus momentum can live for a time $ x^+_\text{lifetime} \sim \frac{1}{\abs{\Delta p^-}}  $ where $\Delta p^- = p^- - p^-_{\text{on-shell}}$.}
\begin{equation}
    x^+_\gamma \sim \frac{1}{\abs{q^-}} = \frac{2q^+}{Q^2}.
\end{equation}
This means that fluctuations into other particles during the interaction are suppressed by the factor
$ x^+_{\text{interaction}} /x^+_{\gamma} \sim Q^2/( 2q^+ P^-_n) = Q^2/W^2$, and hence
it is much more likely that the photon has split before the interaction.
This also holds in general for other types of particles interacting with the target as long as the minus momentum of the particle is not too large. Thus the probe, in this case a virtual photon, sees the target as an instantaneous shock wave.
This high-energy condition is usually written in terms of the Bjorken $x$ variable
\begin{equation}
    \label{eq:x}
    x = \frac{Q^2}{2 P_n \vdot q} = \frac{Q^2}{ W^2 + Q^2- m_n^2}  \approx \frac{Q^2}{W^2 + Q^2}.
\end{equation}
We can see that the high-energy limit $W^2 \gg Q^2$ is equivalent to a small Bjorken~$x$.

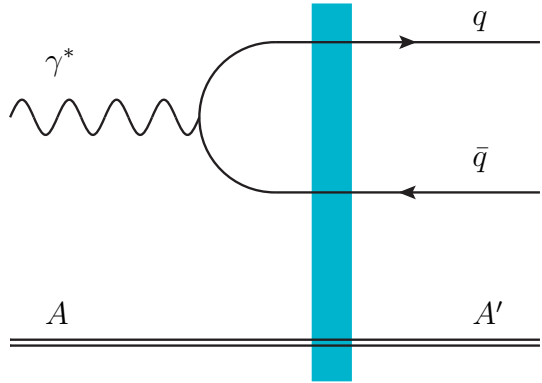
\begin{figure}[t]
    \centering
    \begin{overpic}[width=0.5\textwidth]{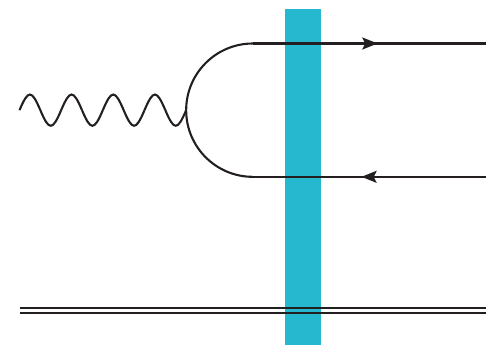}
    \put (10,57) {$\gamma^*$}
    \put (10,12) {$A$}
    \put (85,12) {$A'$}
    \put (85,65) {$q$}
    \put (85,40) {$\bar q$}
   \end{overpic}
    
    \caption{Factorization in the dipole picture at leading order.
    The blue rectangle depicts the nonperturbative interaction with the target.
    }
    \label{fig:dipole_picture_LO}
\end{figure}

The suppression of the Fock state fluctuations during the interaction allows us to factorize the process into three different parts~\cite{Nikolaev:1990ja,Nikolaev:1991et,Mueller:1993rr,Mueller:1994gb,Mueller:1994jq}:
\begin{enumerate}
    \item The photon fluctuates into the Fock state $n$.
    \item The Fock state $n$ interacts with the target.
    \item After the interaction, the Fock state $n$ forms the final state $X$.
\end{enumerate}
At leading order in perturbation theory, this happens by the photon going into the quark-antiquark dipole which interacts with the target, shown in Fig.~\ref{fig:dipole_picture_LO}. This is where the name dipole picture comes from. 
This factorization can be applied very generally to processes in the high-energy limit, and it allows us to write matrix elements as~\cite{Bjorken:1970ah}
\begin{equation}
\label{eq:high-energy_factorization}
\begin{split}
    &\bra{X,A'(x^+=+\infty)}  \hat{S} -1 \ket{ \gamma^*,A(x^+=-\infty) } \\
    &= \sum_{n, n'} \int \dd{[\ps]_n} \dd{[\ps]_{n'}} \bra{X(+\infty)} \ket{n'} \times
    {\bra{n' A'} }  \hat{S} -1 \ket{ n A } \times
    {\bra{n}\ket{\gamma^* (-\infty) }}  
\end{split}
\end{equation}
where $\hat{S}$ denotes the $S$-matrix, and  $n$ and $n'$ correspond to the same Fock state but with possibly different momenta and quantum numbers. 
Here the phase space integrals are defined as
\begin{equation}
\label{eq:PS_momentum}
    \dd{[\ps]_n} = \prod_{i \in n} \dd{\widetilde k_i}
\end{equation}
where 
\begin{equation}
\label{eq:integral_measure_momentum}
    \dd{\widetilde k} 
    = \frac{\dd{k^+} \dd[2]{\kt}}{2k^+ (2\pi)^3} \theta(k^+).
\end{equation}
From this factorization, we see that the nonperturbative interaction with the target is simplified to 
${\bra{n' A'}}  \hat{S} - 1 \ket{ n A } $.
This matrix element is universal in the sense that it is independent from the rest of the process. 
The factors $\bra{X(+\infty)} \ket{n'}, {\bra{n}\ket{\gamma^*(-\infty)} }  $ are in contrast dependent on the initial and final states, and they are most conveniently given by light-cone wave functions which can be calculated using light-cone perturbation theory discussed in Sec.~\ref{sec:lcpt}.

The high-energy limit also simplifies the cross section. The differential element of the cross section for the scattering process $\gamma^* + A \to X + A'$ can be written as
\begin{equation}
    \dd{\sigma^{\gamma^* A \to X A'}}
    = \frac{1}{2 W^2} \dd{[\ps]_{X}}  \dd{[\ps]_{A'}} (2\pi)^4 \delta^4(q + p_A - p_X - p_{A'}) \abs{\mathcal{M}^{\gamma^* A \to X A'}}^2.
\end{equation}
In practice, it can be hard to measure the outgoing nucleus $A'$.
For this reason, one usually considers quantities where the phase space of the nucleus $A'$ has been integrated over. This leads to
\begin{equation}
\begin{split}
    \frac{\dd{\sigma^{\gamma^* A \to X A'}}}{\dd{[\ps]_X}}
    =& \frac{1}{2 W^2} \frac{1}{2 p_{A'}^-} (2\pi) \delta(q^+  - p_X^+ )   \abs{\mathcal{M}^{\gamma^* A \to X A'}}^2 \\
    = & \frac{1}{(2 W^2)^2} 2 q^+ \delta( q^+  - p_X^+ )   \abs{\mathcal{M}^{\gamma^* A \to X A'}}^2
\end{split}
\end{equation}
where the corrections to this equation are of the order $\order{\frac{1}{W^2}}$.
It is also customary to redefine the invariant amplitude such that the factor $1/(2W^2)$ is included in the amplitude. This will be discussed in more detail in Sec.~\ref{sec:eikonal_approximation}.

\section{Light-cone perturbation theory }
\label{sec:lcpt}

Calculating the light-cone wave functions for the initial and final particles can be done in the so-called light-cone perturbation theory~\cite{Kogut:1969xa,Bjorken:1970ah}
which is similar to canonical quantization in old-fashioned perturbation theory.
The main idea is that the fields are quantized at equal light-cone times $x^+$ instead of the equal times $t$ used in the standard formulation of the canonical quantization. This means that the fields satisfy commutation relations such as 
\begin{equation}
    \left[\hat \phi(x^+, x^-, \xt), \hat \pi(y^+ = x^+, y^-, \yt)\right] 
    = i \delta^2(\xt - \yt) \delta(x^- - y^-)
\end{equation}
in the case of a scalar field $\hat \phi(x)$ and its conjugate 
\begin{equation}
    \hat \pi = \frac{\partial \mathcal{L}}{\partial \left(\partial_+ \hat \phi\right)}.   
\end{equation}
This is called the {light-cone quantization} of the fields.
The role of the Hamiltonian in this quantization procedure is then given by the operator
\begin{equation}
    \hat P^- = \int \dd[2]{\xt} \dd{x^-} \left( \hat \pi \partial_+ \hat \phi -\mathcal{L}\right)
\end{equation}
which corresponds to the minus component of the four-momentum.
In general, this leads to a Hamiltonian that has a similar form as in equal-time quantization in the sense that we can divide the light-cone Hamiltonian $\hat P^- = \hat P^-_0 + \hat V$ into the free-field part $\hat P^-_0$ and the interaction term $\hat V$.
Perturbation theory is then done in the interaction picture where the free particles are eigenstates of the operator $\hat P^-_0$ with the on-shell condition 
$p^- =\frac{m^2 + \pt^2}{2p^+}$. 
As a result of this quantization procedure, light-cone 3-momentum $(p^+, \pt)$ is conserved in the interactions whereas the minus component $p^-$ is not\footnote{For a more detailed analysis of the dynamics in the light cone the reader is referred to Ref.~\cite{Brodsky:1997de}.}.
Deriving perturbation theory then follows in a similar way to the old-fashioned perturbation theory.
\textit{A priori}, it is not clear that the physical results of light-cone quantization should then equal with the equal-time quantization, but considering the path integral formulation one can see that the quantization procedure should not have an effect on physical quantities.

While the results of light-cone perturbation theory agree with the equal-time quantization, the Feynman rules are quite different. First of all, light-cone perturbation theory is \textit{time-ordered}. This means that Feynman diagrams with different time-orderings have to be calculated separately,
which is more in line with the old-fashioned perturbation theory that was used in equal-time quantization until the 1960s~\cite{Schwartz:2014sze}. 
Modern Feynman diagram calculations tend to calculate different time-ordered diagrams simultaneously using the covariant perturbation theory.
Doing this for light-cone perturbation theory Feynman rules would result in the same covariant expressions, but this would defeat the advantages we get by light-cone quantization. 
For example, the time-ordering allows one to track the Fock states at each part of the process, and different parts of the process can also be combined using the light-cone wave functions. 
That being said, light-cone perturbation theory also has the major disadvantages of having to calculate much more Feynman diagrams for a single process and losing the explicit Lorentz invariance~\cite{Brodsky:1992mp}.
These are the main reasons why covariant perturbation theory is preferred in almost all calculations.

The main reason for doing light-cone perturbation theory in the high-energy limit is simple: 
as the interaction with the target is instantaneous in the light-cone time $x^+$, it is independent of the minus momenta $p^-$.
The integrals over the minus components of the momenta in covariant perturbation theory can then be done using the residue theorem,
with the different residues corresponding to the time-ordered Feynman diagrams in light-cone perturbation theory~\cite{Kovchegov:2012mbw}.
This leads to describing the initial and final states in terms of the light-cone wave functions as in Eq.~\eqref{eq:high-energy_factorization}.
Thus, starting directly from light-cone perturbation theory one avoids the necessary step of using the residue theorem to calculate the integrals over the minus momenta.

To give the reader a concrete idea of light-cone perturbation theory, we provide a short derivation of the light-cone Feynman rules for calculating light-cone wave functions following Ref.~\cite{Brodsky:1997de}.
To start, consider the following matrix element
\begin{equation}
    \bra{ f(x^+_f)} \ket{ i(x^+_i)} = \mel**{f}{ U(x^+_f, x^+_i)}{i} 
\end{equation}
where the states 
are in the interaction picture and at the light-cone time $x_0^+$ they match the states in the Heisenberg picture: $\ket{n(x_0^+)} \equiv \ket{n}$.
The time-evolution operator in the interaction picture for light-cone quantization is then given by
\begin{equation}
    U(x^+_f, x^+_i) = e^{i \hat P^-_0 x^+_f} e^{-i \hat P^- (x^+_f -x^+_i)} e^{-i \hat P^-_0 x^+_i},
\end{equation}
analogously to equal-time quantization. The light-cone Hamiltonian $\hat P^-$ has been divided into the free and interaction parts $\hat P^- = \hat P^-_0 + \hat V$, and we write the free states as eigenstates of the free part of the Hamiltonian, $\hat P^-_0 \ket{n} = p^-_n \ket{n}$.
These states are normalized as
\begin{equation}
    \braket{n(p^+, \pt)}{n(k^+, \kt)} = 2p^+(2\pi)^3 \delta(p^+- k^+) \delta^2(\pt -\kt).
\end{equation}

The separation of the Hamiltonian into free and interaction parts allows us to write
\begin{equation}
\begin{split}
    \bra{ f(x^+_f)} \ket{ i(x^+_i)} 
    &= e^{ i p^-_f x^+_f - i p^-_i x^+_i} \mel**{f}{e^{-i \hat P^- (x^+_f -x^+_i)}}{i}.
\end{split}
\end{equation}
As $ P^- \geq 0$ for all Fock states, the operator $\hat P^-$ is positive and we can use the residue theorem to rewrite
\begin{equation}
\label{eq:spectral_integral}
    e^{-i \hat  P^- \Delta x^+ } = 
    \int_{-\infty}^{\infty} \frac{\dd{\epsilon}}{2\pi} \frac{i}{\epsilon - \hat P^- + i \delta } e^{-i \epsilon \Delta x^+}
\end{equation}
where $\delta >0$ is an infinitesimal number taken to zero after the integration. Here we have assumed $\Delta x^+ >0$; in the case $\Delta x^+ < 0$ we would have $\frac{-i}{\epsilon - \hat P^- - i \delta}$ instead.

We will then expand $\frac{1}{\epsilon - \hat P^-_0 - \hat V + i\delta}$ as powers of $\hat V$ in order to do perturbation theory, which leads to the expression
\begin{equation}
\begin{split} 
    \mel**{f}{ \frac{1}{\epsilon - \hat P^- + i \delta }}{i}
    =& \mel**{f}{ \frac{1}{\epsilon - \hat P^-_0 + i \delta } + \frac{1}{\epsilon - \hat  P^-_0 + i \delta } \hat V  \frac{1}{\epsilon - \hat P^-_0 + i \delta } +\ldots }{i}\\
    =& \frac{1}{\epsilon - p^-_i + i \delta } \mel**{f}{ 1 + \frac{1}{\epsilon - \hat P^-_0 + i \delta } \hat V   +\ldots}{i}.
\end{split}
\end{equation}
The remaining operators $\frac{1}{\epsilon - P^-_0 + i\delta}$ can be written in terms of their spectral representation
\begin{equation}
    \frac{1}{\epsilon - \hat P^-_0 + i\delta} 
    = \sum_n \int \dd{[\ps]_n} \ket{n} \frac{1}{\epsilon - p_n^- + i \delta} \bra{n}
\end{equation}
so that the original inner product becomes
\begin{equation}
    \begin{split}
         \bra{ f(x^+_f)} \ket{ i(x^+_i)}
         =& \int_{-\infty}^\infty \frac{\dd{\epsilon}}{2\pi} \frac{ i }{\epsilon - p^-_i + i \delta } e^{ i p^-_f x^+_f - i p^-_i x^+_i}  e^{-i \epsilon (x_f^+ -x_i^+)} \\
         & \times 
         \left[ 
         \braket{f}{i} + \frac{1}{\epsilon - p_f^- + i \delta} \mel{f}{\hat V}{i} \right.\\
         & \left. + \sum_n \int \dd{[\ps]_n}  \frac{1}{\epsilon - p_f^- + i \delta} \mel{f}{\hat V}{n} 
         \frac{1}{\epsilon - p_n^- + i \delta} \mel{n}{\hat V}{i} + \ldots
         \right].
    \end{split}
\end{equation}

At this point, we note that we are usually interested in inner products where $x^+_i \to -\infty$. Using the identity
\begin{equation}
\begin{split}
\label{eq:exp_identity}
    \lim_{T \to \infty} \frac{e^{-i T p }}{p + i \delta} 
    = -2i\pi \delta(p)
\end{split}
\end{equation}
we can write
\begin{equation}
    \label{eq:in_wf_matrix_elements}
    \begin{split}
         \bra{ f(x^+_f)} \ket{ i(-\infty)}
         =&  e^{ i (p^-_f-p^-_i) x^+_f } \\
         &\times \left[ 
         \braket{f}{i} +\frac{1}{p^-_i - p_f^- + i \delta} \mel{f}{\hat V}{i} + \ldots
         \right] \\
         \equiv& e^{ i (p^-_f-p^-_i) x^+_f } \left[ \braket{f}{i} +  2p^+_i (2\pi)^3 \delta(p^+_f - p^+_i) \delta^2( \pt_f -\pt_i) 
          \Psi^{i \to f} \right]
    \end{split}
\end{equation}
where 
\begin{equation}
\begin{split}
    &2p^+_i (2\pi)^3 \delta(p^+_f - p^+_i) \delta^2( \pt_f -\pt_i)  \Psi^{i \to f} \\
    &=
    \frac{1}{p^-_i - p_f^- + i \delta} \left[ \mel{f}{\hat V}{i} 
    + \sum_n \int \dd{[\ps]_n} 
     \mel{f}{\hat V}{n}
    \frac{1}{p^-_i - p_{n}^- + i \delta} \mel{n}{\hat V}{i} 
    +\ldots \right]
\end{split}
\end{equation}
and $\Psi^{i \to f}$ is the light-cone wave function for the process $ i \to f$. 
Note that we leave the non-interacting case $\bra{f} \ket{i} $ out of the definition of the wave function. The delta functions come from the conservation of the light-cone 3-momentum $(p^+, \pt)$, and by definition they are not part of the light-cone wave function $\Psi^{i \to f}$.

As a side note, when calculating elements of the scattering matrix we also take the final state to be asymptotic so that $x^+_f = + \infty$. Using the identity~\eqref{eq:exp_identity} again, this leads to
\begin{equation}
\begin{split}
    &\!\!\!\! 
    \mel{f}{\hat S}{i}
    = \braket{f(+\infty)}{i(-\infty)} 
    = \braket{f}{i}  \\
    &+ 2\pi i \delta(p_i^- - p_f^-) 
    \left[ \mel{f}{\hat V}{i} 
    + \sum_n \int \dd{[\ps]_n} 
     \mel{f}{\hat V}{n}
    \frac{1}{p^-_i - p_{n}^- + i \delta} \mel{n}{\hat V}{i} 
    +\ldots \right] \\
    \equiv &  \braket{f}{i} + (2\pi)^4 \delta^4(p_i - p_f ) i \mathcal{M}^{i \to f}
\end{split}
\end{equation}
where $\hat S = \hat 1 + i \hat T$ is the $S$-matrix and $\mathcal{M}^{i \to f}$ is the scattering amplitude for the process $i \to f$.

From this derivation of the light-cone perturbation theory we can read the corresponding Feynman rules. There are some minor variations of the rules corresponding to a different metric and normalization of the light-cone wave function in the literature~\cite{Kogut:1969xa, Kovchegov:2012mbw,Lappi:2016oup,Brodsky:1997de}.
With the conventions of this work, we end up with the following set of rules:
\begin{enumerate}
    \item Draw all of the possible $x^+$-ordered Feynman diagrams corresponding to the process $i \to f$.
    \item Assign an on-shell momentum for each line from left to right such that the light-cone 3-momentum $(p^+, \pt)$ is conserved.
    \item For each vertex, assign a matrix element 
    \begin{equation}
        \mel{\oout}{\hat V}{\iin} =  (2\pi)^3 \delta(p_\oout^+ - p_\iin^+) \delta^2(\pt_\oout -\pt_\iin) \Gamma^{\iin \to \oout}   
    \end{equation}
    where $p_\iin$ is the sum of momenta flowing into the vertex and $p_\oout$ out of the vertex, and $\Gamma^{\iin \to \oout}$ is the Feynman rule for the vertex.
    \item \label{enum:energy_demoninator} For each intermediate state, assign an energy denominator
    \begin{equation}
        \frac{1}{ p^-_i - p^-_\text{intermediate}  + i \delta}
    \end{equation}
    where $p^-_i$ is the sum of minus components for the initial state and $p^-_\text{intermediate}$ for the intermediate state.
    The final state is considered to be an intermediate state.
    \item For each internal line, i.e. lines that are not part of the initial or final state, sum over the helicities and integrate over the momenta $p$ with the phase factors $\dd{\widetilde p}$ from Eq.~\eqref{eq:integral_measure_momentum}.
    \item Include the required symmetry factors for identical particles and the additional factor $(-1)$ for fermion loops and fermion lines beginning and ending in the initial state.
\end{enumerate}
This gives us the expression $2p_i^+  (2\pi)^3 \delta(p_i^+ - p_f^+) \delta^2(\pt_i -\pt_f) \Psi^{i \to f}$ from which one can read the light-cone wave function.
Scattering amplitudes are calculated analogously: leaving out the energy denominator for the final state in Rule~\ref{enum:energy_demoninator} and setting $p_f^- = p_i^-$, one gets the expression 
$  (2\pi)^3 \delta(p_i^+ - p_f^+) \delta^2(\pt_i -\pt_f)\mathcal{M}^{i \to f}$.

The vertices $\Gamma^{\iin \to \oout}$ for QCD are listed in Refs.~\cite{Kovchegov:2012mbw,Lappi:2016oup}. 
This also includes \textit{instantaneous interactions} which can be thought of as additional 4-point interactions between quarks and gluons that appear in the light-cone quantization. These Feynman rules use the light-cone gauge for gluons,
\begin{equation}
    A^+ =0,
\end{equation}
which is a convenient gauge in light-cone perturbation theory. One advantage of this gauge is that it decouples the ghosts from the other particles, meaning that the ghosts can be integrated out trivially.

These rules for calculating the light-cone wave function assume that the initial state is asymptotic at $x_i^+ = - \infty$ and the final state is not an observed state but rather a state that will take part in further scattering processes, such as the interaction with the target in Eq.~\eqref{eq:high-energy_factorization}. These are the standard light-cone wave functions that are usually considered and we will denote them by $\Psi_\iin$ to emphasize that they correspond to the asymptotic \textit{incoming} particle.
In practice, we also need the wave function for the actual observed outgoing state at $x_f^+ = + \infty$, in which case the initial state is not asymptotic but instead a part of a scattering process. To calculate this final-state wave function corresponding to the matrix element $\braket{i(x_i^+)}{f(+\infty)}$, we note that we can repeat our previous derivation for the incoming-state wave function such that the only difference is in Eq.~\eqref{eq:spectral_integral} where the sign of $i \delta$ is now different. This leads to
\begin{equation}
    \label{eq:out_wf_matrix_elements}
    \begin{split}
         \braket{ i(x_i^+)}{ f(+\infty)}
         =&  e^{ i (p^-_i-p^-_f) x^+_i } \\
         &\times \left[ 
         \braket{i}{f} +\frac{1}{p^-_f - p_i^- - i \delta} \mel{i}{\hat V}{f} + \ldots
         \right] \\
         \equiv&  e^{ i (p^-_i-p^-_f) x^+_i } \left[\braket{i}{f} + 2p^+_f (2\pi)^3 \delta(p^+_f - p^+_i) \delta^2( \pt_f -\pt_i) 
          \Psi^{f \to i}_\oout \right]
    \end{split}
\end{equation}
where 
\begin{equation}
\begin{split}
    &2p^+_f (2\pi)^3 \delta(p^+_f - p^+_i) \delta^2( \pt_f -\pt_i)  \Psi^{f \to i}_\oout \\
    &=
    \frac{1}{p^-_f - p_i^- - i \delta} \left[ \mel{i}{\hat V}{f} 
    + \sum_n \int \dd{[\ps]_n} 
     \mel{i}{\hat V}{n}
    \frac{1}{p^-_f - p_{i}^- - i \delta} \mel{f}{\hat V}{i} 
    +\ldots \right].
\end{split}
\end{equation}
From this one can read the Feynman rules for calculating the outgoing-state wave function $\Psi_\oout^{f \to i}$. The only difference to the incoming state is in Rule~\ref{enum:energy_demoninator} where we have now $-i \delta$ instead of $+ i \delta$. The wave function for the matrix element $\braket{f(+\infty)}{i(x_i^+)}$ is then given by the complex conjugate $\left(\Psi^{f \to i}_\oout\right)^*$.

One modification of the above rules concerns self-energy corrections to the asymptotic state. The LSZ reduction formula states that the self-energy corrections should be amputated from the light-cone wave function, and instead they introduce factors $\sqrt{Z_n}$ for the asymptotic state~\cite{Bjorken:1970ah}. This means that the asymptotic state can be written as, in the case of the incoming state, 
\begin{equation}
    \ket{i(-\infty)}  = \sqrt{Z_i} 
    \left[
    \ket{i} +
    \sum_{ n}  \int \dd{ [\ps]_n } 2p_i^+  (2\pi)^3 \delta(p_i^+ - p_n^+) \delta(\pt_i -\pt_n) \Psi^{i \to n}_\iin \ket{n}
    \right]
    .
\end{equation}
Similar factors appear for the outgoing state.

Finally, we would like to mention that there are several different normalization conventions for the light-cone wave functions which correspond to different phase space integration measures. In this thesis, we use the conventions of Eqs.~\eqref{eq:PS_momentum} and \eqref{eq:integral_measure_momentum}. 
These differ from Articles~\cite{relativistic,heavy_long,light,heavy_trans} which follow the notation from Ref.~\cite{Kowalski:2006hc}. There the integration measure for the plus momenta was defined as $\frac{\dd{z_i}}{4\pi}$ instead of $\frac{\dd{z_i}}{4 \pi z_i} = \frac{\dd{k_i^+}}{4 \pi k_i^+}  $, where $z_i = k_i^+ / q^+$ are the plus-momentum fractions of the particles. This introduces an additional factor $ \prod_{i \in n} \frac{1}{\sqrt{z_i}} $ into the wave functions.

\section{Eikonal approximation}
\label{sec:eikonal_approximation}

\begin{figure}[t]
    \centering
\begin{overpic}[width=0.5\textwidth]{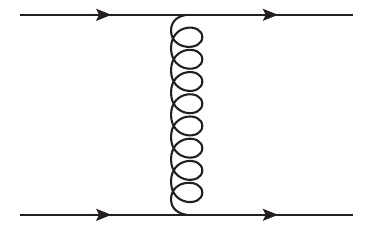}
 \put (20,65) {$q, h_q, c_q$}
 \put (70,65) {$q', h_{q'}, c_{q'}$}
 \put (20,10) {$P, h_P, c_P$}
 \put (70,10) {$P', h_{P'}, c_{P'}$}
 \put (60,45){\vector(0,-1){30}}
 \put (65,30){$k$}
\end{overpic}
    \caption{$t$-channel for the quark-quark scattering.}
    \label{fig:qq_scattering}
\end{figure}
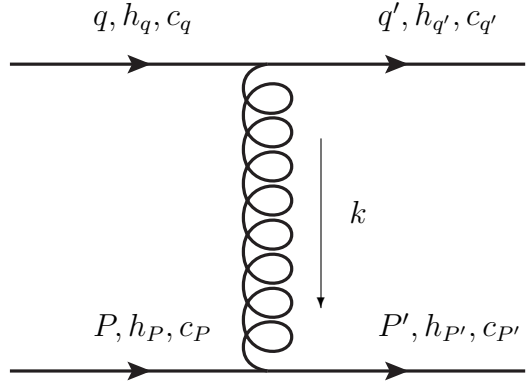

The high-energy limit can be used to also simplify the interaction with the target.
To understand this, consider a quark-quark scattering process where the ``probe'' quark has a momentum $q$ and the ``target'' quark has a momentum $P$ as shown in Fig.~\ref{fig:qq_scattering}\footnote{This example can be found in Ref.~\cite{Kovchegov:2012mbw}.
However, we do not assume that the individual components $q^+$ and $P^-$ are large as we wish to present a boost-invariant motivation for the eikonal approximation.
We also allow for non-zero transverse components $\qt$ and $\Pt$ which is generally the case if one considers multiple gluon exchanges. 
}.
The two quarks collide with a high longitudinal momentum such that the center-of-mass energy $s = (q+P)^2 \approx 2 q^+ P^- \gg m_q^2, m_P^2, \qt^2, \Pt^2$ is large. 
This also means that the interaction happens mainly through the $t$-channel in Fig.~\ref{fig:qq_scattering} as the $s$-channel is suppressed by $1/s$.
Demanding that the initial and final particles are on-shell one can show that the plus and minus momentum exchange is small compared to the transverse momentum exchange:
\begin{equation}
\label{eq:k_momentum_transfer}
    \begin{aligned}
        k^+ &= (q-q')^+ = (P'-P)^+ \approx \frac{q^+}{s} ( \kt^2 + 2 \kt \vdot \Pt), \\
        k^- &= (q-q')^- = (P'-P)^- \approx \frac{P^-}{s} (-\kt^2 +2 \kt \vdot \qt). 
    \end{aligned}
\end{equation}
We can then assume that $q^{\prime +} = q^+$ and $P^{\prime -} = P^-$ as the corrections are suppressed by $\kt^2/s \ll 1$.

Using the Feynman rules from covariant perturbation theory, the scattering amplitude is given by
\begin{equation}
    \label{eq:quark-quark_scattering}
    i \mathcal{M} = g_s^2 t^a_{c_{q'} c_q} t^a_{h_P h_{P'}} \frac{1}{\kt^2} \bar u_{h_q'}(q') \gamma^\mu u_{h_q}(q) \bar u_{h_P'}(P') \gamma^\nu u_{h_P}(P) D_{\mu \nu}(k)
\end{equation}
where $D_{\mu \nu}(k)$ is the gluon propagator. 
To stay consistent with the rest of this thesis,
we consider the gluon propagator in the light-cone gauge where
\begin{equation}
    D^{\mu \nu}(k) = g^{\mu \nu} - \frac{n^\mu k^\nu + n^\nu k^\mu}{n \vdot k}
    = 
    - \left(
    \begin{matrix}
        0 & 0 & 0 & 0 \\
        0 & 2 \frac{k^-}{k^+} & \frac{k^1}{k^+} & \frac{k^2}{k^+} \\
        0 & \frac{k^1}{k^+} & 1 & 0 \\
        0 &  \frac{k^2}{k^+} & 0 & 1 
    \end{matrix}
    \right)
\end{equation}
with $n^\mu = \delta^{\mu - }$. 

To calculate the scattering amplitude \eqref{eq:quark-quark_scattering} we also need to specify a basis for the spinors.
It turns out that in light-cone perturbation theory calculations it is convenient to use a basis where quantities are boost invariant, and with this motivation in mind we choose to use  Lepage--Brodsky basis~\cite{Lepage:1980fj}:
\begin{equation}
\begin{aligned}
\label{eq:LB_spinors}
    u_{+}(\vec k) &=
    \frac{1}{\sqrt{2(E+k^3)}} 
    \begin{pmatrix}
    E+k^3+m \\
    k_R \\
    E+k^3-m\\
    k_R
    \end{pmatrix}
    &
    u_{-}(\vec k) &=
    \frac{1}{\sqrt{2(E+k^3)}} 
    \begin{pmatrix}
    -k_L \\
    E+k^3+m \\
    k_L\\
    -(E+k^3-m)
    \end{pmatrix}
    \\
    v_{+}(\vec k) &=
    \frac{1}{\sqrt{2(E+k^3)}} 
    \begin{pmatrix}
    -k_L \\
    E+k^3-m \\
    k_L\\
    -(E+k^3+m)
    \end{pmatrix}
    \hspace{-5.8pt}
    &
    v_{-}(\vec k) &=
    \frac{1}{\sqrt{2(E+k^3)}} 
    \begin{pmatrix}
    E+k^3-m \\
    k_R \\
    E+k^3+m\\
    k_R
    \end{pmatrix}
\end{aligned}
\end{equation}
where $k_L = k^1-ik^2 = \abs{\kt} e^{-i \varphi_\kt}$ and $k_R = k^1 + i k^2 = \abs{\kt} e^{i \varphi_\kt}$.
Longitudinal boosts on these spinors do not mix the helicity states but only change the momenta so that
\begin{equation}
    S(\Lambda) u_h(\vec k) = u_h (\Lambda \vec k)
\end{equation}
where $S(\Lambda)$ is the action of the longitudinal boost $\Lambda$ in the spinor representation of the Lorentz group.
This is the main reason why this basis is ubiquitously used in the light-cone formalism.
The spinors~\eqref{eq:LB_spinors} are also the eigenstates of the so-called \emph{light-cone helicity} which corresponds to the helicity in the ``infinite-momentum frame''~\cite{Soper:1972xc}.

With the basis~\eqref{eq:LB_spinors} one can find the following scaling relations for the spinor elements:
\begin{equation}
\begin{aligned}
    \bar u(q') \gamma^+ u(q) &\sim q^+ \\
    \bar u(q') \gamma^- u(q) &\sim q^- \\
    \bar u(q') \gamma^i u(q) &\sim \qt^i
\end{aligned}
\end{equation}
and similarly for $\bar u(P') \gamma^\nu u(P)$. 
This allows us to find the dominating terms in the high-energy limit so that Eq.~\eqref{eq:k_momentum_transfer} becomes
\begin{equation}
    i \mathcal{M} \approx -g_s^2
    t^a_{c_{q'} c_q} t^a_{h_P h_{P'}} 
     \frac{1}{\kt^2} \bar u_{h_q'}(q') \gamma^+ u_{h_q}(q) \bar u_{h_P'}(P') \gamma^i u_{h_P}(P) D^{- i}(k)
\end{equation}
where the other terms are suppressed by $1/s$. Explicitly evaluating the terms remaining terms this leads to\footnote{Strictly speaking, for the ``target'' one has to use a different basis of spinors so that the sign of $k^3$ in Eq.~\eqref{eq:LB_spinors} is swapped. This is allowed as one can use different spinors bases for different spinors in the process.}
\begin{equation}
    i \mathcal{M} = 2s \delta_{h_q h_q'} \delta_{h_P h_P'} g_s^2 t^a_{c_{q'} c_q} t^a_{h_P h_{P'}} 
    \frac{1}{\kt^2}.
\end{equation}
We can now read several things from this expression for the scattering amplitude that apply very generally to particle scattering in the high-energy limit:
\begin{enumerate}
    \item The invariant amplitude is proportional to the center-of-mass energy $s$.
    \item The helicities of the target and the probe are conserved.
    \item The momentum transfer is dominated by the transverse component of the 4-momentum. From Eq.~\eqref{eq:k_momentum_transfer} we see especially that the plus momentum of the probe is conserved up to corrections in $1/s$, i.e. $q^+ \approx q^{\prime +}$.
    \item The invariant amplitude depends only on the transverse momentum exchange $\kt^2$.
\end{enumerate}
In practice, the interaction with the target is not this simple. However, the properties noted above are quite general, and it can be shown that they are also satisfied for gluon targets.
These properties then form the basis of the \textit{eikonal approximation}\footnote{Similarly to diffraction, the term ``eikonal'' also has its origin in optics. From Greek \foreignlanguage{greek}{εἰκών} (``image''), the eikonal approximation refers to the assumption that light encountering an object travels in a straight line, forming the ``image'' of the object. This is generally valid as long as the wavelength of the light is much smaller than the object, and an analogous thing is true in high-energy physics: a photon with a ``wavelength'' $x^-_\gamma \sim 1/q^+$ much smaller than the size of the target $x^-_A \sim 1/P^+_N = 2P^-_N/m_N^2$ scatters eikonally. }
in the high-energy limit which states that the interaction with the target can be written as
\begin{equation}
    \label{eq:eikonal_approximation}
    i \mathcal{M}^{n + A \to n' + A'}
    \approx 
    2 s \times 2 p_n^+ (2\pi) \delta(p_n^+ - p_n^{\prime +}) \delta_{h_n h_n'} \delta_{h_A h_A'} S(\kt)
\end{equation}
where $n$ can be a quark, an antiquark or a gluon, and $S(\kt)$ is a color matrix that depends only on the transverse momentum transfer $\kt = \pt_n' - \pt_n$. 
The fact that $S(\kt)$ depends only on the momentum transfer and not on the individual momenta $\pt_n$ and $\pt_n'$ suggests that we should take a Fourier transform in the transverse plane. Writing
\begin{equation}
    \int \frac{\dd[2]{\pt_n} \dd[2]{\pt_n'}}{(2\pi)^4} e^{i \pt_n \vdot \xt_n} e^{-i \pt_n' \vdot \xt_n'} S(\kt) 
    = \delta^2(\xt_n - \xt_n') 
    \int \frac{\dd[2]{\kt}}{(2\pi)^2} e^{-i \kt \vdot \xt_n}  S(\kt) 
\end{equation}
we see that in the position space the transverse coordinates do not change during the interaction.
This can be understood by noting that the interaction with the target is instantaneous and thus the change in the transverse coordinate has to be very small.

Combining the eikonal approximation~\eqref{eq:eikonal_approximation} with the factorization in Eq.~\eqref{eq:high-energy_factorization}, we can write the invariant amplitude as
\begin{equation}
    \label{eq:general_amplitude}
    i\mathcal{M}^{\gamma^* A \to X A'}
    =\sum_n \int \dd{[\widetilde \ps]_n}  2 q^+  (2\pi) \delta(q^+ -p_X^+)
    e^{-i \bt_n \vdot \Deltat}
    \widetilde \Psi^{\gamma^* \to n}_\iin
    \left(\widetilde \Psi^{X \to n}_\oout\right)^\ast \bigl( 1 -  S^{(n)}_A(\xt_i) \bigr)
\end{equation}
where $\widetilde \Psi$ denotes the Fourier transform of the light-cone wave function, $S^{(n)}_A(\xt_i)$ the eikonal approximation for the interaction of the Fock state $n$ with the target $A$,
and the integration measure in the mixed space is given by
\begin{equation}
    \dd{[\widetilde \ps]_n} = \prod_{i \in n} \left( \dd[2]{\xt_i} \frac{\dd{k^+_i}}{(2\pi) 2k^+_i} \right).
\end{equation}
The Fourier-transformed wave functions are defined by 
\begin{equation}
    \begin{split}
    &e^{i \bt_n \vdot \sum_{i \in n} \pt_{i}}\widetilde \Psi^{n \to m}(\pt_{i}, z_{in};\xt_{i}, z_{im})\\
    &= \int \prod_{i \in m} 
    \left( \frac{\dd[2]{\kt_{i}} }{(2\pi)^{2}} \right)
    (2\pi)^2 \delta^2\left(\sum_{i \in m} \kt_{i} - \sum_{i \in n} \pt_{i}\right)
        e^{\sum_{i \in m} i \kt_{i} \cdot \xt_i} 
        \Psi^{n \to m}(\pt_{i}, z_{in};\kt_{i}, z_{im})
    \end{split}
\end{equation}
where $\bt_n = \sum_{i \in m} z_{im} \xt_{i}$ is the impact parameter.
Here we have denoted the momenta for the state $n$ by $\pt_i$ and $\pt^+_i = z_{in} q^+ $, and for the state $m$ by $\kt_i$ and $\kt^+_i = z_{im} q^+ $.
Using the momentum fractions $z$ instead of the plus momenta $p^+$ makes the boost invariance of the wave functions more explicit. 
It should be noted that this mixed space of coordinates $(\xt_i, z_i)$ is very convenient for the eikonal approximation as these do not change during the interaction with the target.
Also, here we have taken the factor $e^{-i \bt \vdot \sum \pt_i}$ out from the Fourier-transformed wave function as it turns out that with this definition the wave function $\widetilde \Psi$ depends only on the dipole sizes $\xt_{ij} = \xt_i - \xt_j$ and the relative momenta $\Pt_{ij} \equiv z_{jn} \pt_i - z_{in} \pt_j $ but not on the impact parameter $\bt$ or the total transverse momentum $\sum_i \pt_{i}$.
This makes the dependence on the transverse momentum transfer $\Deltat = \sum_{i \in X} \pt_i$ in Eq.~\eqref{eq:general_amplitude} explicit, and thus the dependence on the Mandelstam variable $t \approx - \Deltat^2$ is simple to calculate.

We have also rescaled the amplitude by the common factor $2s = 2 W^2$ in Eq.~\eqref{eq:eikonal_approximation}. With the rescaled invariant amplitude from Eq.~\eqref{eq:general_amplitude}, the cross section now reads
\begin{equation}
    \label{eq:general_cross_section}
    \begin{split}
        \frac{\dd{\sigma^{\gamma^* A \to X A'}}}{\dd{[\ps]_X}}
        =&  2 q^+ (2\pi) \delta( q^+  - p_X^+ )   \abs{\mathcal{M}^{\gamma^* A \to X A'}}^2.
    \end{split}
\end{equation}
Eqs.~\eqref{eq:general_amplitude} and \eqref{eq:general_cross_section} hold in general in the high-energy limit, with the possible modification that in Eq.~\eqref{eq:general_amplitude} we left out the non-interaction matrix element $\braket{n}{X}$ for the case $X = n$ out for simplicity (see Eq.~\eqref{eq:out_wf_matrix_elements}).
These equations will be used to calculate exclusive vector meson production in Ch.~\ref{sec:vector_meson_production} and inclusive diffraction in DIS in Ch.~\ref{sec:DDIS}.

\chapter{Dipole-target scattering amplitude}
\label{sec:dipole-target_scattering_amplitude}

To calculate any production amplitude for the process $\gamma^* + A \to X + A'$ in the dipole picture one still needs to understand the nonperturbative interaction with the target. At  leading order, this interaction happens with the quark-antiquark dipole and the target, and it is described by the dipole-target scattering amplitude or the \textit{dipole amplitude} for short. It turns out that interactions with even higher-order Fock states consisting of quarks and gluons can be given in terms of the dipole amplitude at certain limits, which will be discussed in Sec.~\ref{sec:rapidity_evolution} in more detail. Thus, a thorough understanding of the dipole amplitude is important for an accurate description of processes in the dipole picture.

\section{Target as a classical color field}
\label{sec:MV}

While the eikonal limit~\eqref{eq:eikonal_approximation} significantly simplifies the interaction with the target, it is too general to give an actual model for the dipole amplitude.
For this we need to consider the actual physical situation of the scattering and take input from QCD.
The main idea is that at high energies the gluon distribution starts to dominate in the target and we can thus neglect the quark contribution. 
The second idea is to note that when the target is moving at a high velocity, it gets Lorentz-contracted such that the gluon field density $\mu^2$ is very high, $\mu^2 \gg \lqcd^2$. This means that one is generally in the weak-coupling region $\as(\mu^2) \ll 1$ which allows us to treat the gluons as classical color fields. 
This is the basic starting point in the McLerran--Venugopalan model for high-energy scattering~\cite{McLerran:1993ka,McLerran:1993ni,McLerran:1994vd}.
We can then model the target as a color field $A_\cl$ that is solved from the classical Yang-Mills equation
\begin{equation}
    \label{eq:CYM}
    [D_\mu, F^{\mu \nu}_\cl] = J^\nu
\end{equation}
where $D_\mu = \partial_\mu - i g A_\mu^\cl$ is the covariant derivative, $F_\cl^{\mu \nu} = [D^\mu, D^\nu]$ is the field strength tensor and $J^\nu(x)$ is the color current.
As the target is moving in the minus direction with a large momentum,
we can model the current as
\begin{equation}
    J^\nu(x) = \delta^{- \nu} \rho(x^+, \xt)
\end{equation}
so that only the minus component of the current is relevant. Here $\rho(x^+, \xt) = t^a \rho^a(x^+, \xt)$ is the color charge density of the target, and the high-energy limit ensures that it is independent of the coordinate $x^-$ and sharply peaked for the time of the interaction in $x^+$.
These assumptions allow us to solve the field $A^\mu_\cl$ from Eq.~\eqref{eq:CYM}, with the solution
\begin{equation}
    \label{eq:classical_color_field}
    \begin{aligned}
        \boldsymbol{\nabla}^2 A^-_\cl(x^+, \xt) &=  -\rho(x^+, \xt) \\
        A^+_\cl(x^+, \xt) &=  A^i_\cl(x^+, \xt) = 0
    \end{aligned}
\end{equation}
in the light-cone gauge $A^+ = 0$. 
We note that this solution is not unique as there is still some gauge freedom left. 
Often 
it is more convenient to write a solution such that only the more physical transverse components of the gluon field remain, as the minus component $A^-$ is not actually a dynamical field but a boundary condition for the gluon field~\cite{McLerran:1993ni}.
However, for the purpose of this section it is simpler to work with the solution in Eq.~\eqref{eq:classical_color_field}.

The solution~\eqref{eq:classical_color_field} for the target color field allows us to calculate the interaction with the target. This is done with the equations of motion in the background field $A_\cl$. For a quark field $\psi$ this is given by
\begin{equation}
    (i\slashed{D} - m) \psi = 0.
\end{equation}
Note that the high-energy limit guarantees that during the interaction with the target the coordinates $(x^-, \xt)$ are roughly constant, and we are only interested in the change in the plus direction. Neglecting the other derivative terms and also the mass of the quark, this leads to the equation
\begin{equation}
    \gamma^+ \partial_+ \psi = ig A_\cl^- \gamma^+ \psi.
\end{equation}
The solution for this equation is given in terms of a {Wilson line}
\begin{equation}
    \psi(x^+_f, \xt) = V(x^{ +}_f, x^+_i, \xt)  \psi(x^+_i, \xt),
\end{equation}
\begin{equation}
    V(x^+_f, x_i^+, \xt) = \mathcal{P} \exp(ig \int_{x_i^+}^{x_f^+} \dd{y^+} A_\cl^{a-}(y^+, \xt) t^a)
\end{equation}
where $\mathcal{P}$ denotes path-ordering for the integral and $t^a$ are the color matrices in the fundamental representation. Similarly, for conjugate fields $\overline \psi$ one gets the Hermitean conjugate $V^\dag(x^+_f, x_i^+ \xt)$, and for gluon fields an adjoint Wilson line
\begin{equation}
    \label{eq:adjoint_wilson}
    U(x^{ +}_f, x_i^+,  \xt) = \mathcal{P} \exp(ig \int_{x_i^+}^{x^+_f} \dd{y^+} A_\cl^{a-}(y^+, \xt) T^a)
\end{equation}
where the color matrices $T^a$ are now in the adjoint representation.

The connection to the dipole amplitude is that the eikonal interaction between a quark-antiquark pair and the target is given in terms of the Wilson lines as
\begin{equation}
    \begin{split}
    &\mel{q_a(x^+_f, \xt_0') \bar q_b (x^+_f, \xt_1')}{\hat S_A}{ q_c(x^+_i, \xt_0) \bar q_d(x^+_i, \xt_1)} \\
    &= \delta^2(\xt_0' - \xt_0) \delta^2(\xt_1' - \xt_1)
    V_{ac}(x^+_f, x^+_i, \xt_0) V_{db}^\dag(x^+_f, x^+_i, \xt_1)
    \end{split}
\end{equation}
where 
the subscripts refer to the color indices of the particles.
This means that the particles only get color-rotated during the interaction with the target.
In diffractive scattering the initial and final states are color singlets, so that in total the expression for the dipole amplitude leads to
\begin{equation}
    \label{eq:dipole_amplitude_from_Wilson_lines}
    \begin{split}
    &\frac{\delta^{ab}}{\sqrt{N_c}} \frac{\delta^{cd}}{\sqrt{N_c}} 
    \mel{q_a (\xt_0') \bar q_b (\xt_1') }{ 1- \hat S_A}{q_c (\xt_0) \bar q_d(\xt_1)}\\
    &= \delta^2(\xt_0 - \xt_0' )  \delta^2(\xt_1 - \xt_1' )\left\{ 1 - \frac{1}{N_c} \Tr[V(\xt_0) V^\dag(\xt_1)] \right\}.
    \end{split}
\end{equation}
Here we have also taken $x^+_f \to + \infty$, $x^+_i \to -\infty$ as the dependence of the Wilson lines on the light-cone times $x^+_f$, $x^+_i$ is very slow, arising from the fact that $A_\cl^-(x^+)$ is highly suppressed for times $x^+ \neq 0$.
A similar equation can be derived for any Fock state interacting with the target.

Using Eq.~\eqref{eq:dipole_amplitude_from_Wilson_lines} still requires that we know the target color density $\rho$ to solve for the field $A_\cl$.
This is, however, a nonperturbative quantity and needs to be modeled. One very successful model is the Gaussian approximation~\cite{McLerran:1993ka,McLerran:1993ni,McLerran:1994vd} which assumes that the target is a linear combination of all possible color configurations with a Gaussian weight. This corresponds to taking an average of expressions like Eq.~\eqref{eq:dipole_amplitude_from_Wilson_lines} with
\begin{equation}
    \label{eq:dipole_path_integral}
    \left\langle 1 - \frac{1}{N_c} \Tr[V(\xt_0) V^\dag(\xt_1)]  \right\rangle
    = \int \mathcal{D} \rho \, \W[\rho] \left\{ 1 - \frac{1}{N_c} \Tr[V(\xt_0) V^\dag(\xt_1)]  \right\}
\end{equation}
where
\begin{equation}
    \label{eq:gaussian_approximation}
    \W[\rho] =  \exp(- \int_{-\infty}^\infty \dd{x^+} \int \dd[2]{\xt} \frac{\Tr[ \rho(x^+, \xt)^2]}{\mu^2(x^+, \xt)})
\end{equation}
is the weight for the color density $\rho$, and $\mu^2$ is the average color charge squared per unit volume in $\dd{x^+} \dd[2]{\xt}$ and unit color.
The motivation for such a model is the central limit theorem which states that averages from probability distributions with a finite variance tend to a Gaussian distribution. 
It is also useful to note that the Gaussian approximation leads to the following correlator for
the  color densities $\rho$:
\begin{equation}
    \langle \rho^a(x^+, \xt) \rho^b(y^+, \yt) \rangle 
    =  \delta^{ab} \mu^2(x^+, \xt) \delta(x^+ - y^+) \delta^2(\xt- \yt).
\end{equation}
This states that the color densities at different light-cone times and transverse coordinates are not correlated, which is a natural assumption if one considers the target as a collection of point-like color charges.

With the Gaussian approximation Eq.~\eqref{eq:gaussian_approximation}, the path integral in Eq.~\eqref{eq:dipole_path_integral} can be evaluated to give the dipole amplitude an expression in terms of the density $\mu^2(x^+, \xt)$.
Assuming that the density $\mu^2(x^+, \xt)$ varies very slowly in terms of the transverse coordinate $\xt$, this can be written as~\cite{Iancu:2003xm}
\begin{equation}
    \label{eq:MV_dipole}
    N_{01} \equiv \left\langle 1 - \frac{1}{N_c} \Tr[V(\xt_0) V^\dag(\xt_1)]  \right\rangle 
    \approx 1- \exp( - \frac{\xt_{01}^2 \Qs^2}{4} \ln\frac{1}{\Lambda^2 \xt_{01}^2})  .
\end{equation}
Here $\Lambda $ is an infrared regulator and 
$\Qs^2 =  \as \cf \int \dd{x^+} \mu^2(x^+, \btvar)$
is the so-called \textit{saturation scale}. 
The variable $\btvar = \frac{1}{2}(\xt_0 + \xt_1)$ is the average of the quark and antiquark coordinates,
and hence the saturation scale depends on the transverse density profile of the target.
This also means that the saturation scale is enhanced by the mass number of the target, as the density has only a slight dependence on the mass number but the radius behaves like $R \sim A^{1/3}$.
This suggests that the saturation scale is enhanced by $\qs^2 \sim A^{1/3}$ for heavy nuclei.

Several things can be noted from the form of the dipole amplitude~\eqref{eq:MV_dipole}.
For small dipole sizes $\xt_{01}^2$ the dipole amplitude behaves like $N_{01} \approx \frac{1}{4}  \xt_{01}^2 \Qs^2$, but for large dipoles the saturation scale tames the growth to the \textit{black-disk limit} $N_{01} \to 1$.
This behavior is important for the unitarity of the process:
if the black-disk limit were violated, the unitarity of the $S$-matrix would be broken~\cite{Iancu:2003xm}.
It is also useful to note that the dipole amplitude from the MV model satisfies the form of the eikonal approximation in Eq.~\eqref{eq:eikonal_approximation}: it depends only on the transverse coordinates and color indices of the scattering particles, and the mixed-space coordinates $(\xt_i, z_i)$ and 
the helicities are conserved in the interaction.

\section{High-energy evolution of the dipole amplitude}
\label{sec:rapidity_evolution}

So far, the dependence on the center-of-mass energy $W^2$ has actually dropped out of Eqs.~\eqref{eq:general_amplitude} and \eqref{eq:general_cross_section}.
This is a little bit puzzling, as generally cross sections tend to increase with energy. 
It turns out that indeed the dipole amplitude should have an energy dependence, and this will be important for higher-order equations to be finite.
This energy dependence is related to the larger phase space available for gluon emission.
This can be seen explicitly in next-to-leading order calculations where the emission of slow gluons with $z_g \ll 1$ starts to dominate. Resumming these gluon emissions leads to the JIMWLK equation for the energy dependence of the dipole amplitude.

The JIMWLK equation can be derived by considering the weight $\W[\rho]$ of the target's color configuration at some rapidity $Y$ which is related to the energy of the system, $W^2 \sim e^Y$.
Considering then a Fock state $n$ interacting with the target, we can write the interaction using Eq.~\eqref{eq:dipole_path_integral} as
\begin{equation}
    \langle  \hat{\mathcal{O}} \rangle = \int \mathcal{D} \rho \, \W[\rho] \,\mathcal{O} [\rho]
\end{equation}
where the operator $\hat {\mathcal{O}}$ consists of the Wilson lines for the interacting Fock state $n$.
The Fock state $n$ may emit gluons with a momentum fraction $z_g$.
It turns out that the gluon emission is enhanced by $1/z_g$ for gluons with a small momentum fraction, with the integral over $z_g$ diverging at $z_g \to 0$. 
However, at small $z_g$ the assumptions for the validity of the eikonal approximation break down as at some point the invariant mass of this $n+ g$ state becomes comparable with the energy:
\begin{equation}
    M_{n+g}^2 \approx  \frac{\kt_g^2}{z_g} \gtrsim W^2.  
\end{equation}
We should then limit the $z_g$-integral by some cut-off $\zmin \sim 1/W^2 \sim e^{-Y}$ so that we are only working in a region where the eikonal approximation is valid.
The idea is that the slow gluons with $z_g < \zmin$ are then defined as a part of the target.

\begin{figure}
        \centering
        \begin{align*}            
        &
        \begin{array}{l}
            \begin{overpic}[width=0.40\textwidth]{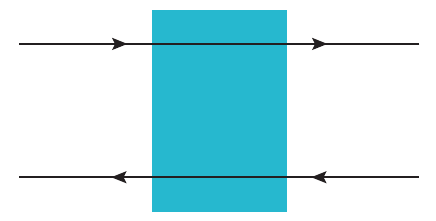}
                \put (39,23) {$Y +\delta Y$}
            \end{overpic}
        \end{array}
        -
        \begin{array}{l}
            \begin{overpic}[width=0.40\textwidth]{diags/jimwlk_0.pdf}
                \put (48,23) {$Y $}
            \end{overpic}
        \end{array}
        \\
        =&
    \begin{array}{l}
        \includegraphics[width=0.14\textwidth]{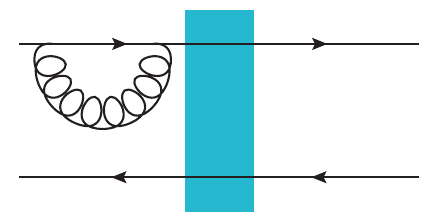}
    \end{array}
        +   
    \begin{array}{l}
        \includegraphics[width=0.14\textwidth]{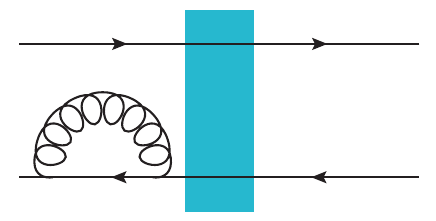}
    \end{array}
    +
    \begin{array}{l}
        \includegraphics[width=0.14\textwidth]{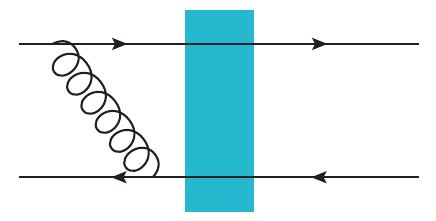}
    \end{array}
        +
    \begin{array}{l}
        \includegraphics[width=0.14\textwidth]{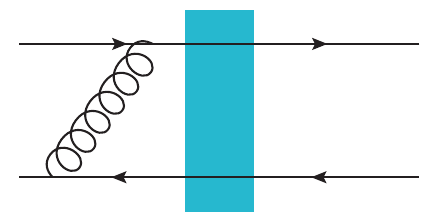}
    \end{array}
        +
    \begin{array}{l}
        \includegraphics[width=0.14\textwidth]{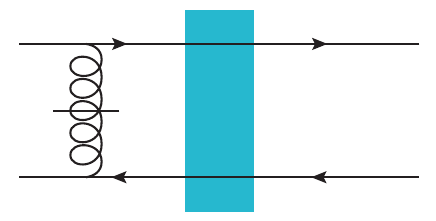}
    \end{array}
    \\
        +&
    \begin{array}{l}
        \includegraphics[width=0.14\textwidth]{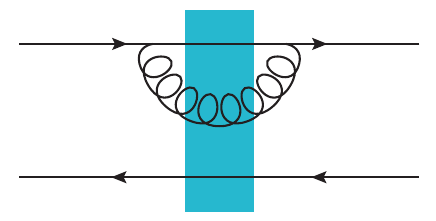}
    \end{array}
        +   
    \begin{array}{l}
        \includegraphics[width=0.14\textwidth]{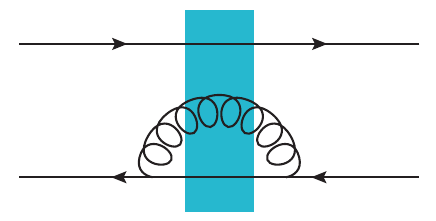}
    \end{array}
        +
    \begin{array}{l}
        \includegraphics[width=0.14\textwidth]{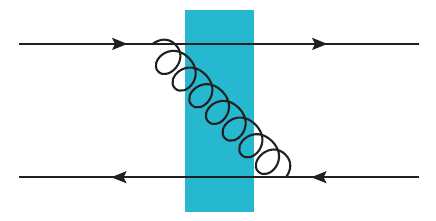}
    \end{array}
        +
    \begin{array}{l}
        \includegraphics[width=0.14\textwidth]{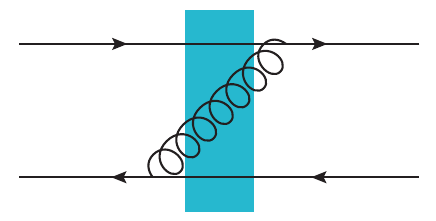}
    \end{array}
    \\
        +&
    \begin{array}{l}
        \includegraphics[width=0.14\textwidth]{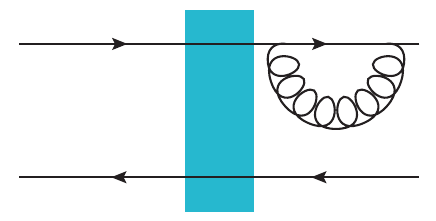}
    \end{array}
        +   
    \begin{array}{l}
        \includegraphics[width=0.14\textwidth]{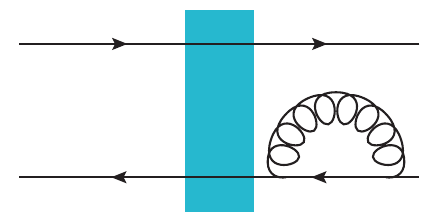}
    \end{array}
    +
    \begin{array}{l}
        \includegraphics[width=0.14\textwidth]{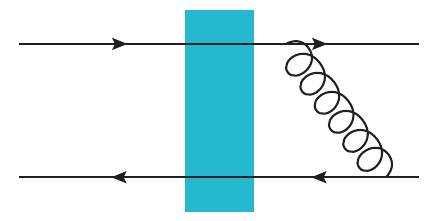}
    \end{array}
        +
    \begin{array}{l}
        \includegraphics[width=0.14\textwidth]{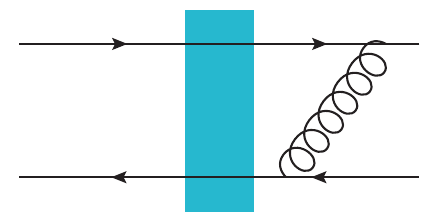}
    \end{array}
        +
    \begin{array}{l}
        \includegraphics[width=0.14\textwidth]{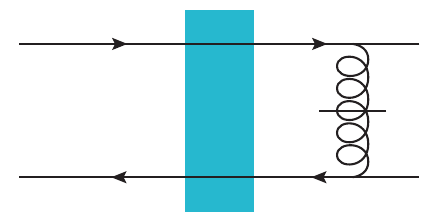}
    \end{array}
        \end{align*}
        \caption{
            JIMWLK equation for a quark-antiquark dipole shown schematically in terms of the Feynman diagrams involving a slow gluon with $z_g \ll 1$.
            Here $Y+ \delta Y$ and $Y$ refer to the rapidity at which the target probed.
        }
        \label{fig:JIMWLK}
\end{figure}

Demanding that the physical cross sections do not depend on the exact value of the cut-off $\zmin$ means that the weight $\W[\rho]$ has to have a dependence on the cut-off and thus also on the rapidity $Y$.
This rapidity dependence can be calculated perturbatively by considering the difference between $\W[\rho]$ at rapidities $Y$ and $Y+ \delta Y$ where $\delta Y$ is small. 
For a quark-antiquark pair this then leads to an equation like shown in Fig.~\ref{fig:JIMWLK}, but this can be generalized to more general Fock states. Taking $\delta Y \to 0$ one ends up with the JIMWLK equation~\cite{Iancu:2000hn,Iancu:2001ad,Iancu:2001md,Ferreiro:2001qy,Jalilian-Marian:1996mkd,Jalilian-Marian:1997jhx,Jalilian-Marian:1997qno}
\begin{equation}
    \label{eq:jimwlk_W}
    \partial_Y \W_Y[\alpha]  = -\jimwlk \W_Y[\alpha]
\end{equation}
where the color field has been written in terms of $\partial^2 \alpha = \rho$. The JIMWLK Hamiltonian $\jimwlk$ is defined as 
\begin{equation}
    \jimwlk = - \frac{\as}{2} \int \dd[2]{\xt}\dd[2]{\yt} \frac{\partial}{\partial \alpha^a(\xt)  }  \eta^{ab}_{\xt \yt} \frac{\partial}{\partial \alpha^b(\yt)}
\end{equation}
where 
\begin{equation}
    \eta^{ab}_{\xt \yt} =  \frac{4}{g^2 \pi^2} \int \dd[2]{\zt} \mathcal{K}(\xt,\yt,\zt) \left[ \bigl(1- U(\zt)  U^\dag(\xt) \bigr) \bigl(1 -U(\yt) U^\dag(\zt)\bigr) \right]^{ab}
\end{equation}
and $U$ is the adjoint Wilson lines from Eq.~\eqref{eq:adjoint_wilson}.
The kernel $\mathcal{K}$ in $\eta^{ab}_{\xt\yt}$ can be written as
\begin{equation}
    \mathcal{K}(\xt, \yt, \zt) = \frac{(\zt - \xt) \vdot (\zt-\yt)}{(\zt-\xt)^2 (\zt-\yt)^2}
\end{equation}
and it is related to the probability of an emission of a gluon with a coordinate $\zt$ from a color dipole with coordinates $\xt$ and $\yt$.
Rapidity evolution for the expectation value of the operator $\hat {\mathcal{O}}$ can then be written as
\begin{equation}
    \label{eq:jimwlk_o}
    \partial_Y \langle \hat {\mathcal{O}} \rangle 
    =  -  \left\langle \jimwlk \hat {\mathcal{O}} \right\rangle
\end{equation}
which follows from the hermiticity of $\jimwlk$.

An important special case of Eq.~\eqref{eq:jimwlk_o} is the case of a quark-antiquark dipole scattering off the target, which corresponds to
\begin{equation}
    \hat {\mathcal{O}} = \hat S_{01} = \frac{1}{N_c} \Tr \left[ V(\xt_0) V^\dag(\xt_1) \right].
\end{equation}
The JIMWLK equation~\eqref{eq:jimwlk_o} in this case reads
\begin{equation}
    \label{eq:jimwlk_dipole}
    \partial_Y \langle \hat S_{01} \rangle =  
     \frac{\as \cf}{\pi^2} \int \dd[2]{\xt_2} \frac{\xt_{10}^2}{\xt_{20}^2\xt_{21}^2} 
     \left\langle  \hat S_{012}- \hat S_{01}  \right\rangle
\end{equation}
where 
\begin{equation}
    \hat S_{012} 
= \frac{1}{\nc \cf} \Tr \left[ V(\xt_0) t^a V^\dag(\xt_1) t^b \right] \left[ U(\xt_2) \right]^{ba}
\end{equation}
is the Wilson line operator for a $q \bar q g$ Fock state. Using Fierz identities for the Wilson lines it is possible to write this as
\begin{equation}
    \hat S_{012} = 
    \frac{\nc}{2 \cf} \left( \hat S_{02} \hat S_{12}-\frac{1}{\nc^2} \hat S_{01}  \right)
\end{equation}
so that Eq.~\eqref{eq:jimwlk_dipole} becomes
\begin{equation}
    \label{eq:jimwlk_dipole2}
    \partial_Y \langle \hat S_{01} \rangle =  
     \frac{\as \nc}{ 2\pi^2} \int \dd[2]{\xt_2} \frac{\xt_{10}^2}{\xt_{20}^2\xt_{21}^2} 
     \left\langle  \hat S_{02} \hat S_{12}- \hat S_{01}   \right\rangle.
\end{equation}
This differential equation alone does not form a closed system as one then needs to know how the product of two dipole operators $\langle \hat S \hat S \rangle$ evolves. This can also be calculated using the JIMWLK equation, but its evolution equation will then contain operators with even more Wilson lines.
This leads to an infinite system of coupled differential equations called the \textit{Balitsky hierarchy}~\cite{Balitsky:1995ub}.
In practice, this infinite set of differential equations has to be truncated at some point.
One way to do this is to use the Gaussian approximation~\eqref{eq:gaussian_approximation} for the distribution of the target color field configurations, which allows higher-order Wilson line operators to be written in terms of the dipole operator~\cite{Kovchegov:2008mk}.
Another way is to use
the mean-field approximation $\langle \hat S \hat S \rangle \approx \langle \hat S \rangle \langle \hat S \rangle $, which is valid for example in the large-$\nc$ limit.
Using the mean-field approximation to Eq.~\eqref{eq:jimwlk_dipole2} leads to 
\begin{equation}
    \label{eq:bk}
    \partial_Y \langle \hat S_{01} \rangle =  
     \frac{\as \nc}{2 \pi^2} \int \dd[2]{\xt_2} \frac{\xt_{10}^2}{\xt_{20}^2\xt_{21}^2} 
     \left(
     \langle  \hat S_{02} \rangle \langle \hat S_{12} \rangle-  \langle \hat S_{01}  \rangle \right)
\end{equation}
which is the famous BK equation~\cite{Balitsky:1995ub,Kovchegov:1999yj}. 
It is the differential equation for the rapidity evolution of the dipole amplitude $N_{01} = 1 - \langle \hat S_{01} \rangle$.
No analytical solutions of the BK exist because of its nonlinear nature, but it can be solved numerically if the dipole amplitude at some initial rapidity $\Ybkzero$ is given.
While the JIMWLK equation~\eqref{eq:jimwlk_dipole2} is formally the correct evolution equation for the dipole amplitude, usually the simpler BK equation~\eqref{eq:bk} is used in numerical calculations.
This is because the differences between the JIMWLK and BK equations are numerically small, much less than the simple estimate $1/\nc^2 \approx 10 \%$ from the large-$\nc$ approximation~\cite{Kovchegov:2008mk,Lappi:2020srm}.

\begin{figure}
    \centering
    \includegraphics[width=0.68\textwidth]{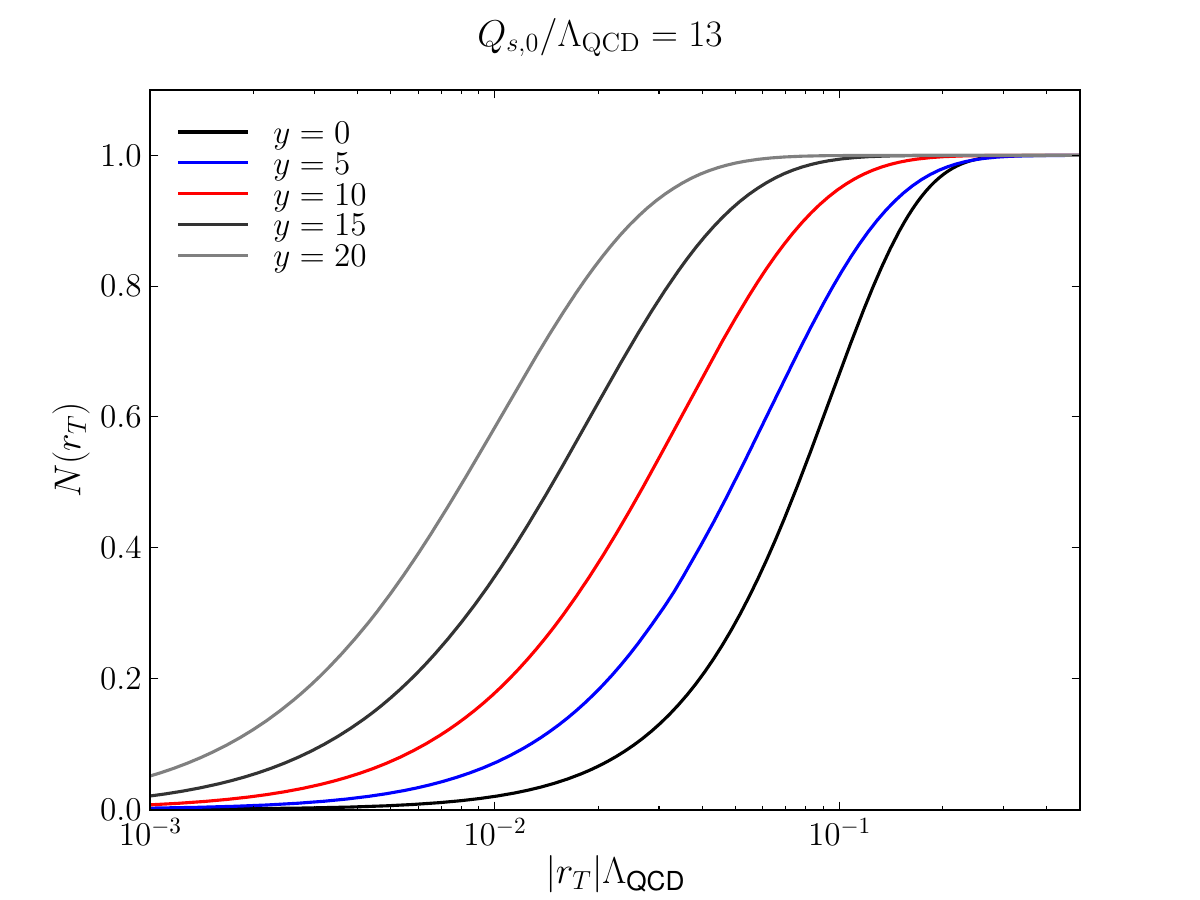}
    \caption{The  BK-evolved dipole amplitude for different values of the rapidity $Y$ as a function of the dipole size $r_T$, with the MV model~\eqref{eq:MV_dipole} as the initial condition. Rapidity increases from right to left.
    Figure from Ref.~\cite{Mantysaari:2015uca}.
    Reprinted with permission from H. Mäntysaari.
    }
    \label{fig:dipole_amplitude_evolution}
\end{figure}

The BK equation depends on both the dipole sizes $\xt_{ij}$ and the impact parameter $\btvar = \frac{1}{2} (\xt_i + \xt_j)$.
This impact parameter dependence of the BK equation is problematic as it makes the evolution very sensitive to the infrared region, which has to be remedied by including confinement effects~\cite{Golec-Biernat:2003naj,Berger:2010sh,Berger:2011ew,Berger:2012wx}.
For this reason, one usually neglects the impact parameter dependence in the BK evolution such that the dipole amplitudes $N_{ij} = 1- \langle \hat S_{ij} \rangle$ are evaluated at the same impact parameter and only the dependence on the dipole sizes remains.
While this assumption should be valid for heavy nuclei where the color density varies only slowly in $\btvar$, for protons this is less justified.
However, for the time being it is not known how to implement the impact parameter dependence of the evolution rigorously and thus it is neglected for the dipole amplitudes considered in this thesis.

The nonlinearity of the BK equation~\eqref{eq:bk} is crucial:
without nonlinear effects it reduces to the BFKL equation. The nonlinear effects ensure that, with the fixed impact parameter, the black-disk limit $N_{01} \leq 1$ is satisfied which is important for the unitarity of the $S$-matrix. 
In Fig.~\ref{fig:dipole_amplitude_evolution}, we show how the BK evolution changes the dipole amplitude: essentially, the dipole amplitude increases in rapidity, and for large dipole sizes $\xt_{01}$ it saturates to one.
This saturation region of the target has gained the name \textit{color-glass condensate}, and 
it is a prediction of the BK equation that it will be eventually reached at sufficiently high energies.
However, direct evidence of saturation has not been found at the energies available at the current experimental facilities. It is expected that the enhancement of saturation effects in heavy nuclei will allow us to probe the saturation region in the future~\cite{AbdulKhalek:2021gbh}.
The general framework of treating the target as a mixture of classical color fields that evolve with the JIMWLK equation~\eqref{eq:jimwlk_W} is usually referred to as the color-glass condensate effective field theory, and this is very commonly used for calculations in the dipole picture.

\section{Numerical fits for the dipole amplitude}

For an actual calculation in the dipole picture, one needs a specific model for the dipole amplitude $N_{01} = 1-S_{01}$ that is suitable for numerics.
The general way to do this is to take some simple model for the dipole amplitude with free parameters that are fitted to the data.

Several models exist in the literature, some of which are more physically motivated than others.
One of the first models used for the dipole amplitude is the Golec-Biernat--Wüsthoff model~\cite{Golec-Biernat:1998zce,Golec-Biernat:1999qor}
\begin{equation}
    S_{01}(\rt, Y = \log 1/x) = \sigma_0 \exp(-\frac{\left(\rt \times \SI{1}{GeV}\right)^2}{4 } \left(\frac{x_0}{x} \right)^{\lambda/2})
\end{equation}
which has been integrated over the impact parameter $\btvar$, and the constants $\sigma_0$, $x_0$ and $\lambda$ are free parameters.
This is purely a phenomenological model inspired by saturation and geometric scaling at HERA, but it is quite successful in describing the data~\cite{Golec-Biernat:1998zce,Golec-Biernat:1999qor}.
It essentially introduces the energy dependence of the saturation scale $\Qs^2 \sim  \left(\frac{x_0}{x} \right)^{\lambda/2}$ as a power, which for small dipoles matches the power-like energy dependence of the cross section in the region where saturation effects are not relevant.

Another widely used model is the \textit{impact parameter saturation} (IPsat) model~\cite{Kowalski:2003hm,Kowalski:2006hc}
\begin{equation}
    S_{01}(\rt, \btvar, Y= \ln 1/x) = \exp( - \frac{\pi}{2 \nc} \rt^2 \as(\mu^2) xg(x,\mu^2) T(\btvar) )
\end{equation}
which depends on the gluon parton distribution function (PDF) $xg(x,\mu^2)$ and the transverse profile of the target $T(\btvar)$. The form of the gluon PDF is fitted to the data, and it satisfies the Dokshitzer--Gribov--Lipatov--Altarelli--Parisi  (DGLAP) evolution~\cite{Gribov:1972ri,Dokshitzer:1977sg,Altarelli:1977zs} for the dependence on the factorization scale $\mu^2 \sim 1/\rt^2$.
This model relates the dipole amplitude directly to the gluon PDF, motivated by the idea of the interaction as a two-gluon exchange in Fig.~\ref{fig:pomeron_LO}.
However, it is not clear how the dipole amplitude and the gluon PDF are related exactly beyond this leading-order picture.

A drawback of the two models mentioned above is that they do not satisfy the correct high-energy evolution given by the JIMWLK or BK equations.
A more physically motivated model is to fit the initial condition of the dipole amplitude to the data at some initial rapidity $\Ybkzero$ and then evolve the dipole amplitude to higher rapidities~\cite{Albacete:2009fh,Albacete:2010sy,Lappi:2013zma}.
This ``initial condition fit + rapidity evolution'' is also the only one of these models that can be consistently used at NLO calculations where the large logarithms of the BK equation start to appear, and therefore we will mainly focus on this model in this thesis.
It is interesting to note that the IPsat model is essentially orthogonal to this approach: there the dependence on the Bjorken $x$, and thus energy, is fitted to the data and the dependence on the dipole sizes is predicted by the DGLAP evolution.
The BK-evolved approach instead fits the dependence on the dipole sizes at the initial rapidity to the data, and the dependence on the energy is then a prediction from the BK evolution.

All of these models have some freedom in the parametrization of the dipole amplitude.
This freedom has to be constrained by the data, which is usually done by fitting the parameters to the HERA structure function data~\cite{H1:2009pze,H1:2012xnw,H1:2015ubc,H1:2018flt} because of its high precision.
The high-energy factorization then guarantees that this same dipole amplitude can be used also in other calculations.
Because the fitting procedure to the structure function data is so important for numerical calculations, we will briefly consider how the structure functions can be calculated in the dipole picture.

\subsection{Inclusive deep inelastic scattering}
\label{sec:inclusive_DIS}

\begin{figure}[t]
    \centering
    \begin{overpic}[width=0.70\textwidth]{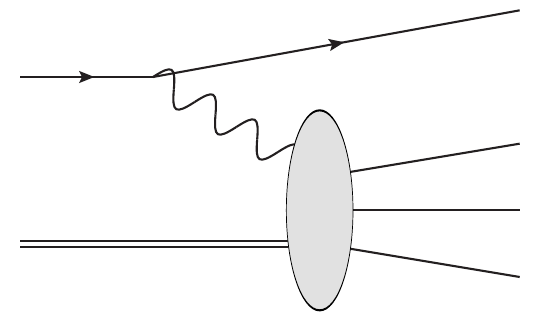}
    \put (37,29) {$\gamma^*(q)$}
    \put (10,49) {$e(k)$}
    \put (85,60) {$e(k')$}
    \put (10,18) {$A(P_A=AP_n)$}
    \put (85,36) {$X(P_X)$}
   \end{overpic}
    
    \caption{Inclusive deep inelastic scattering. The final states $X$ are summed over. 
    }
    \label{fig:DIS}
\end{figure}
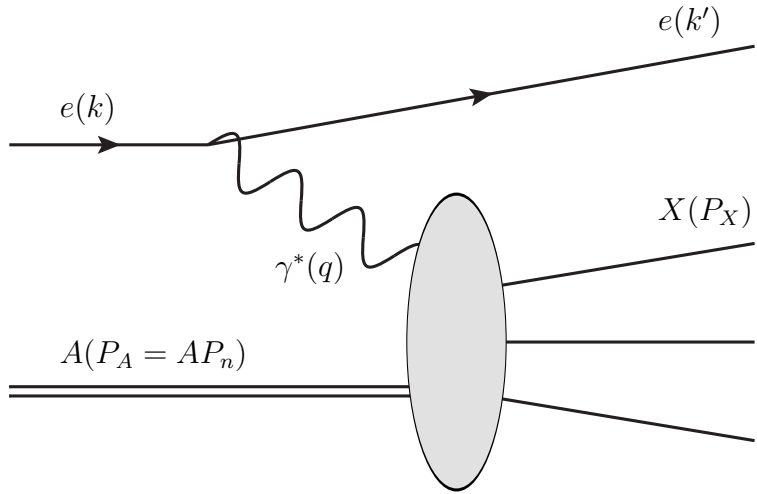

In inclusive DIS we allow any final state in the $\gamma^* + A$ process as shown in Fig.~\ref{fig:DIS}.
This can be calculated in the dipole picture using the framework we have presented in Sec.~\ref{sec:dipole_picture}, and the optical theorem allows 
us to relate the inclusive cross section to the forward elastic scattering amplitude as
\begin{equation}
    \label{eq:inclusive_DIS_cross_section}
    \sigma^{\gamma^* A} = 2 \Im \mathcal{M}^{\gamma^* A \to \gamma^*  A}
\end{equation}
with the scattering amplitude given by Eq.~\eqref{eq:general_amplitude}.
This process is very inviting for measuring the dipole amplitude as everything else in the process is fully perturbative, and as an inclusive process the corresponding cross section is very large.

The cross section for inclusive DIS can be divided into longitudinal and transverse productions based on the photon polarization. These are, however, not directly measurable as the photon polarization itself is not observable. Instead, it is more useful to define the \textit{structure functions}
\begin{align}
\label{eq:Flambda}
    F_\lambda(x,Q^2) &= \frac{Q^2}{4\pi^2 \aem} \sigma^{\gamma_\lambda^* A},\\
    \label{eq:F2} 
    F_2 (x,Q^2)&= F_L(x,Q^2) + F_T(x,Q^2),
\end{align}
which allow us to define the experimentally measured \textit{reduced cross section}
\begin{equation}
    \label{eq:reduced_cross_section}
    \sigma_r(y,x,Q^2) = F_2(x,Q^2) - \frac{y^2}{1+ (1-y)^2} F_L(x,Q^2).
\end{equation}
The reduced cross section depends on the photon virtuality $Q^2$, the Bjorken $x$~\eqref{eq:x}, and the inelasticity $y$ defined as 
\begin{align}
    y &= \frac{2 P_n \vdot q}{2 P_n \vdot k}
    = \frac{W^2 + Q^2 - m_n^2}{s - m_e^2 - m_n^2}
    \approx \frac{W^2 + Q^2}{s}
\end{align}
where $s$ is the center-of-mass energy of the lepton-nucleus system and the momenta are shown in Fig.~\ref{fig:DIS}.
The structure functions only depend on the Bjorken $x$ and the photon virtuality $Q^2$ but not on the inelasticity $y$. 
In principle, the structure functions $F_2$ and $F_L$ could be determined from the reduced cross section by measuring it for different lepton-nucleus energies $s$ and thus for different inelasticities $y$.
In practice, this leads to less accurate data~\cite{H1:2013ktq} and thus it is easier to simply calculate the reduced cross section from the structure functions for data comparisons.

\subsection{Initial condition for the numerical fit}

The BK equation needs a nonperturbative initial condition for the rapidity evolution.
A common ansatz is the MV model in Eq.~\eqref{eq:MV_dipole} which has a physical motivation for the general form of the dipole.
This initial condition at $Y= \Ybkzero$ is often generalized to
\begin{equation}
    S_{01}(\rt, \btvar, \Ybkzero) = \exp[ -\frac{1}{4} \left( \rt^2 \Qszero(\btvar) \right)^\gamma \ln(\frac{1}{\abs{\rt} \lqcd} +e) ]
\end{equation}
where  the anomalous dimension $\gamma$ is also introduced. The MV model predicts $\gamma = 1$ but in actual fits it may be taken as a parameter of the initial condition fitted to the data.
This form is still quite general, and for the fits one needs to specify some form for the impact parameter dependence of the saturation scale $\Qszero(\btvar)$.
The dipole amplitude is usually fitted to the structure function data which only depends on the impact-parameter integrated dipole amplitude,
and then
the simplest assumption is to consider the integration as an overall factor to the dipole amplitude,
\begin{equation}
    \int \dd[2]{\btvar} S_{01}(\rt, \btvar, Y) = \frac{\sigma_0}{2}  S_{01}(\rt,Y),
\end{equation}
which corresponds to assuming that the impact parameter dependence is given by a step function
\begin{equation}
    \Qszero(\btvar) =  \theta( R - \abs{\btvar} ) \Qszero
\end{equation}
where $R$ is the transverse radius of the target. 
With such a model, there are then three constants to fit for the initial condition:
the anomalous dimensions $\gamma$, the saturation scale at the initial rapidity $\Qszero$, and the target transverse area $\sigma_0 /2$.

Such models (with some modifications) have been widely used to fit the dipole amplitude.
For proton targets, there exist several fits for these parameters at LO~\cite{Albacete:2009fh,Albacete:2010sy,Lappi:2013zma} and also at NLO \cite{Beuf:2020dxl} by fitting the initial condition to the HERA structure function data.
At LO, however, these fits suffer from the fact that it is not possible to describe simultaneously massless and massive quark production with the same parameters when using the BK equation to describe the target's evolution in energy~\cite{Albacete:2010sy}. It is not physical for the dipole amplitude to depend on the quark masses as they should be negligible in the high-energy interaction with the target, and thus one would expect that the massless and massive structure functions are given by the same dipole amplitude.
This is one of the other reasons why it is important to go beyond the leading order in the dipole picture to check if this problem persists at higher orders. Indeed, the situation at NLO is already quite different as will be discussed in Sec.~\ref{sec:massive_quark_structure functions}.

\subsection{NLO fit with the massless structure function data }
\label{sec:nlo_massless_dipole_amplitude}

At the time of writing this thesis, the only dipole amplitude fits done at the full NLO are the ones in Ref.~\cite{Beuf:2020dxl} which have been fitted to the HERA inclusive DIS data~\cite{H1:2009pze,H1:2015ubc} using NLO equations for the structure functions~\eqref{eq:Flambda} with massless quarks. For this reason, these are the dipole amplitudes also used in Articles~\cite{heavy_long,light,heavy_trans} when predicting exclusive vector meson production from protons at NLO. There is some more freedom in the fitting procedure at NLO compared to the leading order, and as such we will go through the most important details of the fits in order to clarify the differences between them.

\begin{enumerate}
    \item 
    Two different data sets were used for the fitting. The first is the full HERA data set, and the second one is pseudodata consisting only of the light-quark contribution where the massive quark contribution was subtracted from the full HERA data using a prediction with the IPsat parametrization from Ref.~\cite{Mantysaari:2018nng}. The light-quark pseudodata is physically better motivated as the calculation of the structure functions in Ref.~\cite{Beuf:2020dxl} uses only light quarks.
    \item Three different versions of the BK evolution are used for the energy evolution of the dipole amplitude. These are called \textit{kinematically constrained BK} (KCBK)~\cite{Beuf:2014uia}, \textit{resummed BK} (ResumBK)~\cite{Iancu:2015joa,Iancu:2015vea}, and \textit{target rapidity BK} (TBK)~\cite{Ducloue:2019ezk}. The main reason for the different forms of the BK evolution is that they are different approximations of the full NLO BK evolution.
    To be completely consistent with the perturbation theory one should use the NLO BK equation in the NLO calculation, but because of its numerical complexity this is not really feasible in an already demanding numerical fit. The three different BK evolutions used give in general a good approximation for the NLO BK equation \cite{Hanninen:2021byo}. In KCBK, a kinematical constraint is introduced that forces an explicit time ordering between subsequent gluon emissions. In ResumBK, large single and double transverse logarithms that occur at higher orders are resummed into the kernel of the BK equation. In TBK, one uses the target rapidity instead of the projectile rapidity in the evolution. The target rapidity is calculated from the projectile rapidity $Y$ with the transformation
    \begin{equation}
        \label{eq:target_rapidity}
        \eta = Y - \max \left\{0, \ln(\frac{1}{\rt^2 Q_0^2}) \right\}
    \end{equation}
    where $Q_0^2 \equiv \SI{1}{GeV^2}$ is the transverse scale of the target that is used to regulate large dipole sizes.
    \item 
    Two different schemes for the running of the coupling constant $\as$ are used. In both cases the dependence on the dipole size is given by
    \begin{equation}
        \label{eq:as_running}
        \as(\rt^2) = \frac{4 \pi}{\beta_0 \ln \left[ \left( \frac{\mu_0^2}{\lqcd^2} \right)^{1/c} + \left( \frac{4 C^2}{\rt^2 \lqcd^2} \right)^{1/c} \right]^c}
    \end{equation}
    where $\beta_0 = (11N_c - 2 N_F)/3$, $N_F = 3$ and $\lqcd = \SI{0.241}{GeV}$. The constants $\mu_0 = 2.5 \lqcd$ and $ c = 0.2$ regulate the running of the coupling in the infrared region. The constant $C^2$ is a free parameter determined from the fit and it controls how coordinate scales are related to momentum scales. From Fourier analysis its predicted value is $C^2 = e^{-2\gamma_E}$~\cite{Kovchegov:2006vj}, but keeping it as a free parameter allows for absorbing some nonperturbative or higher-order contributions. The two different schemes for the running of the coupling are then related to what dipole sizes $\rt$ are used in the coupling constant~\eqref{eq:as_running} for the $q \bar q g$ state. In the \textit{parent dipole} scheme the choice is $\rt^2 = \xt_{01}^2$
    which corresponds to the transverse size of the quark-antiquark dipole.
    The other scheme is called \textit{Balitsky+smallest dipole}, and in this scheme one uses the Balitsky prescription~\cite{Balitsky:2006wa} for the running of the coupling when evolving the dipole amplitude.
    For the impact factor
    the smallest dipole $\rt^2 = \min\{ \xt_{01}^2, \xt_{20}^2, \xt_{21}^2 \}$ is used  instead. 
    The reason for this is that it is not clear how to use the Balitsky prescription in general kinematics, and the smallest dipole can be thought of as an approximation of the Balitsky prescription.
    \item 
    Two different starting points for the BK evolution of the dipole amplitude are used, namely $\Ybkzero = 0$ and $\Ybkzero = \ln 1/0.01$. The later starting point for the BK evolution, $\Ybkzero = \ln 1/0.01$, is typically more used in fitting the dipole amplitude
   as it is not clear if the assumptions for deriving the MV model or the BK evolution are valid for larger values of $x$ (and hence smaller values of $Y$).
\end{enumerate}

These different setups have four free parameters. Three of them are related to the MV model: the saturation scale at the initial rapidity $\Qszero$, the anomalous dimension $\gamma$, and the transverse area of the proton $\sigma_0/2$. The final fit parameter is the constant $C^2$ controlling the running of the coupling constant. With this fitting procedure, one finds a set of these parameters for each of the different setups, resulting in $2 \times 2 \times 3 \times 2 = 24$ different fits.
All of these fits describe the structure function data used in the fit extremely well, and as such one cannot distinguish between the fits based on this data alone.
This changes, however, when one also takes into account the massive quark structure function data.

\subsection{Structure functions with massive quarks at NLO}
\label{sec:massive_quark_structure functions}

With the expressions for the massive quark structure functions at NLO now available~\cite{Beuf:2021qqa,Beuf:2021srj,Beuf:2022ndu}, we can study how including the quark mass affects predictions for structure functions at NLO.
It is especially important to see if both the total and charm production cross sections can be described by the same dipole amplitude, which was not possible at leading order if the evolution of the dipole amplitude is given by the BK equation~\cite{Albacete:2010sy}.
This was the main motivation for Article~\cite{structure} where the massive quark structure functions were calculated at NLO using the dipole amplitude fits described in Sec.~\ref{sec:nlo_massless_dipole_amplitude}.

Some words should be said about the numerical evaluation of the massive quark structure functions
as
the NLO equations including the quark mass are numerically quite demanding. 
They involve multi-dimensional integrals with a high number of dimensions, and
getting these integrals to converge with a reasonable amount of integration points is quite tricky. 
To do this one has to be especially careful with the cancellation of possible numerical singularities. For example, numerical integration of the generalized Bessel functions
\begin{equation}
    \label{eq:generalized_bessel}
    \begin{split}
    \mathcal{G}^{(a;b)}_\text{(x)} =& \int_0^\infty \frac{\dd{u}}{u^a} \exp(-u \left[ \overline Q^2_{(\textrm{x})}+m^2 \right]-\frac{|\xt_{3;(\textrm{x})}|^2}{4u}) \\
    &\times \int_0^{u/\omega_{(\textrm{x})}} \frac{\dd{t}}{t^b} \exp( -t \omega_{(\textrm{x})} \lambda_{(\textrm{x})} m^2 -\frac{|\xt_{2;(\textrm{x})}|^2}{4t} ) 
    \end{split}
\end{equation}
introduced in Refs.~\cite{Beuf:2021qqa,Beuf:2021srj,Beuf:2022ndu} is demanding in the limit $\abs{\xt_{2;\text{(x)}}}^2 \to 0$,
as the $t$-integral develops a singularity at $t \to 0$ if $\abs{\xt_{2;\text{(x)}}}^2 = 0$. This problem is not severe if $b=1$, as then the divergence is only logarithmic, but the power-like divergence in the case $b=2$ leads to numerical instabilities.
The convergence can be improved by subtracting the singular part in such a way that the remaining integral can be done analytically. A suitable subtraction is
using
the limit $\lambda_{\text{(x)}} \to 0$ as then
the integrand has the same behavior at $t \to 0$, and we can calculate the integral analytically:
\begin{equation}
    \label{eq:generalized_bessel_integrated}
\begin{split}
    \mathcal{G}^{(a;2)}_\text{(x)}(\lambda_\text{(x)} \to 0) =&
    \frac{2^{2+a}}{ |\xt_{2;(\textrm{x})}|^2   } 
    \left( \frac{\overline Q^2_{(\textrm{x})}+m^2 }{|\xt_{3;(\textrm{x})}|^2 +\omega_{(\textrm{x})}|\xt_{2;(\textrm{x})}|^2 } \right)^{\frac{a-1}{2}} \\
    & \times K_{a-1}\left( \sqrt{ \left( \overline Q^2_{(\textrm{x})}+m^2  \right)
    \left(  |\xt_{3;(\textrm{x})}|^2 +\omega_{(\textrm{x})}|\xt_{2;(\textrm{x})}|^2  \right)
    } \right).
\end{split}
\end{equation}
Such a subtraction leads to a much better convergence of the integrals which is crucial for a numerical implementation of the structure functions. Additionally, one can reduce the integration dimension by noting that with the change of variables $t \to y=t\omega_\text{(x)}/u$ the $u$-integral can be done analytically, leading to the expression
\begin{equation}
    \label{eq:generalized_bessel_simplified}
    \begin{split}
    \mathcal{G}^{(a;b)}_\text{(x)} 
     =&  \int_0^{1} \frac{\dd[]{y}}{y^{\frac{1}{2}(2-a+b)}} 2^{a+b-1} \omega_{(\textrm{x})}^{b-1} \left( \frac{y  \lambda_{(\textrm{x})} m^2 +  \overline Q^2_{(\textrm{x})}+m^2}{y |\xt_{3;(\textrm{x})}|^2 +\omega_{(\textrm{x})}|\xt_{2;(\textrm{x})}|^2 } \right)^{\frac{1}{2}(a+b-2)}\\
&\times     K_{a+b-2} \left(\sqrt{\frac{1}{y} \left( y \lambda_{(\textrm{x})} m^2 +  \overline Q^2_{(\textrm{x})}+m^2\right) \left(y |\xt_{3;(\textrm{x})}|^2 +\omega_{(\textrm{x})}|\xt_{2;(\textrm{x})}|^2  \right)} \right).
    \end{split}
\end{equation}
The subtraction in the numerical implementation is then done by subtracting Eq.~\eqref{eq:generalized_bessel_simplified} for $b=2$ at the integral level, and adding the integrated result Eq.~\eqref{eq:generalized_bessel_integrated} to the rest of the calculation where the additional $t$- and $u$-integrals are not present.

\begin{table}[t]
    \begin{tabular}{c||c|c|c|c|c|c|c|c|c}
         \# &Data &\begin{tabular}[c]{@{}c@{}}BK\\equation\end{tabular} & $\as$ & $\Ybkzero$ & \begin{tabular}[c]{@{}c@{}}$m_c$\\$[\mathrm{GeV}]$\end{tabular} & $\chi^2_\mathrm{c}/N$ &  \begin{tabular}[c]{@{}c@{}}$m_b$\\$[\mathrm{GeV}]$\end{tabular} & $\chi^2_\mathrm{b}/N$ & $\chi^2_\mathrm{tot}/N$   \\
         \hline 
         1 & Light-q & ResumBK & PD &  0 & 1.42 & 1.86 & $4.83$ & $1.37$ & 1.25\\
         2 & Light-q & KCBK   & PD & 0 & 1.49 & 2.55 & $4.96$ & $1.58$ & 1.23 \\
         3 & Light-q & TBK & BSD & 0 & 1.29 & 1.02 & $5.04$ & $1.12$ & 1.83 
    \end{tabular}
    \caption{
    Dipole amplitude fits from Ref.~\cite{Beuf:2020dxl} that were found to be compatible with the massive quark structure function data from HERA.
    Here ``light-q'' refers to only using the light-quark pseudodata to fit the dipole amplitudes, and
    ``PD'' and ``BSD'' are the \textit{parent dipole} and \textit{Balitsky+smallest dipole} running coupling schemes for $\as$.
    The optimal masses for the charm and bottom quarks are shown along with the $\chi^2/N$ values obtained with the optimal mass for the corresponding heavy quark production data. 
    The values $\chi^2_\mathrm{tot}/N$ refer to the total structure functions which contain both the light and heavy quarks.
    Table from Article~\cite{structure}.
    }
    \label{table:fits}
    \end{table}

\begin{figure}[t]
    \centering
    \begin{subfigure}{0.49\textwidth}
        \centering
        \includegraphics[width=\textwidth]{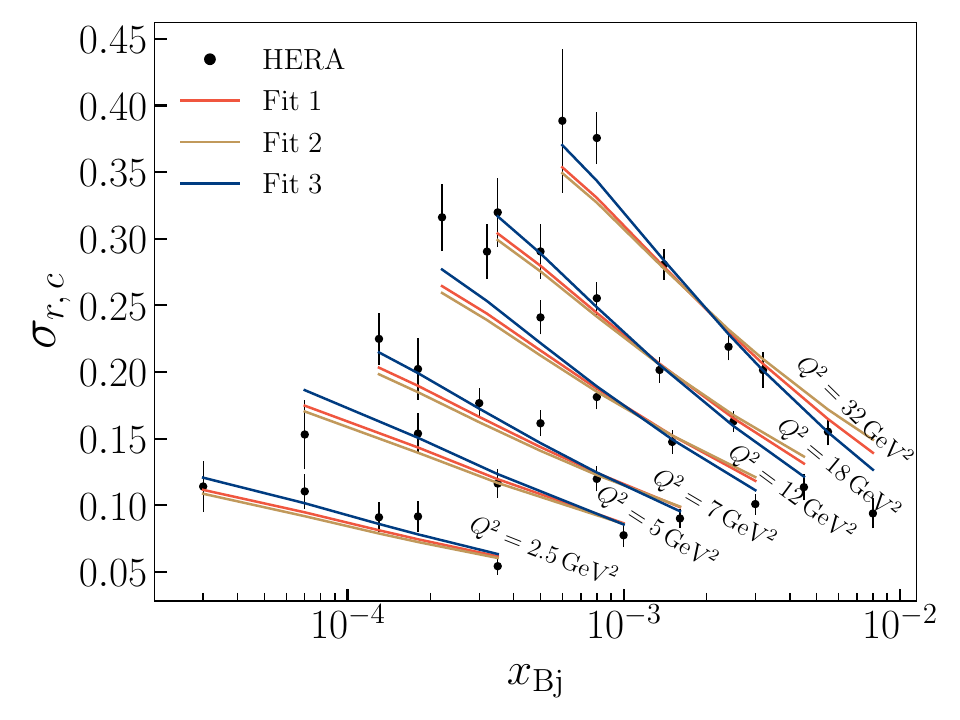}
        \caption{ Charm reduced cross section }
        \label{fig:charm-production}
    \end{subfigure}
    \begin{subfigure}{0.49\textwidth}
        \centering
        \includegraphics[width=\textwidth]{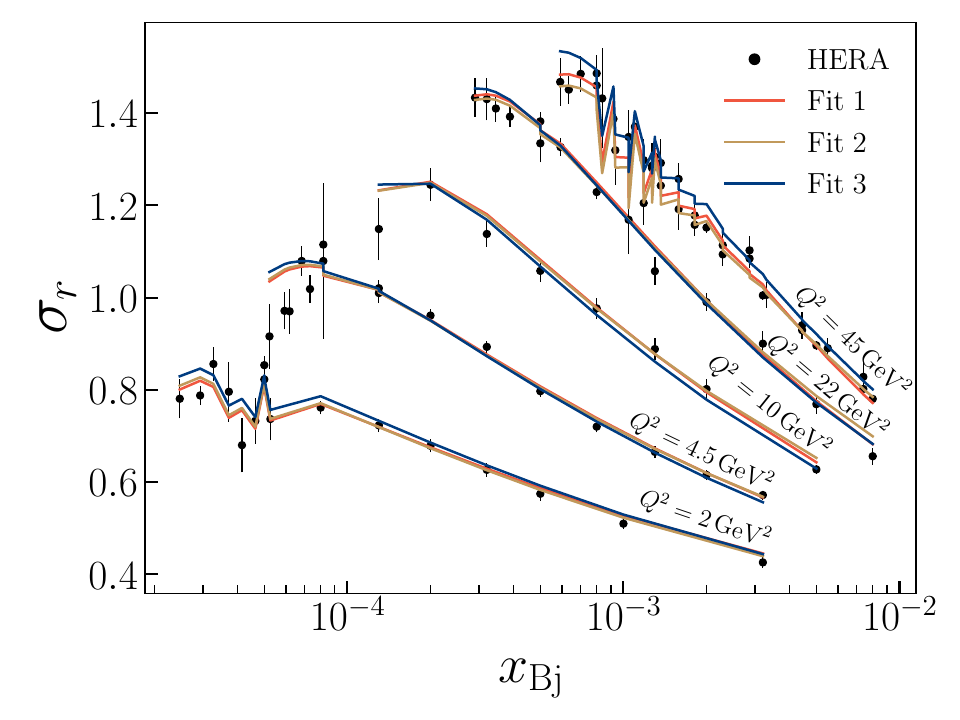}
        \caption{ Total reduced cross section }
        \label{fig:inclusive-production}
    \end{subfigure}
      \caption{Reduced cross section with massive quarks using the three dipole amplitude fits found to be compatible with the HERA data~\cite{H1:2009pze,H1:2012xnw,H1:2015ubc,H1:2018flt}. The reduced cross sections are plotted as a function of Bjorken $x$ for various values of the photon virtuality. 
      Note that the jumps in the lines for the total reduced cross section are not a result of numerical uncertainty, but rather they are a consequence of depicting points for different values of inelasticity $y$.
      Figures from Article~\cite{structure}, reproduced under the license CC BY 4.0.
      }
        \label{fig:massive-structure-functions}
\end{figure}

    Even after this, the remaining expressions are still numerically demanding.
    For this reason, instead of trying to perform fits to the charm data and inclusive data with quark masses included 
    it was more feasible to first see if the fits to the massless quark structure function data can also be used to describe the massive data.
This was the motivation for Article~\cite{structure}  where we calculated the total and charm quark structure functions at NLO with the NLO dipole amplitude fits described in Sec.~\ref{sec:nlo_massless_dipole_amplitude}. Out of the 24 fits only three were found to be compatible with both the charm quark and inclusive reduced cross section data from HERA~\cite{H1:2009pze,H1:2015ubc,H1:2018flt}, listed in Table~\ref{table:fits}.
Results were also compared to the bottom quark production data from HERA~\cite{H1:2018flt} but due to the large data uncertainties this does not provide further constraints for the fits.
The masses of the charm and bottom quarks were allowed to vary within reasonable limits, and the $\chi^2/N$ values are listed for the optimal mass. 
Results with these three dipole amplitude fits are shown in Fig.~\ref{fig:massive-structure-functions} where a good agreement with the data is found. It should be emphasized that, apart from varying the charm quark mass, these dipole amplitudes are not fitted to the charm quark data and thus these are genuine predictions using the previously obtained dipole amplitude fits.

Some comments can be made about these three fits. First, they have all been fitted to the light-quark pseudodata. 
This is expected as in the fitting procedure only the massless quark structure functions were calculated, and thus fitting to the total reduced cross section which also includes the heavy quark contribution would  overestimate the results. 
Second, these fits start the BK evolution at the earlier rapidity $\Ybkzero = 0$.
This can also be understood from the fitting procedure, as  
for both values of the initial rapidity $\Ybkzero$ 
the dipole amplitude in the structure function is calculated down to the factorization rapidity $\Yif = 0$, and the evolution of the dipole amplitude is frozen for values $\Yif < Y < \Ybkzero$. 
This is not entirely consistent as it leads to double counting in the rapidity region $\Yif < Y < \Ybkzero$.
The contribution from this region, however, should be small, but it is important to remember that this procedure leads to a slight overestimation of the cross section. It would be more consistent to set $\Yif = \Ybkzero$ in which case no such ambiguity arrives.

The charm quark data seems to naturally disqualify the fits that are not consistent based on these two conditions.
We are then still left with $6 = 3 \times 2$ fits with three different forms for the BK evolution and two different running coupling schemes. It is interesting to note that of these six fits only three are compatible with the charm quark production data, and these three fits all correspond to the different BK evolutions. Also,
both of the running coupling schemes are present in this set of fits, with different BK evolutions seeming to prefer different schemes.
While we do not have a clear reason for the different schemes preferred, 
we note that massive quark production data 
probes the dipole amplitude more in the perturbative region $\rt^2 \lqcd^2 \ll 1$ compared to the light quark production.
This explains why the heavy quark data provides more constraints for the dipole amplitude.

These fits show that it is possible to describe both inclusive and charm production data simultaneously with the same dipole amplitudes at NLO even when using a BK evolution for the energy dependence of the dipole amplitude. This is not possible at leading order~\cite{Albacete:2010sy}, even with an approximative NLO BK evolution~\cite{Ducloue:2019jmy}, showing the importance of the full NLO calculation including the mass effects also in the impact factor.
The NLO corrections modify both the evolution of the dipole amplitude and the mass dependence of the impact factor in such a way that together these effects allow for a precise description of both the charm quark and inclusive production data simultaneously.
This is important for the consistency of calculations in the dipole picture, as the dipole amplitude should be universal and independent of the quark mass in the high-energy limit.

\begin{figure}[t]
    \centering
    \begin{subfigure}{0.49\textwidth}
        \centering
        \includegraphics[width=\textwidth]{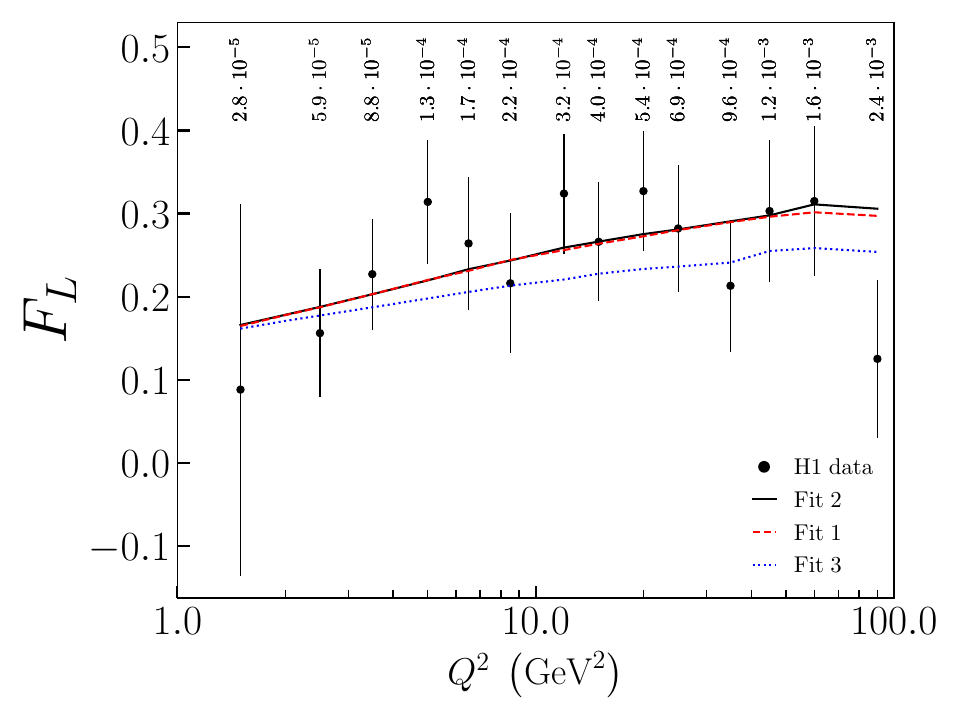}
        \caption{ HERA kinematics. }
        \label{fig:F_L_HERA}
    \end{subfigure}
    \begin{subfigure}{0.49\textwidth}
        \centering
        \includegraphics[width=\textwidth]{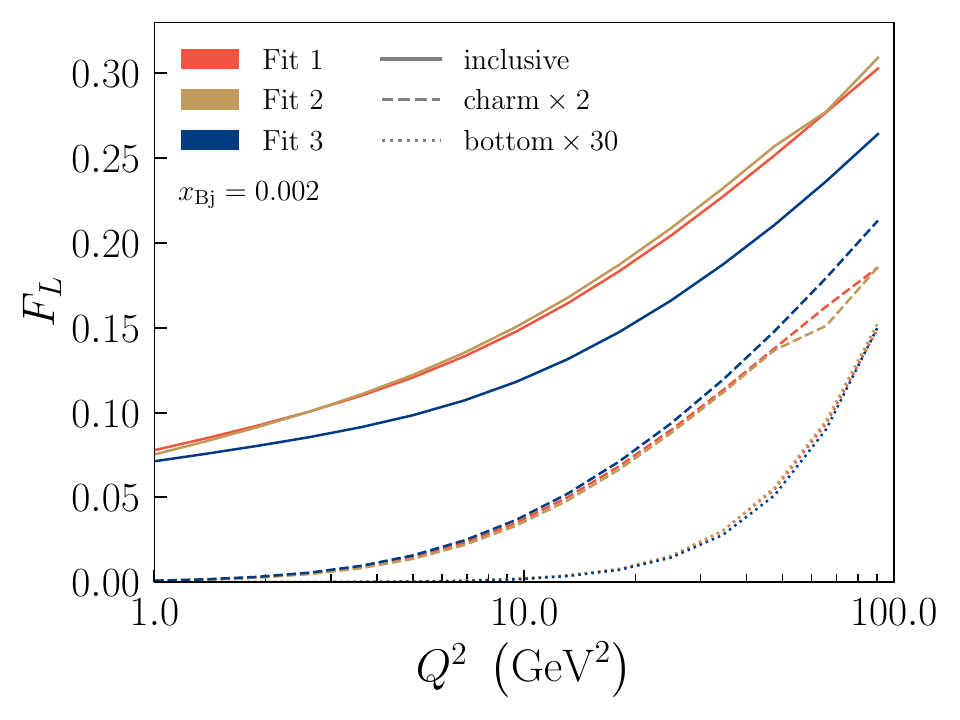}
        \caption{ EIC kinematics. }
        \label{fig:F_L_EIC}
    \end{subfigure}
      \caption{
        Longitudinal structure function $F_L$ as a function of the photon virtuality $Q^2$ using the three dipole amplitude fits found to be compatible with the HERA data~\cite{H1:2009pze,H1:2012xnw,H1:2015ubc,H1:2018flt}.\\
        \textbf{Left:} Comparison to the HERA data~\cite{H1:2013ktq}. The different values of Bjorken $x$ for the data points are shown. \\
        \textbf{Right:} Predictions for the EIC with a constant Bjorken $x$. 
        The inclusive case is shown along with the charm and bottom quark structure functions multiplied by factors of $2$ and $30$ for visibility.
        Figure from Article~\cite{structure}, reproduced under the license CC BY 4.0.
      }
        \label{fig:F_L}
\end{figure}

With just the charm and inclusive reduced cross section data it is not possible to distinguish between the three remaining fits, and thus data comparisons for other processes are required to show differences between the fits. 
For example, 
while the longitudinal structure function $F_L$ is not completely independent from the reduced cross section, it is generally more sensitive to the saturation region and can therefore give additional constraints for the dipole amplitude.
As shown in Fig.~\ref{fig:F_L}, the current HERA data~\cite{H1:2013ktq} for $F_L$ is not enough to show differences between the three dipole amplitude fits, but in
the EIC kinematics Fit 3 with the TBK evolution leads to different predictions.
Diffractive processes are also more sensitive to the nuclear structure, and hence they are good candidates for a more precise determination of the dipole amplitude. 
In the future, one should do a global fit for the dipole amplitude using all of the available data for different processes to determine precisely the fit parameters with uncertainty estimates. 
In addition to the proton dipole amplitude,
it would be interesting to fit also the nuclear dipole amplitude directly to the data, which is not possible with the HERA data that is only for proton targets.
With the EIC data in the future it will be possible to fit the dipole amplitude for heavy nuclei independently from protons~\cite{AbdulKhalek:2021gbh}.

\chapter{Exclusive vector meson production}
\label{sec:vector_meson_production}

In an exclusive scattering process all of the produced particles are measured,
and in a photon-nucleus scattering this means that the photon is essentially transformed into the produced particles.
For single-particle final states,
most likely particles produced in this way are vector mesons as they have the same quantum numbers $J^{PC} = 1^{--}$ as the photon, 
corresponding to a pomeron exchange in the terminology of the Regge theory.
Exclusive vector meson production amounts to a significant amount, about $10 \%$, of all diffractive processes~\cite{H1:1994ahk}.

\begin{figure}[t]
\centering
\begin{overpic}[width=0.8\textwidth]{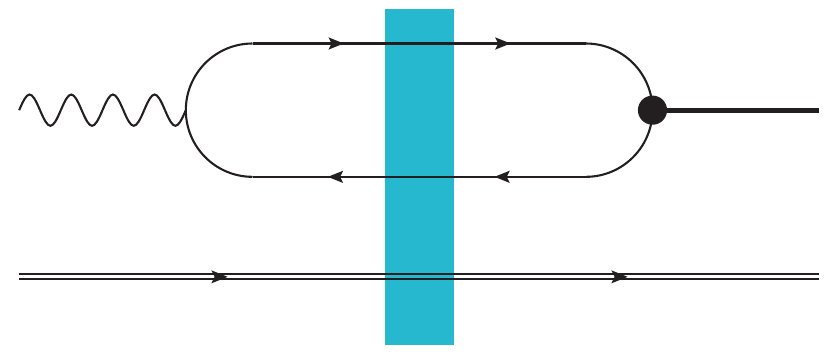}
 \put (5,34) {$\gamma^*$}
 \put (90,32) {$V$}
 \put (5,12) {$A$}
 \put (90,12) {$A'$}
 \put (14,18) {$\bt$}
 \put (18,18){\vector(0,1){8}}
 \put (18,18){\vector(0,-1){8}}
 \put (30,29) {$\xt_{01}$}
 \put (36,29){\vector(0,1){7}}
 \put (36,29){\vector(0,-1){7}}
 \put (36,39) {$\xt_0, z_0$}
 \put (36,18) {$\xt_1, z_1$}
\end{overpic}
    \caption{Exclusive vector meson production at leading order in the dipole picture. The blue rectangle depicts the instantaneous interaction with the target.}
    \label{fig:exclusive-vm-production-LO}
\end{figure}
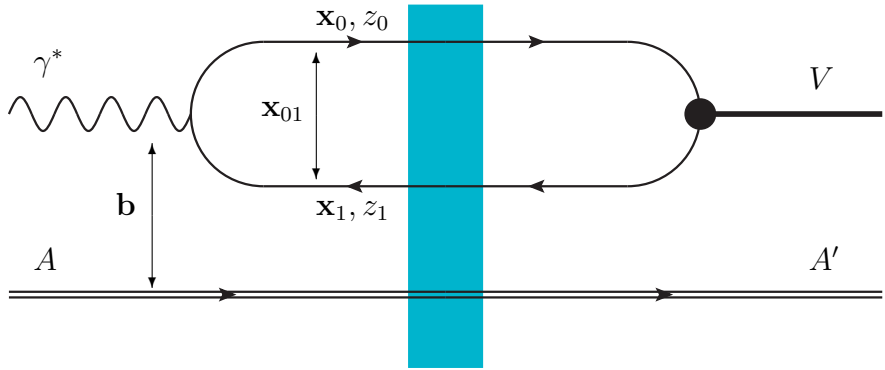

At leading order in the dipole picture, exclusive vector meson production can be described by the Feynman diagram in Fig.~\ref{fig:exclusive-vm-production-LO}, and the corresponding invariant amplitude reads
\begin{equation}
\label{eq:VM_production_amplitude_LO}
\begin{split}
    i\mathcal{M}_{\lambda_\gamma \lambda_V} =& \int \dd[2]{\xt_{01}} \dd[2]{\bt} \int_0^1 \frac{\dd{z_0} \dd{z_1}}{ (4\pi)^2 z_0 z_1} (4\pi) \delta(1-z_0-z_1) \\
    &\times e^{-i \bt \vdot \Deltat}  \widetilde \Psi_{\lambda_\gamma}^{\gamma^* \to q \bar q}(\xt_{01}, z_i) 
    \left(\widetilde \Psi_{\lambda_V}^{V \to q \bar q }(\xt_{01},z_i)\right)^\ast
    \left( 1- \dipole_{01} \right)
\end{split}
\end{equation}
where the notations for the wave functions and the variables are explained in Sec.~\ref{sec:eikonal_approximation}.
The dipole amplitude depends on the rapidity variable in the process according to the JIMWLK equation in Sec.~\ref{sec:rapidity_evolution}, and usually in the leading-order calculations the rapidity is chosen as $Y = \ln(1/\xpom)$ where
\begin{equation}
    \xpom = \frac{q \vdot (P_n - P_n')}{q \vdot P_n} = \frac{Q^2 + M_V^2 -t}{W^2+Q^2-m_n^2}
\end{equation}
is the momentum fraction carried by the pomeron in the high-energy limit, and the momentum-transfer squared is given by $t = (P_n - P_n')^2 \approx - \Deltat^2$.
Note that here the impact parameter $\bt = z_0 \xt_0 + z_1 \xt_1$ is defined as the center-of-mass position in the transverse plane, which is the Fourier conjugate of the momentum transfer $\Deltat$.
In the literature, it is common to write the amplitude in terms of the average of the transverse coordinates, $\btvar = \frac{1}{2} (\xt_0 + \xt_1) = \bt + \left(\frac{1}{2}-z_0 \right) \xt_{01}$, and assume that the dipole amplitude $N\left(\xt_{01}, \btvar \right) = \left\langle 1-\dipole_{01} \right \rangle$ does not depend on the angle between $\xt_{01}$ and $\btvar$. This simplifies numerical calculations  and was also done in Article~\cite{relativistic}.

Experimentally, exclusive vector meson production is a very clean event, as the vector meson is the only particle produced. Then only the decay products of the vector meson need to be measured, which can be done very accurately by measuring the decay to a muon pair for example. 
For sufficiently high energies, $\xpom \lesssim 0.01$, there exists data from electron-proton collisions at HERA~\cite{ZEUS:1998xpo,ZEUS:2002wfj,ZEUS:2004yeh,H1:2005dtp,H1:2009cml,H1:2013okq} and nuclear collisions at LHC~\cite{ALICE:2012yye,ALICE:2013wjo,ALICE:2019tqa,ALICE:2021gpt,LHCb:2021bfl,CMS:2016itn,CMS:2023snh}. The HERA data has been measured for various different center-of-mass energies $W$ and photon virtualities $Q^2$, providing accurate data for both light and heavy vector mesons. At the LHC, the measurements have been done in nucleus-nucleus collisions where one of the nuclei emits the photon. In practice, measuring exclusive vector meson production in nuclear collisions requires a high impact parameter between the nuclei, which renders the photons to be quasi-real with a virtuality $Q^2 \approx 0$\footnote{Virtual photons have a lifetime $\sim 1/Q$, which means that for impact parameters higher than the nuclear radii only photons with $Q^2 \approx 0$ contribute.}. 
This has consequences for the perturbativity of the process, as 
the perturbative scale for exclusive vector meson production is given by $Q^2 + M_V^2$. Thus, for light vector mesons one cannot expect the process to be perturbative for small photon virtualities, whereas for heavy mesons the meson mass provides a perturbative scale. This allows us to compute exclusive heavy vector meson production also in the LHC kinematics, but for light vector mesons we have only the HERA data to compare to.
In the future, more data will be expected to come from the future EIC where exclusive vector meson production will be measured in electron-nucleus collisions~\cite{Accardi:2012qut,Aschenauer:2017jsk,AbdulKhalek:2021gbh}. This will also allow measurements of production from heavy nuclei for non-zero photon virtualities.

In addition to measuring the dependence on the energy and the photon virtuality, exclusive vector meson production also allows for the measurement of the momentum transfer $t = -\Deltat^2$. Measuring the momentum transfer dependence gives us information about the impact parameter dependence of the interaction with the target, as is shown explicitly in Eq.~\eqref{eq:VM_production_amplitude_LO} by the Fourier term $e^{-i \bt \vdot \Deltat}$.
Measuring the momentum-transfer dependent cross section also allows us to consider \textit{coherent} and \textit{incoherent} vector meson production separately. In the Good-Walker approach to diffraction~\cite{Good:1960ba}, these are defined as:
\begin{align}
\label{eq:vm_cross_section}
    \frac{\dd \sigma}{\dd \abs{t}} &= \frac{1}{4\pi} \langle \abs{\mathcal{M}}^2 \rangle
    =\left(\frac{\dd \sigma}{\dd \abs{t}} \right)_\text{coherent} +  \left(\frac{\dd \sigma}{\dd \abs{t}} \right)_\text{incoherent} \\
\label{eq:vm_cross_section_coh}
    \left(\frac{\dd \sigma}{\dd \abs{t}} \right)_\text{coherent} &= \frac{1}{4\pi} \abs{\langle \mathcal{M} \rangle}^2 \\
\label{eq:vm_cross_section_incoh}
    \left(\frac{\dd \sigma}{\dd \abs{t}} \right)_\text{incoherent} &= \frac{1}{4\pi}\left(  \langle \abs{\mathcal{M}}^2 \rangle -  \abs{\langle \mathcal{M} \rangle}^2 \right).
\end{align} 
Note that the factor $1/(4\pi)$ in Eq.~\eqref{eq:vm_cross_section} is a convention that depends on the definition of the amplitude, and a different convention was used in Articles~\cite{
relativistic,
heavy_long,
light,
heavy_trans}. 
Here $\langle \cdots \rangle$ denotes the average over the target configurations described in Sec.~\ref{sec:MV}, and thus
the difference between the coherent and the total cross section is in the final configuration of the nucleus: total production allows any color neutral configuration of the target, whereas in coherent production the final state is in the same configuration as the initial state. Incoherent production then corresponds to a ``variance'' of the amplitude and it measures fluctuations in the target configuration.

\begin{figure}
    \centering
    \includegraphics[width=0.7\textwidth]{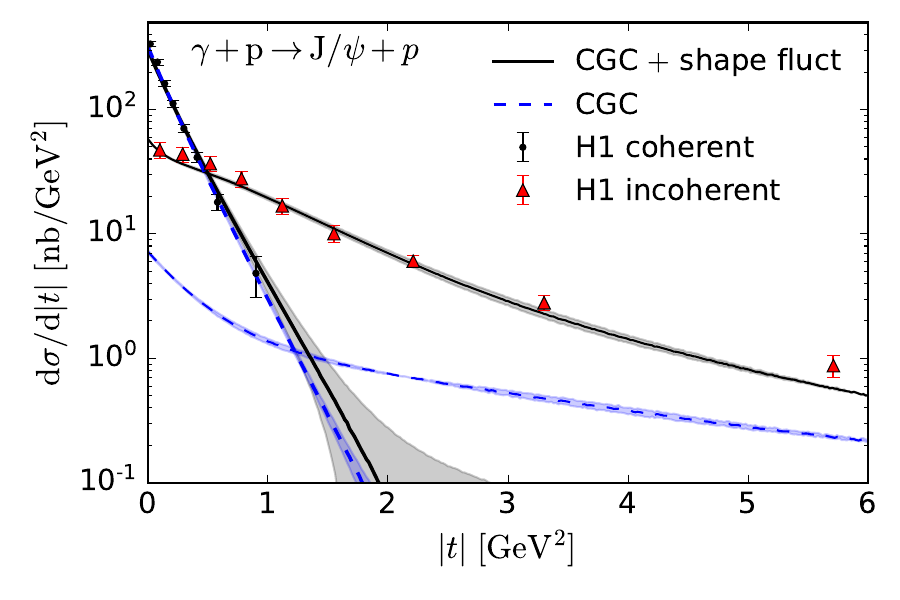}
    \caption{ HERA data~\cite{H1:2013okq} for coherent and incoherent $\jpsi$ production with predictions from a CGC model.  Note that including shape fluctuations (in black solid line) is important for incoherent production as it measures target fluctuations. 
    Figure from Ref.~\cite{Mantysaari:2022sux}:
    H. Mäntysaari, F. Salazar and B. Schenke, \textit{Nuclear geometry at high energy
    from exclusive vector meson production}, \textit{Phys. Rev. D} 106 (2022) no. 7 074019, DOI: \href{https://doi.org/10.1103/PhysRevD.106.074019}{10.1103/PhysRevD.106.074019}. Reproduced under the license CC BY 4.0.
    }
    \label{fig:coherent_vs_incoherent}
\end{figure}

Coherent and incoherent production are relevant at different scales for the momentum exchange $t$ as shown in Fig.~\ref{fig:coherent_vs_incoherent}. In general, coherent production is more important for low values of $\abs{t}$ where the data is also the most accurate. 
The $t$-dependence of the coherent cross section has been experimentally found to be well described by
\begin{equation}
    \label{eq:vm_production_t_dependence}
   \frac{\dd \sigma}{\dd \abs{t}}
   \approx e^{-b_\text{eff} \abs{t}} \frac{\dd \sigma}{\dd \abs{t}} (t=0)
\end{equation}
where $b_\text{eff}$ is a parameter that should be fitted to the experimental data~\cite{ZEUS:2002wfj, ZEUS:2005bhf, H1:2009cml, ZEUS:1998xpo}. It can be understood as the effective transverse area of the meson-target system.
This phenomenological model for the $t$-dependence of coherent production allows one to estimate the $t$-integrated coherent cross section from the differential cross section at $t=0$, as was done in Articles~\cite{heavy_long, light, heavy_trans} to avoid additional modeling for the impact parameter dependence of the dipole amplitude.
Coherent production also allows one to use the target-averaged dipole amplitude that also appears in inclusive DIS. For this reason, only the coherent production was considered in Articles~\cite{relativistic,heavy_long,light,heavy_trans}.

Some notes should also be made about the difference in the dipole amplitude used in  exclusive vector meson production and inclusive DIS. First of all, also the real part of the invariant amplitude
(i.e. the imaginary part of the dipole amplitude)
contributes to exclusive vector meson production as opposed to inclusive DIS which depends only on the imaginary part of the invariant amplitude
(i.e. the real part of the dipole amplitude)
by the optical theorem. 
Thus, if one wishes to use the dipole amplitude determined from inclusive DIS, one has to account for the real part of the production amplitude by other means. This contribution can be estimated to be small in the high-energy limit, and using Regge theory it is possible to write it as~\cite{Gribov:1968ie,Nemchik:1996cw}
\begin{equation}
    \Re \mathcal{M} = \Im \mathcal{M} \times  \tan(\frac{\pi}{2} \delta )
\end{equation}
where 
\begin{equation}
    \delta = \frac{\partial}{\partial (1/\xpom)} \Im \mathcal{M}.
\end{equation}

Another difference between the dipole amplitude in exclusive vector meson production and inclusive DIS is that the dipole amplitude may depend on the minus-momentum exchange in the process.
This means that when taking the average $\langle \cdots \rangle$ over the target configurations in exclusive vector meson production, the initial state $A$ and the final state $A'$ for the target have different momenta. 
This is in contrast to  inclusive DIS where using the optical theorem to calculate the cross section corresponds to forward elastic scattering where the initial and final states have exactly the same momenta.
While the plus-momentum is conserved in the interaction with the target, the transverse and minus components might differ for vector meson production, with the transverse-momentum exchange $t = - \Deltat^2$ and the minus-momentum exchange
\begin{equation}
    \frac{P_n^- - P_n^{\prime -}}{P_n^-} \approx \frac{Q^2+ M_V^2 -t}{W^2 + Q^2 - m_n^2} = \xpom.
\end{equation}
While the $t$-dependence can be understood from the Fourier transform of the dipole amplitude,
\begin{equation}
    \int \dd[2]{\bt} e^{- i \bt \vdot \Deltat} N(\xt_{01}, \bt),
\end{equation}
the minus-momentum exchange is more complicated.
In terms of collinear factorization, this non-zero minus-momentum exchange is related to the fact that structure functions can be written in terms of the standard parton distribution functions while exclusive vector meson production requires \textit{generalized} parton distributions (GPDs)~\cite{Collins:1996fb}.
The difference coming from the minus-momentum exchange can be estimated in certain limits using collinear factorization where it is related to the \textit{skewness} of the GPDs.
In Ref.~\cite{Shuvaev:1999ce}, the difference between maximally skewed GPDs and standard PDFs was calculated, and assuming that the interaction between the quark-antiquark dipole and the target happens through an exchange of two gluons one can estimate how this correction should appear in exclusive vector meson production.
While this derivation of the skewness correction does not directly apply to the dipole picture,
this skewness correction is often included in phenomenological data comparisons by multiplying the production amplitude by the factor~\cite{Martin:1999wb,Kowalski:2006hc}
\begin{equation}
    R_g = \frac{2^{2\delta +3}}{\sqrt{\pi}} 
    \frac{\Gamma\left(\delta + \frac{5}{2}\right)}{\Gamma(\delta +4)}.
\end{equation}

These corrections appear as a general overall factor that increases the cross section. The real-part correction is generally smaller, less than $20\%$, but the skewness correction can vary between $20\%$ and $60\%$~\cite{Mantysaari:2016jaz,Mantysaari:2017dwh}. The corrections also depend on the energy $W$ and the photon virtuality $Q^2$, decreasing for higher energies and lower photon virtualities, i.e. the corrections become numerically less important for smaller $\xpom$.
The real-part and skewness corrections were included for comparisons with the data in Article~\cite{relativistic} but were left out of Articles~\cite{heavy_long,light,heavy_trans} where the main focus was on the calculation and not on data comparisons. 

\section{Vector meson wave function}
\label{sec:wave_function}

The dependence on the produced vector meson is completely determined by the meson wave function. As a nonperturbative quantity, the vector meson wave function is also a major source of theoretical uncertainty for the process.
The uncertainty can be somewhat reduced by considering symmetry relations for the meson wave function. In equal-time quantization, one could use the $\so(3)$ rotational symmetry and spin-parity conservation to write the wave function for the $q \bar q $ state in the rest frame as
\begin{equation}
\label{eq:wf_rf_expansion}
    \Psi_{s \bar s}^\lambda(\vec r) 
    = \sum_{L m_L m_s} 
    \left\langle \frac{1}{2} s, \frac{1}{2} \bar s  \Bigg| 1 m_s  \right\rangle
    \left\langle 1 m_s , L m_L  | 1 \lambda  \right\rangle
    Y_L^{m_L}(\Omega_r) \phi^L(\abs{\vec r}) 
\end{equation}
where $\langle \cdots, \cdots | \cdots \rangle$ are the Clebsch-Gordan coefficients, $Y_L^{m_L}$ spherical harmonics, $\Omega_r$ is the angular part of the 3-vector $\vec r$ and $\phi^L$ is the radial part of the wave function corresponding to the orbital angular momentum $L$. For vector mesons, the only possible values for the orbital angular momentum are $L=0, 2$ which correspond to the S- and D-waves in the spectroscopic notation.
In light-cone perturbation theory, one cannot write the light-cone wave function in this form. This follows from the fact that the $\so(3)$ symmetry is broken by the light-cone quantization such that only the $\so(2)$ rotation symmetry of the transverse plane remains. 
This means that the total angular momentum is not explicitly conserved, but the magnetic quantum numbers, $\lambda$, $m_s$, $m_L$, corresponding to the angular momentum in the transverse plane are conserved. 
This can be used to factorize the dependence on the transverse angle $\varphi$ out of the wave function by
\begin{equation}
    \label{eq:vm_wf_angular_dependence}
        \Psi^\lambda_{s \bar s} (\rt, z) = e^{i \varphi m_L} \phi_{s \bar s}^\lambda(\abs{\rt}, z)
\end{equation}
where $m_L = \lambda - (s + \bar s)$ is the orbital magnetic quantum number and $\phi_{s \bar s}^\lambda(\abs{\rt}, z)$ is the part of the wave function that is independent of the transverse angle.

Parity is not a symmetry in light-cone perturbation theory as it requires also changing the sign of the $x^3$-axis, which corresponds to interchanging $x^+ \leftrightarrow x^-$. 
Instead, one can consider the \textit{mirror parity} defined as $\hat{P}_x = \hat{R}_x(\pi) \hat{P}$, where $\hat{P}$ is the standard parity operator and $\hat{R}_x(\phi)$ corresponds to a rotation around the $x^1$-axis by the angle $\phi$. This definition of the mirror parity corresponds to changing the sign of the $x^1$-axis~\cite{Li:2015zda,Soper:1972xc}. 

To see how the mirror parity and $C$-parity act on a meson, consider an eigenstate of the spin-parity $J^{PC}$ with a definite polarization $\lambda$. We can write this as
\begin{equation}
    \ket{J^{PC}, \lambda} = \int \dd[2]{\xt_0} \dd[2]{\xt_1} \int_0^1 \frac{\dd{z}}{ z (1-z)} \sum_{s \bar s} \Psi^\lambda_{s \bar s}(\xt_{01}, z) \ket{q_s(\xt_0,z) \bar q_{\bar s}(\xt_1,1-z)}
\end{equation}
at leading order.
Charge conjugation interchanges the quark and antiquark such that
\begin{equation}
\begin{split}
    \hat{C}\ket{J^{PC}, \lambda} 
    &= \int \dd[2]{\xt_0} \dd[2]{\xt_1}\int_0^1 \frac{\dd{z}}{ z (1-z)} \sum_{s \bar s} \Psi^\lambda_{s \bar s}(\xt_{01}, z) \ket{ \bar q_s(\xt_0,z)  q_{\bar s}(\xt_1,1-z)} \\
    =& -\int \dd[2]{\xt_0} \dd[2]{\xt_1}\int_0^1 \frac{\dd{z}}{ z (1-z)} \sum_{s \bar s} \Psi^\lambda_{\bar s s}(-\xt_{01}, 1-z) \ket{ q_s(\xt_0,z)  \bar q_{\bar s}(\xt_1,1-z)} \\
    =&  \int \dd[2]{\xt_0} \dd[2]{\xt_1} \int_0^1 \frac{\dd{z}}{ z (1-z)}
    (-1)^{1+m_L} \Psi^\lambda_{\bar s s}(\xt_{01}, 1-z) \ket{ q_s(\xt_0,z)  \bar q_{\bar s}(\xt_1,1-z)}
\end{split}
\end{equation}
which leads to the identity
\begin{equation}
    \label{eq:vm_wf_c-parity}
    C \Psi^\lambda_{s \bar s}(\xt_{01}, z) 
    = (-1)^{1+m_L}  \Psi^\lambda_{\bar s s}(\xt_{01}, 1-z).
\end{equation}
Similarly, for mirror parity we get
\begin{equation}
\begin{split}
    \hat{P}_x\ket{J^{PC}, \lambda} 
    \hspace{-1.5pt}=& \int \dd[2]{\xt_0} \dd[2]{\xt_1}\int_0^1 \frac{\dd{z}}{ z (1-z)} \sum_{s \bar s} \Psi^\lambda_{s \bar s}(\xt_{01}, z) \ket{ q_{-s}(P_x \xt_0,z)  \bar q_{-\bar s}(P_x \xt_1,1-z)} \\
    =& \int \dd[2]{\xt_0} \dd[2]{\xt_1} \int_0^1 \frac{\dd{z}}{ z (1-z)} \sum_{s \bar s} \Psi^\lambda_{-s, - \bar s}(P_x \xt_{01}, z) \ket{ q_{s}( \xt_0,z)  \bar q_{\bar s}( \xt_1,1-z)} \\
    =& \int \dd[2]{\xt_0} \dd[2]{\xt_1} \int_0^1 \frac{\dd{z}}{ z (1-z)} \\
    & \times \sum_{s \bar s} (-1)^{m_L} e^{-2im_L \varphi} \Psi^\lambda_{-s, - \bar s}( \xt_{01}, z) \ket{ q_{s}( \xt_0,z)  \bar q_{\bar s}( \xt_1,1-z)}
\end{split}
\end{equation}
where $P_x \rt$ corresponds to mirroring the vector $\rt$ around the $x^2$-axis so that its $x^1$-component is flipped.
Noting that mirror parity acts as~\cite{Li:2015zda}
\begin{equation}
    \hat{P}_x\ket{J^{PC}, \lambda}  =  (-1)^J P \ket{J^{PC}, -\lambda},
\end{equation} this leads to
\begin{equation}
    \label{eq:vm_wf_P-parity}
    (-1)^J P \Psi^{-\lambda}_{s \bar s}(\xt_{01}, z) 
    = (-1)^{m_L} e^{-2im_L \varphi}  \Psi^\lambda_{-s, - \bar s }(\xt_{01}, z).
\end{equation}

Substituting the spin-parity $J^{PC} = 1^{--} $ of the vector meson in Eqs.~\eqref{eq:vm_wf_c-parity} and \eqref{eq:vm_wf_P-parity} leads to a set of relations between different components of the light-cone wave function which somewhat restricts the degrees of freedom. 
It should be mentioned that these relations assume the sign convention
$ \boldsymbol{\varepsilon}^\lambda = \frac{1}{\sqrt{2} }( -\lambda , -i) $
for the polarization vectors which is consistent with the Condon-Shortley convention for spherical harmonics.
Also, it is assumed that the spinors are eigenstates of the $x^3$-component of the total angular momentum operator $\hat J_3 = \hat  L_3 +  \hat S_3$, and the quark and antiquark spinors are related by charge conjugation, $v_s(\vec k) = -i \gamma^2 u_s(\vec k)$. These assumptions about the spinors are true for most spinor bases used in the literature, and especially the Lepage--Brodsky basis~\eqref{eq:LB_spinors} satisfies these.

Using the angular dependence of the meson wave function~\eqref{eq:vm_wf_angular_dependence}, it is possible to show that vector meson production is highly suppressed unless the polarizations of the photon and meson are the same~\cite{Mantysaari:2020lhf}. In fact, if one assumes that the dipole amplitude $N(\xt_{01}, \bt)$ does not depend on the angle between $\xt_{01}$ and $\bt$, the leading-order production amplitude for differing polarizations vanishes. Therefore the contribution from polarization-changing components is usually ignored, and one considers only the case $\lambda_\gamma = \lambda_V$.

\subsection{Relativistic corrections to the heavy vector meson wave function}
\label{sec:heavy_VM_wave_function}

When talking about heavy vector mesons, one usually means heavy {quarkonia} states such as $\jpsi$ or $\Upsilon$. The main advantage of these particles is that they can be treated as nonrelativistic states such that the relative velocity $v$ of the quark and antiquark is small, $v \ll 1$. There are multiple ways to describe a nonrelativistic state mathematically. For example, potential models using the Schrödinger equation have been quite successful in explaining qualitatively the existing quarkonium states~\cite{Eichten:1979ms,Barik:1980ai,Quigg:1977dd,Buchmuller:1980su}, and by solving the Schrödinger equation one can obtain a rest-frame wave function that can be used for other calculations. 

Another possibility is to use the effective field theory of nonrelativistic quantum chromodynamics (NRQCD) which has been developed for describing quarkonium states~\cite{Bodwin:1994jh}. The main idea of NRQCD is to expand quantities as a power series of the velocity of the heavy quark $v$, such that the nonperturbative physics is described by universal long-distance matrix elements (LDMEs) that appear both in the decay and production of quarkonia. The LDMEs can then be related to the rest-frame wave function and its derivatives at the origin. The leading-order approach in NRQCD is to treat the quark-antiquark pair moving at zero velocity, meaning that the rest-frame wave function is a delta function in momentum space, $\Psi(\vec k) \sim (2\pi)^3 \delta^3 (\vec k )$, or equivalently a constant in position space, $\widetilde \Psi(\vec r) \sim 1$. These wave functions are proportional to the LDME $\langle \mathcal{O}_1 \rangle$ which can be determined e.g. from the leptonic width of the corresponding quarkonium state.

Higher-order terms in NRQCD can be included order by order in terms of the heavy quark velocity,
which can also be used to add relativistic corrections to the vector meson wave function.
Assuming that the meson wave function is peaked around $\vec k = 0$, the standard expectation value
\begin{equation}
    \int \frac{\dd[3]{k}}{(2\pi)^3} k^{2n} \abs{\Psi(\vec k)}^2 = \langle k^{2n} \rangle
\end{equation}
also suggests that
\begin{equation}
    \int \frac{\dd[3]{k}}{(2\pi)^3} k^{2n} \Psi(\vec k) \sim \langle k^{2n} \rangle \int \frac{\dd[3]{k}}{(2\pi)^3}\Psi(\vec k)
\end{equation}
which in the position space corresponds to
\begin{equation}
    \nabla^{2n} \widetilde \Psi(0)
    \sim \langle k^{2n} \rangle \widetilde \Psi(0).
\end{equation}
Writing then the wave function $\widetilde \Psi(\vec r)$ as a Taylor series, we can note that each term in the series is suppressed by the velocity $v$ as
\begin{equation}
    \label{eq:NRQCD_expansion}
    \widetilde \Psi(\vec r) = 
    \underbrace{\widetilde \Psi(0)}_{\mathcal{O}(v^0)}
    + \underbrace{r^i \partial_i   \widetilde \Psi(0) }_{\mathcal{O}(v^1)}
    + \underbrace{\frac{1}{2} r^i r^j \partial_i \partial_j   \widetilde \Psi(0) }_{\mathcal{O}(v^2)}
    + \mathcal{O}(v^3).
\end{equation}
This series can then be truncated at the desired point to include corrections of the order $\mathcal{O}(v^n)$, and after this the wave function uncertainty is reduced to a finite number of unknown constants corresponding to derivatives of the wave function at the origin $\partial^n \widetilde \Psi(0)$. These unknown constants can then be written in terms of the universal LDMEs of NRQCD~\cite{Bodwin:2007fz}.

In addition to the suppression of higher orders in the expansion~\eqref{eq:NRQCD_expansion}, the non-dominant orbital angular momentum and spin  components are velocity-suppressed.
For example, for the lowest-energy heavy vector mesons $\jpsi$ and $\Upsilon$ the dominant spin component is the S-wave, and correspondingly the D-wave is suppressed by $v^2$~\cite{Bodwin:1994jh}. The D-wave is suppressed even further in the decay and production of these particles, as the D-wave component has to be combined with terms proportional to $\vec k^2$ to give a non-zero result. This means that the D-wave component is in total suppressed by $v^4$ in the production of $\jpsi$ and $\Upsilon$. The situation is similar for non-dominant spin components which are proportional to $e^{i \varphi m_L}$ according to Eq.~\eqref{eq:vm_wf_angular_dependence}. These have to be combined similarly with terms proportional to $k^{\abs{m_L}}e^{-i \varphi m_L}$ to yield a non-zero contribution, which brings an additional suppression of $v^{\abs{m_L}}$.

These ideas of using NRQCD to add relativistic corrections were used in Article~\cite{relativistic} where we considered the order $v^2$ correction to the rest-frame wave function.
At this order, we can neglect the D-wave such that we only have the S-wave contribution, and by rotational symmetry only the first and third terms in Eq.~\eqref{eq:NRQCD_expansion} are non-zero, which corresponds to having two unknown constants in the wave function.
Numerical values for these constants have been determined in Ref.~\cite{Bodwin:2007fz} for $\jpsi$ and in Ref.~\cite{Chung:2010vz} for $\Upsilon$ by considering electromagnetic decays of quarkonia.

The potential models and the NRQCD approach give us the rest-frame wave function which is not the same thing as the light-cone wave function required for calculating processes in light-cone perturbation theory.
The exact relation between the two wave functions is not known as it is highly nontrivial because of the different quantizations leading to the wave functions.
Discarding the differences in the Fock state expansions of the states corresponding to different quantization schemes, one can treat the differences between the two wave functions arising from the differences in the spinors and the variables. 

For the rest-frame wave function, it is more convenient to use the standard 3-momentum or -position coordinates as variables. The standard spinor basis for the rest frame is the Bjorken--Drell basis~\cite{Bjorken:1965sts}:
\begin{align}
    \label{eq:BD_spinors}
    u_s =& \sqrt{E + m}
    \left( 
    \begin{matrix}
        \xi_s \\
        \frac{\vec \sigma \vdot \vec k}{E+m} \xi_s
    \end{matrix}
    \right)
    &
    v_s =& \sqrt{E + m}
    \left( 
    \begin{matrix}
        \frac{\vec \sigma \vdot \vec k}{E+m} \chi_s \\
        \chi_s
    \end{matrix}
    \right)
\end{align}
where
\begin{align}
    \xi_\uparrow =&
    \left( 
    \begin{matrix}
        1 \\ 0
    \end{matrix}
    \right)
    &
    \xi_\downarrow =&
    \left( 
    \begin{matrix}
        0 \\ 1
    \end{matrix}
    \right)
    &
    \chi_\uparrow =&
    \left( 
    \begin{matrix}
        0 \\ -1
    \end{matrix}
    \right)
    &
    \chi_\downarrow =&
    \left( 
    \begin{matrix}
        1 \\ 0
    \end{matrix}
    \right).
\end{align}
Here $E$ is the energy of the particle and $\sigma_i$ are the Pauli matrices.
They satisfy the relations $v_s(\vec k ) = -i \gamma^2 u_s(\vec k)^* $ and $\chi_s = i \sigma_2 \xi_s^*$ which are useful when considering the conservation of $C$-parity.
These spinors are also the eigenstates of the spin-operator $\hat S_3$ boosted to the particle's rest frame~\cite{Bogolyubov:1975ps}.
This is the reason why the Bjorken--Drell basis is sometimes called the \emph{spin} basis, and it also allows us to use the conservation of angular momentum to describe the wave function as a combination of eigenstates of the operators $\hat{L}_3$ and $\hat{S}_3$. This leads to the decomposition of the wave function in terms of components with specific $L$ and $S$ quantum numbers in Eq.~\eqref{eq:wf_rf_expansion}. 

As described in Sec.~\ref{sec:eikonal_approximation}, in light-cone perturbation theory the convenient basis for the spinors is the Lepage--Brodsky basis~\eqref{eq:LB_spinors}. This is the reason why this choice for the spinors is usually also made for describing the light-cone wave function. 
To describe the light-cone wave function in terms of the rest-frame wave function correctly, one needs to correct for this difference in the spinor basis. 
This can be done by a simple change of the basis in the vector space of the spinors.
Note that we can write the Lepage--Brodsky spinors in the form of Eq.~\eqref{eq:BD_spinors} using the following 2-spinors\footnote{In fact, it is possible to write \textit{any} choice of the spinors in the Bjorken--Drell form with suitably chosen 2-spinors. This follows from the fact that the solutions for the Dirac equations $(\slashed{k}-m)u(k) = 0$ and $(\slashed{k}+m)v(k)=0$ form  2-dimensional vector spaces.}:
\begin{align}
    \xi_+(\vec k) =&
    N
    \left( 
    \begin{matrix}
        E+k^3+m \\ 
        k_R
    \end{matrix}
    \right)
    &
    \xi_-(\vec k) =&
    N
    \left( 
    \begin{matrix}
        -k_L \\ 
        E+k^3+m
    \end{matrix}
    \right)
    \\
    \chi_+(\vec k) =&
    N
    \left( 
    \begin{matrix}
        k_L \\ 
        -(E+k^3+m)
    \end{matrix}
    \right)
    &
    \chi_-(\vec k) =&
    N
    \left( 
    \begin{matrix}
        E+k^3+m \\ 
        k_R
    \end{matrix}
    \right)
\end{align}
where $N = \frac{1}{\sqrt{2(E+m)(E+k^3)}}$ is a normalization factor.
The change for the spinor basis can then be written as
\begin{equation}
    \Psi_{h \bar h} = \sum_{s= \uparrow, \downarrow} \sum_{\bar s= \uparrow, \downarrow} \xi^\dag_h \xi_s \chi^\dag_{\bar s} \chi_{\bar h} \Psi_{s \bar s}.
\end{equation}
This procedure is also called the \textit{Melosh rotation}~\cite{Melosh:1974cu}.
It should be noted that the Melosh rotation preserves the symmetry relations~\eqref{eq:vm_wf_angular_dependence}, \eqref{eq:vm_wf_c-parity}, \eqref{eq:vm_wf_P-parity} between the different components of the wave functions as both the Lepage--Brodsky and Bjorken--Drell bases satisfy the assumptions required for these relations.

The Melosh rotation is sometimes thought of as a boost to the ``infinite-momentum frame''~\cite{Beyer:1998xy,Hufner:2000jb}. To see this, note that a longitudinal boost with a rapidity $Y$ acts on the Bjorken--Drell spinors as
\begin{equation}
    \exp( \frac{1}{4} [\gamma^0, \gamma^3] Y ) u_s(\vec k) = \sqrt{E' + m}
    \left(
    \begin{matrix}
        \xi'_s \\ \frac{\vec \sigma \vdot \vec k'}{E' + m} \xi_s'
    \end{matrix}
    \right)
\end{equation}
where 
\begin{align}
    E' =& E \cosh Y + k^3 \sinh Y,
    &
    k^{3\prime} =& E \cosh Y + k^3 \sinh Y,
    &
    \kt' =& \kt,
\end{align}
correspond to the boosted energy and momenta and the 2-spinor is transformed into
\begin{equation}
    \xi_s' = \frac{1}{\sqrt{(E+m)(E'+m)}} 
    \left[ \cosh(\frac{Y}{2}) (E'+m)-\sinh(\frac{Y}{2})  \sigma \vdot \vec k' \sigma^3\right] 
    \xi_s.
\end{equation}
We can now consider the quark to be moving with a very high momentum $k^3$, and we wish to boost it closer to the rest frame.
This corresponds to taking $Y \to -\infty$ which then changes the form of the 2-spinors to match the Lepage--Brodsky basis by $\xi'_\uparrow \to \xi_+(\vec k')$ and $\xi'_\downarrow \to \xi_- (\vec k')$.
This means that the spinors in the Lepage--Brodsky basis can be thought of as the Bjorken--Drell spinors boosted to the ``infinite-momentum frame''.
It should be stressed, however, that such an infinite boost is not \textit{required} for the Melosh rotation.
Instead, we view it as a mathematical transformation between the two different spinor bases without any physical meaning.
One is free to choose the spinor basis as one wishes, and in this case the convenient bases for the rest frame and the light cone simply happen to be different.

The other correction one has to make when going from the rest-frame wave function to the light-cone wave function is the change in the variables.
Essentially, one has to change $k^3 \to k^+$, which is nontrivial as in the rest-frame the total energy is not conserved and in the light cone it is the minus component of the momentum that is not conserved.
A common approach is to assume the conservation of the plus-momentum, which in the rest frame of the quark-antiquark pair leads to the expression
\begin{equation}
    k^3 = M_{q \bar q} \left( z - \frac{1}{2} \right)
\end{equation}
where
\begin{equation}
    M_{q \bar q}^2 = \frac{\kt^2 + m^2}{z(1-z)} 
\end{equation}
is the invariant mass of the quark-antiquark pair.
This change of variables is also known as the \textit{Terentev substitution}~\cite{Terentev:1976jk}.
The main problem with this is that it is impossible to conserve both the energy and the plus momentum in the wave function, or in other words the center-of-mass energy of the quark-antiquark pair $M_{q \bar q}$ does not agree with the mass of the meson $M_V$, and thus this relation cannot be truly exact. 
In Article~\cite{relativistic}, this appears through the fact that the leptonic widths for longitudinal and transverse modes differ depending on whether it is the $M_{q \bar q}$ or $M_V$ that appears in the equations. 
The difference between the two masses can be seen as corrections from higher-order Fock states, related to the fact that the vector meson is the (approximate) eigenstate of the full Hamiltonian and the quark-antiquark pair only of the free Hamiltonian. Thus, we can treat the ambiguity in changing the variables as a higher-order correction that would need to be remedied if one were to consider corrections higher order in both velocity and $\as$.

This combination of the Melosh rotation and the Terentev substitution is a common way to get a light-cone wave function from the rest-frame wave function. In Refs.~\cite{Krelina:2018hmt,Krelina:2019egg,Cepila:2019skb,Babiarz:2019sfa,Babiarz:2020jkh,Babiarz:2023ebe} it has been used for a potential-model wave function, and in Article~\cite{relativistic} we used it for the NRQCD-based wave function to include the relativistic $v^2$ corrections.
From the explicit form of this \emph{NRQCD expansion} wave function one can note several things as a consistency check. First, the symmetry relations~\eqref{eq:vm_wf_angular_dependence}, \eqref{eq:vm_wf_c-parity} and \eqref{eq:vm_wf_P-parity} are satisfied without imposing them directly. Second, the non-dominant spin components have the additional suppression of $v^{\abs{m_L}}$ as explained previously.

\begin{figure}
	\centering
    \begin{subfigure}{0.49\textwidth}
        \centering
        \includegraphics[width=\textwidth]{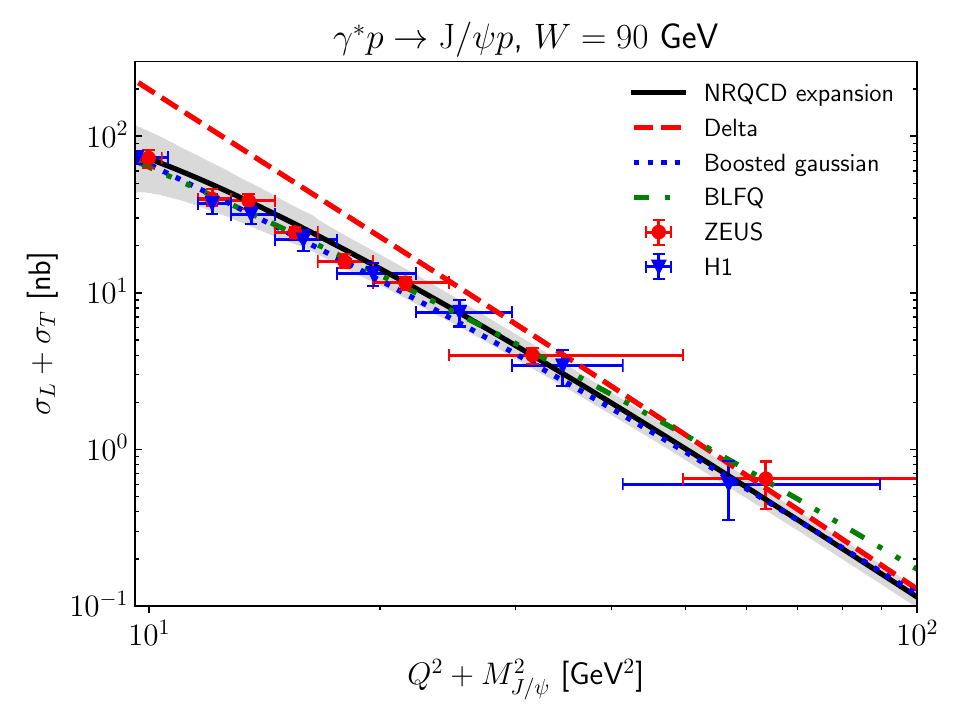}
        \caption{Total exclusive $\jpsi$ production. }
        \label{fig:jpsi_production_Q2_different_wf}
    \end{subfigure}
    \begin{subfigure}{0.49\textwidth}
        \centering
        \includegraphics[width=\textwidth]{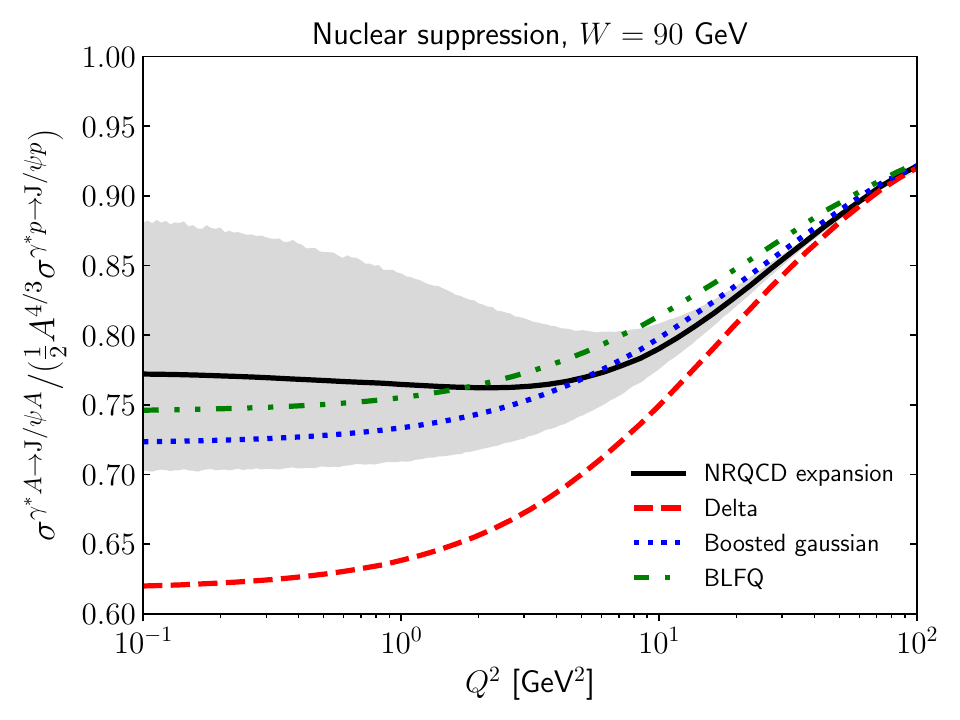}
        \caption{ Nuclear suppression for $\jpsi$ production.}
        \label{fig:jpsi_nuclear_suppression}
    \end{subfigure}
	  \caption{
   Exclusive $\jpsi$ production as a function of the photon virtuality $Q^2$ compared to the HERA data~\cite{H1:2005dtp,ZEUS:2004yeh} for different wave functions. 
   \emph{Delta} is the fully nonrelativistic limit for the wave function and \emph{NRQCD expansion} includes the $v^2$ relativistic corrections.
   \emph{Boosted gaussian}~\cite{Kowalski:2006hc} and \emph{BLFQ}~\cite{Li:2017mlw} are phenomenological wave functions fitted to the leptonic width of $\jpsi$ and the charmonium mass spectrum, respectively. 
   The IPsat parametrization from Ref.~\cite{Mantysaari:2018nng} was used for the dipole amplitude.
   Figures from Article~\cite{relativistic}, reproduced under the license CC BY 4.0.
   }
        \label{fig:jpsi_LO}
\end{figure}

The results of including these $v^2$ corrections in NRQCD  to $\jpsi$ production  at  leading order in $\as$ are shown in Fig.~\ref{fig:jpsi_production_Q2_different_wf}.
The agreement with the HERA data is extremely good, although the error band coming from the uncertainties of the LDMEs is quite large. 
Nevertheless, comparing the results to the fully nonrelativistic case we see that these relativistic corrections are important for small photon virtualities near $Q^2=0$.
For large photon virtualities the relativistic corrections lose their importance, in agreement with explicit calculations in the limit $Q^2\to \infty$~\cite{Hoodbhoy:1996zg}.
The dependence of relativistic effects on the photon virtuality can be understood by noting that the dipole sizes probed in the process are roughly $\rt^2 \sim 1/(Q^2 + M_V^2)$, which means that for small virtualities larger dipoles become important.
It is this dependence on large dipoles in the meson wave function that is modified by the relativistic corrections as can be seen from the expansion in Eq.~\eqref{eq:NRQCD_expansion}.
It should be noted that for the $\Upsilon$ particle the relativistic corrections are not expected to be important even for low photon virtualities, as the estimates for the average heavy quark velocity are very low~\cite{Chung:2010vz}.

In Fig.~\ref{fig:jpsi_nuclear_suppression}, we show estimates for nuclear suppression of $\jpsi$ production in the EIC kinematics.
Nuclear suppression is defined as the ratio
\begin{equation}
    \label{eq:nuclear_suppression}
   R_A =  \frac{\sigma^{\gamma^* + A \to \jpsi + A}}{c A^{4/3} \sigma^{\gamma^* + p \to \jpsi + p}}
\end{equation}
which can be used as a measure of nonlinear effects in the nucleus.
The nuclear dipole amplitude $N_A(\rt, \btvar, Y) = 1- S_A(\rt, \btvar, Y)$ has been approximated from the proton's dipole amplitude 
$N_p(\rt, \btvar, Y) = 1 - S_p(\rt, \btvar, Y)$, 
$S_p(\rt, \btvar, Y) = \exp( -T_p(\btvar) f(\rt, Y) )$, 
by writing~\cite{Kowalski:2003hm}
\begin{equation}
    \label{eq:nuclear_dipole_amplitude}
    S_A(\rt, \btvar) = \exp(- A T_A(\btvar) f(\rt, Y) )
\end{equation}
where $T_p(\btvar)$, $T_A(\btvar)$ are the transverse density profiles of the proton and the nucleus, and $f(\rt, Y)$ is a function describing the dependence on the dipole size and  rapidity $Y = \ln 1/\xpom$. Noting that $f(\rt, Y) \to 0$ when $\rt \to 0$, we can see that in the limit $Q^2 \to \infty$ the ratio~\eqref{eq:nuclear_suppression} becomes 
\begin{equation}
    R_A(Q^2 \to \infty) = \frac{1}{c} \frac{A^2 \int \dd[2]{\btvar} T_A(\btvar)^2}{ A^{4/3} \int \dd[2]{\btvar} T_p(\btvar)^2}
    \equiv 1
\end{equation}
which defines the normalization constant $c$.
The factor $A^{4/3}$ in Eq.~\eqref{eq:nuclear_suppression} has been chosen to minimize the dependence of the constant $c$ on the mass number $A$.
Note that the effects of the wave function cancel in this limit.
The estimates for nuclear suppression in Fig.~\ref{fig:jpsi_nuclear_suppression} indicate 
that while this cancellation of the wave function is true in the asymptotic limit, for values $Q^2 \lesssim M_V^2$ the wave function effects start to become visible.
Thus, the form of the wave function does not cancel in this ratio, and a naïve nonrelativistic approximation does not give a realistic picture of the saturation effects.
In general, relativistic corrections suppress the contribution from large dipoles, which results in a smaller sensitivity to nonlinear effects.
This is especially visible for smaller photon virtualities where the production is more sensitive to larger dipole sizes.
It is thus important to use a realistic wave function for $\jpsi$ when looking for saturation in heavy nuclei.

\section{Exclusive vector meson production at next-to-leading order}

\begin{figure}
	\centering
    \begin{subfigure}{0.49\textwidth}
        \centering
        \includegraphics[width=\textwidth]{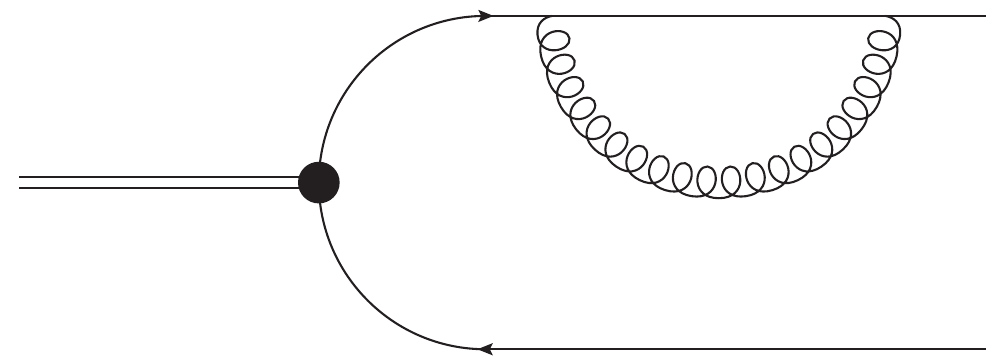}
        \caption{ }
        \label{fig:quark_self-energy}
    \end{subfigure}
    \begin{subfigure}{0.49\textwidth}
        \centering
        \includegraphics[width=\textwidth]{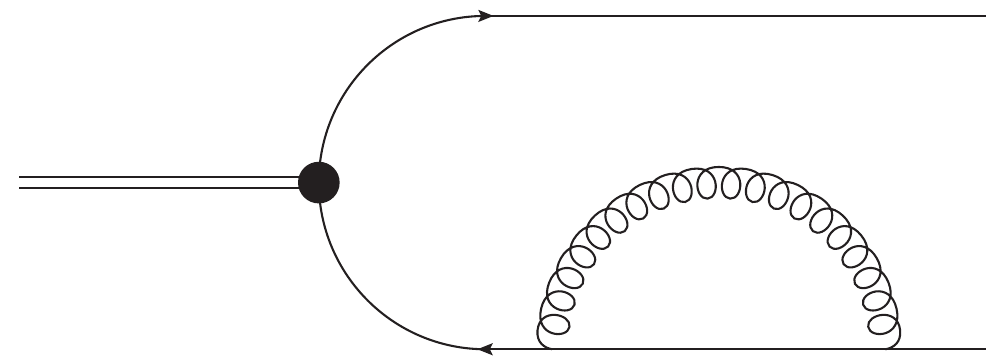}
        \caption{ }
        \label{fig:antiquark_self-energy}
    \end{subfigure}
    \begin{subfigure}{0.49\textwidth}
        \centering
        \includegraphics[width=\textwidth]{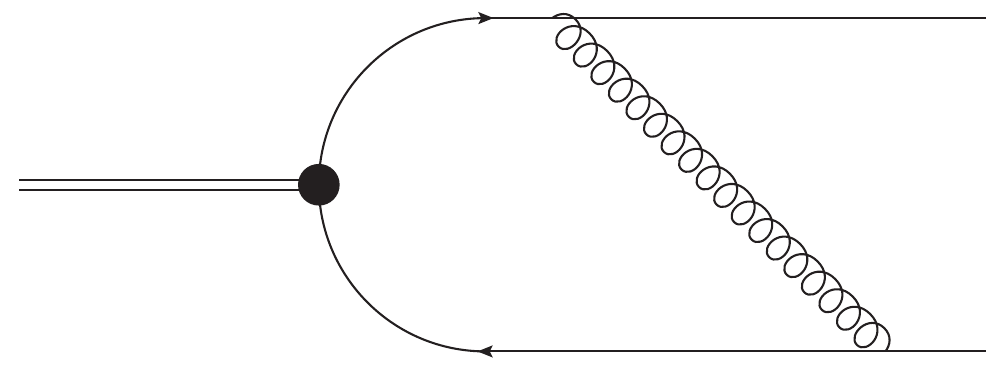}
        \caption{ }
        \label{fig:vertex_correction_1}
    \end{subfigure}
    \begin{subfigure}{0.49\textwidth}
        \centering
        \includegraphics[width=\textwidth]{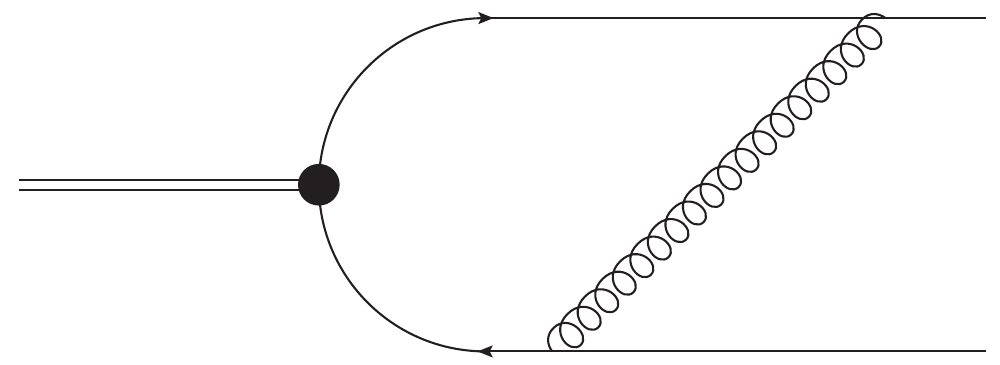}
        \caption{ }
        \label{fig:vertex_correction_2}
    \end{subfigure}
    \begin{subfigure}{0.49\textwidth}
        \centering
        \includegraphics[width=\textwidth]{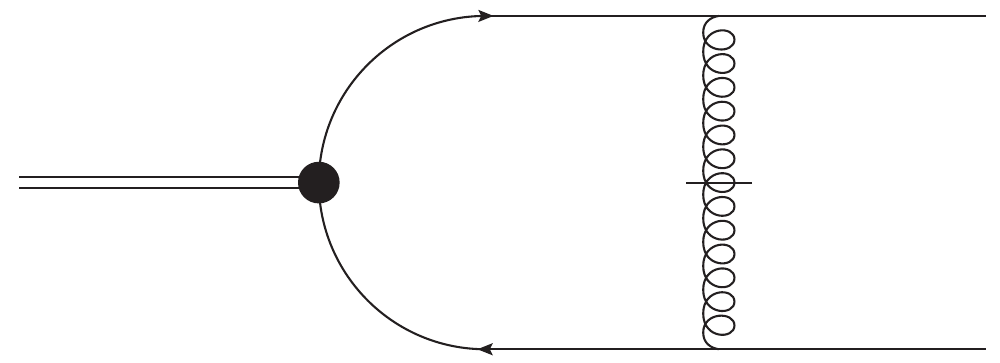}
        \caption{ }
        \label{fig:vertex_correction_virtual}
    \end{subfigure}
	  \caption{The NLO corrections to the meson wave function $\Psi^{V \to q \bar q}$.}
        \label{fig:VM_to_qqbar}
\end{figure}

\begin{figure}
	\centering
    \begin{subfigure}{0.49\textwidth}
        \centering
        \includegraphics[width=\textwidth]{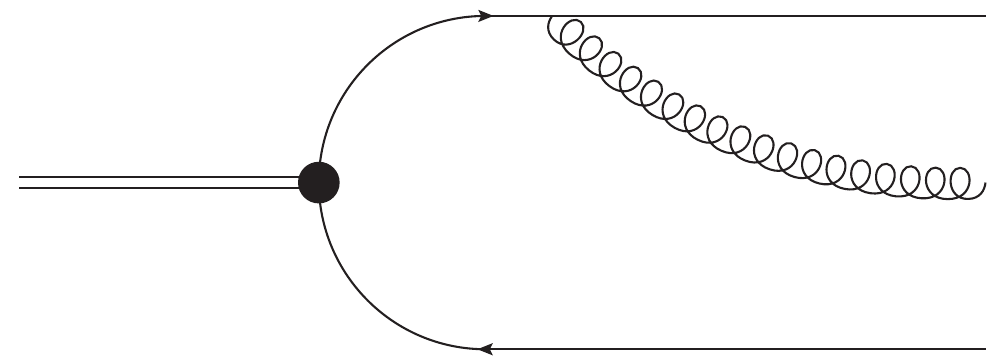}
        \caption{ }
        \label{fig:gluon_emission_quark}
    \end{subfigure}
    \begin{subfigure}{0.49\textwidth}
        \centering
        \includegraphics[width=\textwidth]{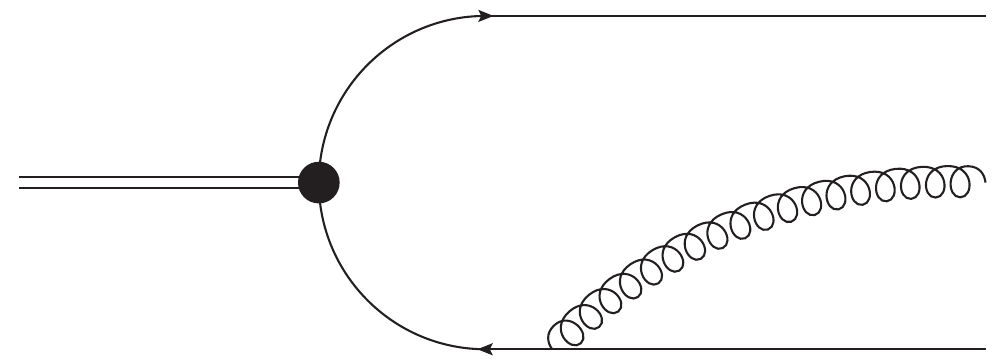}
        \caption{ }
        \label{fig:gluon_emission_antiquark}
    \end{subfigure}

    \begin{subfigure}{0.40\textwidth}
        \centering
        \includegraphics[width=\textwidth]{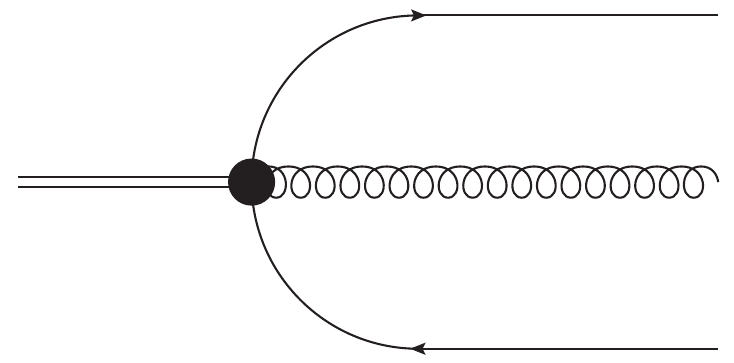}
        \caption{ }
        \label{fig:gluon_emission_nonperturbative}
    \end{subfigure}
	  \caption{Contributions to the meson wave function $\Psi^{V \to q \bar q g}$ at NLO.}
        \label{fig:VM_to_qqbarg}
\end{figure}

At next-to-leading order, there are essentially two modifications that we need to make to the leading-order amplitude Eq.~\eqref{eq:VM_production_amplitude_LO}. First, we need to also include the case where the photon fluctuates into a $q \bar q g$ state that interacts with the target. This can be included in the production amplitude by writing
\begin{equation}
\label{eq:VM_production_amplitude_NLO}
\begin{split}
    i\mathcal{M} =& \int \dd[2]{\xt_{0}} \dd[2]{\xt_1} \int_0^1 \frac{\dd{z_0} \dd{z_1}}{ (4\pi)^2 z_0 z_1} (4\pi) \delta(1-z_0-z_1) \\
    & \times e^{-i \bt \vdot \Deltat} \widetilde \Psi^{\gamma \to q \bar q}(\xt_{01}, z_i)
    \left( \widetilde \Psi^{V \to q \bar q }(\xt_{01},z_i)\right)^*
    \left( 1- \dipole_{01} \right) \\
    &+
    \int \dd[2]{\xt_{0}} \dd[2]{\xt_1} \dd[2]{\xt_2} \int_0^1 \frac{\dd{z_0} \dd{z_1} \dd{z_2}}{ (4\pi)^3 z_0 z_1 z_2} (4\pi) \delta(1-z_0-z_1 - z_2) \\
    &\times e^{-i \bt \vdot \Deltat} \widetilde \Psi^{\gamma \to q \bar qg}(\xt_{i}, z_i) 
    \left(\widetilde \Psi^{V \to q \bar q g }(\xt_{i},z_i)\right)^*
    \left( 1- \tripole_{012} \right).
\end{split}
\end{equation}
Note that here we also include the dependence on the total momentum transfer $\Deltat$ that was missing from Articles~\cite{heavy_long, light, heavy_trans} where only the case $t = -\Deltat^2 = 0$ was considered.
Second, the wave functions $\widetilde \Psi^{\gamma \to q \bar q}$, $\widetilde \Psi^{V \to q \bar q}$ need to be calculated at  next-to-leading order. 
The photon wave functions are perturbative also at NLO and have been calculated in Refs.~\cite{Hanninen:2017ddy,Beuf:2016wdz,Beuf:2017bpd} for massless quarks and in Refs.~\cite{Beuf:2021qqa,Beuf:2021srj,Beuf:2022ndu} for massive quarks. For the meson,
we need to calculate Feynman diagrams in Figs.~\ref{fig:VM_to_qqbar} and \ref{fig:VM_to_qqbarg} where the  $ V \to q \bar q  $ vertex is given by the leading-order wave function. The self-energy diagrams~\ref{fig:quark_self-energy} and \ref{fig:antiquark_self-energy} contribute to the renormalization factor $\sqrt{Z_V}$ of the meson, and the gluon exchange diagrams \ref{fig:vertex_correction_1}, \ref{fig:vertex_correction_2} and \ref{fig:vertex_correction_virtual} are then genuine NLO corrections to the wave function.
For the Fock state $q \bar qg$, the perturbative contribution to the meson wave function  $\Psi^{V \to q \bar qg}$ can be calculated in a similar manner using Diagrams \ref{fig:gluon_emission_quark} and \ref{fig:gluon_emission_antiquark}. Diagram~\ref{fig:gluon_emission_nonperturbative} corresponds to a nonperturbative contribution to the $\Psi^{V \to q \bar qg}$ wave function. In principle, it should be included in the calculation, but it turns out that in the limits considered in Articles~\cite{heavy_long,light,heavy_trans} this nonperturbative contribution can be neglected. 
The required meson wave functions at NLO have been calculated in Ref.~\cite{Escobedo:2019bxn} for heavy vector mesons and in Article~\cite{light} for light vector mesons and will be explained briefly in Secs.~\ref{sec:heavy_VM_NLO} and \ref{sec:light_VM}.

The next-to-leading order calculation also requires that we are careful with the regularization of loop integrals. As light-cone perturbation theory breaks the explicit Lorentz symmetry, a Lorentz invariant regularization scheme such as dimensional regularization cannot be used. 
Instead, a common regularization scheme suitable for light-cone calculations is to consider longitudinal and transverse directions separately. In the transverse direction, one uses dimensional regularization so that the integrals are done in $D-2 = 2- 2\varepsilon$ transverse dimensions. For the longitudinal direction, one introduces a cut-off for gluons' plus-momenta such that 
$k_2^+ > \alpha q^+$ where $\alpha >0$.
This means that we encounter two different kinds of divergences: 
infrared (IR) or ultraviolet (UV) divergences of the transverse momenta when $\varepsilon \to 0$, and divergences in the gluon's plus momentum when $\alpha \to 0$.

Several things need to be considered when summing different NLO contributions to get a finite result in the end. First of all, the two different parts in the production amplitude, Eq.~\eqref{eq:VM_production_amplitude_NLO}, are separately divergent but many of the divergences cancel in their sum. This is because the $q \bar q g$-contribution contains divergent gluon loops similar to Diagrams~\ref{fig:quark_self-energy} and \ref{fig:antiquark_self-energy}, with the exception that the gluon also crosses the shock wave describing the interaction with the target. 
In addition to this the nonperturbative parts of the calculation, i.e. the dipole amplitude and the vector meson wave function, need to be renormalized. The renormalization of the dipole amplitude is done by the BK equation, which leads to the rapidity dependence of the dipole amplitude.
The NLO calculations in Articles~\cite{heavy_long, light, heavy_trans} suggest that this rapidity scale should be chosen as $\Ydip=Y_0 + \ln((1-z_0) \frac{W^2 +Q^2-m_n^2}{Q^2_0})$ in the leading-order part, where $Q^2_0$ is the transverse scale of the target and $\Yif$ is the factorization rapidity scale discussed in Sec.~\ref{sec:nlo_massless_dipole_amplitude}. This differs somewhat from the common choice $\Ypom = \ln 1/\xpom = \ln(\frac{W^2 +Q^2-m_n^2}{Q^2 + M_V^2- t})$ used in the leading-order calculations.
For the meson wave function, only the leading-order wave function corresponding to the $V \to q \bar q$ vertex needs to be renormalized. This is done differently for heavy and light vector mesons and will be discussed in more detail in Secs.~\ref{sec:heavy_VM_NLO} and~\ref{sec:light_VM}.

\subsection{Heavy vector meson production in the nonrelativistic limit}
\label{sec:heavy_VM_NLO}

A framework to include higher-order corrections systematically in $\as$ and heavy quark velocity $v$ has been developed in Ref.~\cite{Escobedo:2019bxn}, where the meson wave functions are expanded as a power series of corrections in the heavy quark velocity $v$ and the coupling constant $\as$ as
\begin{equation}
    \label{eq:NRQCD_expansion_NLO}
                    {\widetilde \Psi^n_{V}} = \sum_{k,l} C^k_{n \leftarrow l}
                     \int_0^1 \frac{\dd{z'}}{4\pi}  \left(\frac{1}{m} \widetilde \nabla\right)^k{\phi^l}(\rt = \mathbf{0},z').
\end{equation}
This equation describes how the light-cone wave function $\Psi_V$ for a Fock state $n$ can be written in terms of the leading-order wave function $\phi$ for the Fock state $l$. The coefficients $C^k_{n \gets l}$ can be calculated perturbatively, and they contain the $\as$ corrections from Feynman diagrams. The derivatives $\widetilde \nabla = \left( \boldsymbol{\nabla}_\rt, 2 m i \left[ z'-\frac{1}{2} \right] \right)$ correspond to relativistic corrections where the order of the term is given by the power $k$, with the exception that non-dominant spin components have additional relativistic suppression as explained in Sec.~\ref{sec:heavy_VM_wave_function}. 
This allows one to expand the wave function as a power series in both $\as$ and $v$, and in a sense it is a generalization of Eq.~\eqref{eq:NRQCD_expansion} to the mixed space including also corrections in $\as$.
The NRQCD estimate for the velocity of the heavy quark is $v \gtrsim \as$~\cite{Bodwin:1994jh}, and numerical estimates confirm that for $\jpsi$ we have roughly $v^2 \sim \as$~\cite{Bodwin:2007fz}. This leads to the estimate 
$1 \gg \as \gtrsim v^2 \gg \as v^2$ for different terms in the expansion, which suggests that the most important correction to the leading order is the NLO correction in the nonrelativistic limit. 
This is also the limit where the coefficients $C^k_{n \gets l} $ have been calculated in Ref.~\cite{Escobedo:2019bxn}.
Note, however, that the importance of the relativistic corrections depends on the kinematical region as discussed in Sec.~\ref{sec:heavy_VM_wave_function}. 

Taking the nonrelativistic limit for the NLO calculation leads to a significant simplification of the calculation. For example, the only nonperturbative part of the vector meson wave function is then the leading-order wave function for the $q \bar q$ state, as the wave functions for the other states are suppressed by $v$ or $\as$. For example, the leading-order wave function for the $q \bar q g$ state corresponding to Diagram~\ref{fig:gluon_emission_nonperturbative} is suppressed by $\as v^2$ in the production amplitude. Also, in the nonrelativistic limit only the integrated leading-order wave function 
\begin{equation}
    \int_0^1 \frac{\dd{z'}}{4\pi} {\phi^{q \bar q}}(\rt = \mathbf{0},z')
\end{equation}
survives, meaning that the nonperturbative physics of vector meson formation is given by a single constant.
It should be mentioned that this is in agreement with the approach where one starts from the nonrelativistic limit in the rest frame as described in Sec.~\ref{sec:heavy_VM_wave_function}.

This nonperturbative constant still needs to be renormalized, which in a general scheme can be written as~\cite{Escobedo:2019bxn}
\begin{equation}
    \int_0^1 \frac{\dd{z'}}{4\pi} {\phi^{q \bar q}}(\mathbf{0},z')
    = \int_0^1 \frac{\dd{z'}}{4\pi} {\phi^{q \bar q}_\text{renorm}}(\mathbf{0},z')
    \left[ 1 - \frac{\as C_F}{2\pi} \left( \frac{1}{\alpha} + f_\text{scheme} \right) \right]
    .
\end{equation}
Choosing the finite part $f_\text{scheme}$ can be done in several different ways. 
The simplest way is to choose $f_\text{scheme} = 0$ as then it turns out that $\phi_\text{renorm}^{q \bar q}$ corresponds to using dimensional regularization in all space-time dimensions~\cite{Escobedo:2019bxn}.
This regularization scheme for the wave function has been used in Refs.~\cite{Bodwin:2007fz,Chung:2010vz} to determine nonperturbative LDMEs for $\jpsi$ and $\Upsilon$, and as such using this scheme allows one to use those results without additional scheme matching. 
This is especially useful if one wants to consider relativistic corrections at leading order on top of the NLO corrections.

Another choice for the scheme is to consider the leptonic width of the meson at NLO~\cite{Escobedo:2019bxn}
\begin{equation}
    \Gamma\left( V \to e^- e^+ \right)
    = \frac{2 N_c e_f^2 e^4}{3\pi M_V}
    \left|\int \frac{\dd[]{z'}}{4 \pi} {\phi^{q \bar q}}(\mathbf{0},z')\right|^2 
             \left[ 1 + \frac{ \alpha_s C_F}{\pi} \left( {\frac{1}{\alpha}}-4 \right) \right]
\end{equation}
and solve the leading-order wave function from this equation.
This corresponds to the choice $f_\text{scheme} = -4 $. 
While this is simpler as it directly relates the nonperturbative leading-order wave function to the physical leptonic width, it is not as convenient in actual calculations.
For example, when taking into account the running of the coupling it is not clear if the running is the same in the leptonic width and the production amplitude.
For these reasons, in Article~\cite{heavy_trans} we recommend using the scheme with $f_\text{scheme} = 0$ for calculating vector meson production. The dependence of the production on the renormalization scheme is mostly mild at high energies, but in certain kinematical regions it can have a significant numerical contribution to the production cross section~\cite{heavy_trans}.

In addition to the leading-order wave function, one also needs to renormalize the mass of the heavy quark.
This was done in the pole mass scheme which is set by demanding that the mass of the quark agrees with the pole of the quark propagator.
The mass renormalization is more complicated in the light-cone quantization compared to the covariant formalism, as the breaking of the Lorentz symmetry brings additional complications. 
This is discussed in detail in Ref.~\cite{Beuf:2022ndu} where the effect of using the pole mass scheme in the photon wave function is calculated using the Lorentz invariance of the photon form factors.
A different approach for mass renormalization is described in Ref.~\cite{Escobedo:2019bxn} where the mass renormalization for the meson wave function is calculated using an approach closer to the covariant formalism. 
We have checked that these two approaches for the pole mass scheme agree, meaning that no additional scheme matching has to be done to use the two wave functions.

\begin{figure}
	\centering
    \begin{subfigure}[t]{0.49\textwidth}
        \centering
        \includegraphics[width=\textwidth]{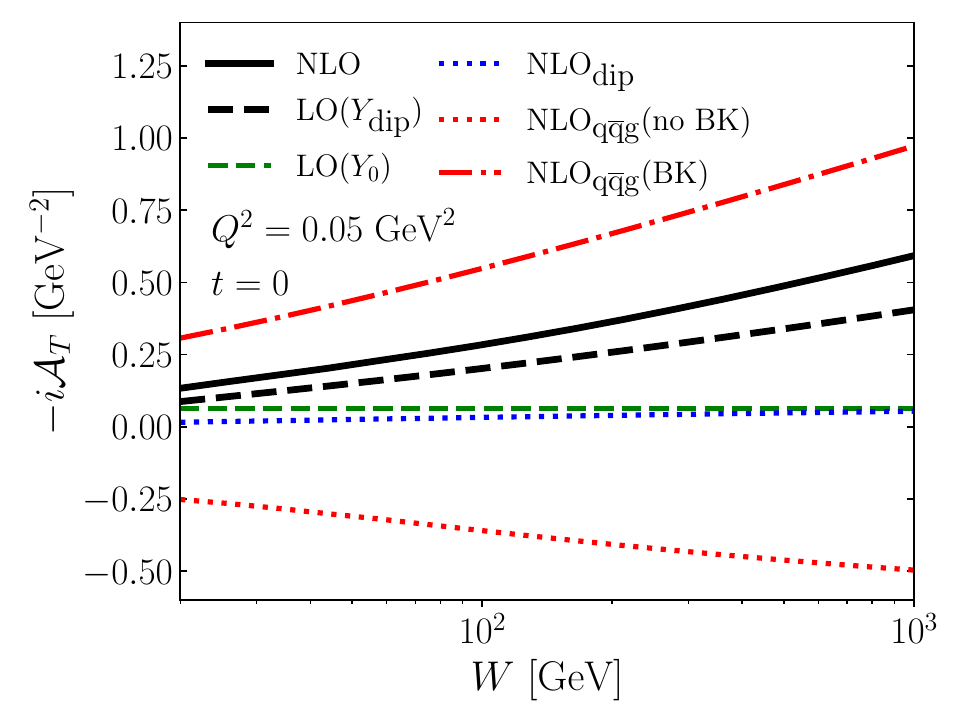}
        \caption{Different contributions to the NLO production amplitude. }
        \label{fig:jpsi_production_amplitude}
    \end{subfigure}
    \begin{subfigure}[t]{0.49\textwidth}
        \centering
        \includegraphics[width=\textwidth]{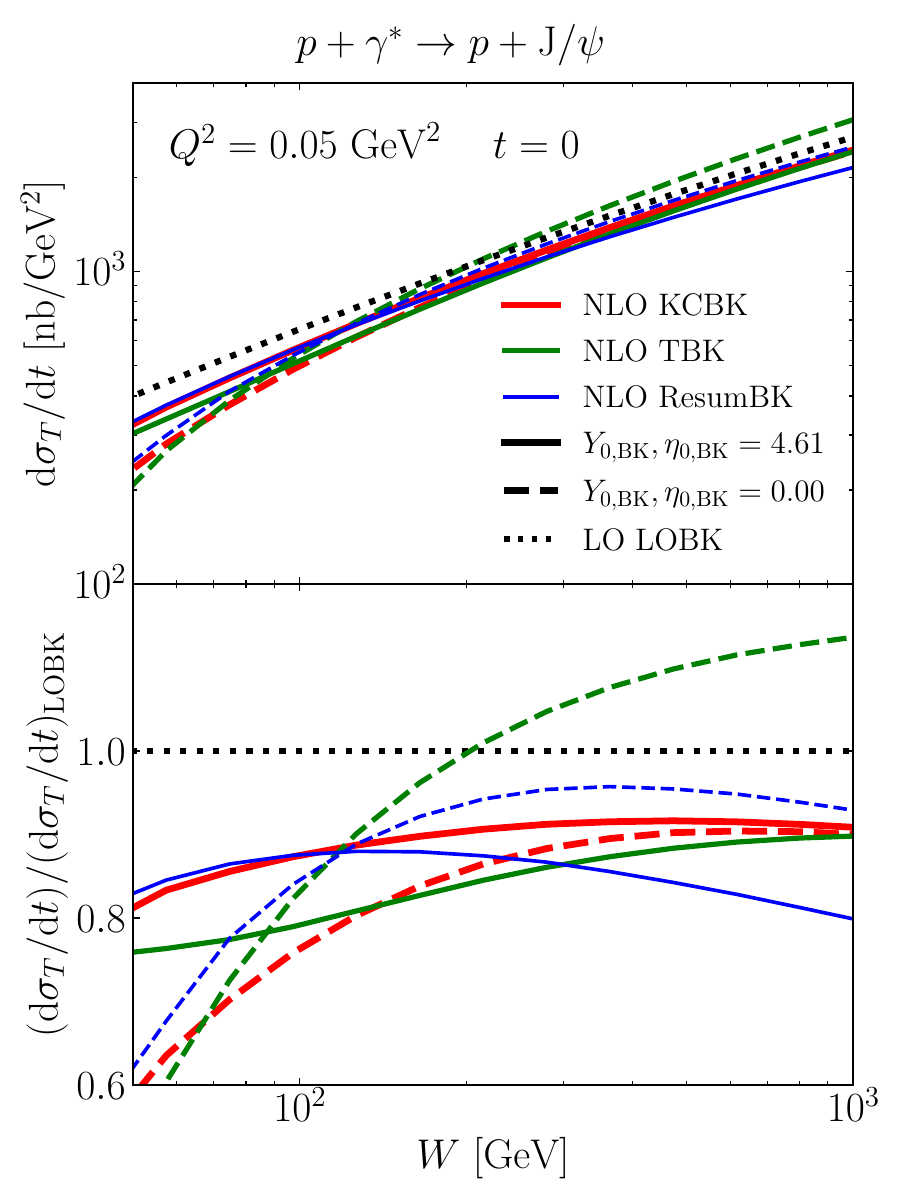}
        \caption{ Upper plot: Differential cross section for $\jpsi$ production. \\
        Lower plot: Ratio of the NLO result to the LO result. }
        \label{fig:jpsi_cross_section}
    \end{subfigure}
	  \caption{
   Tranverse $\jpsi$ production at  next-to-leading order as a function of the center-of-mass energy $W$.
   Figures from Article~\cite{light}, reproduced under the license CC BY 4.0.}
        \label{fig:jpsi_NLO_corrections}
\end{figure}

After these renormalizations and absorbing the remaining rapidity divergence $\ln \alpha$ to the BK evolution of the dipole amplitude, the invariant amplitude for the vector meson production is finite and can be calculated. The final result is complicated and contains high-dimensional integrals, but it is possible to evaluate these expressions numerically. The NLO expressions and their numerical solutions were computed in Article~\cite{heavy_long} for longitudinal production and Article~\cite{heavy_trans} for transverse production.
The main results of these articles  are summarized in Fig.~\ref{fig:jpsi_NLO_corrections}.
The NLO results are shown using the NLO dipole amplitude fits discussed in Sec.~\ref{sec:nlo_massless_dipole_amplitude} that were fitted to the full HERA structure function data with the \textit{Balitsky+smallest dipole} running coupling scheme.
In general, the NLO results are fairly close to the LO result, although there is some dependence on the dipole amplitude used.
This is especially visible at lower energies, $W \lesssim \SI{100}{GeV}$, where the BK evolution is not as dominant. 

Remembering that the dipole amplitudes were fitted to the same HERA structure function data, it is interesting to see differences in vector meson production between the results corresponding to different fits. One reason for this is that vector meson production probes the dipole amplitude at length scales $1/ (Q^2 + M_V^2)$ as opposed to $1/ Q^2$ in inclusive DIS, making vector meson production more sensitive to the behavior of the dipole amplitude at smaller dipoles. 
Another reason for the differences between the different dipole amplitudes is that when calculating the cross section we choose to drop terms of the order $\mathcal{O}(\as^2)$ for consistency with the power counting. To do this, we need to determine the LO part from the result which is not unique. Our choice for this is the LO amplitude calculated with the rapidity $\Ydip$ that is motivated by the NLO part of the calculation. This choice corresponds to resumming large logarithms $\sim \as \ln 1/x$ with the BK equation and including them as a part of the LO result.
The caveat with this approach is that the resulting $\mathcal{O}(\as^2)$ terms may be quite large, which can be seen in Fig.~\ref{fig:jpsi_production_amplitude} as the difference between the ``NLO'' and ``LO$(\Ydip)$'' results.
The effect of dropping the $\mathcal{O}(\as^2)$ contributions is especially large at lower energies where dropping these contributions may result in negative cross sections.
One does not need to do this division of the LO result and the genuine NLO corrections when calculating the inclusive DIS cross section and fitting the dipole amplitude, and dependence on this choice is an additional uncertainty to exclusive vector meson production. If one does not drop the $\mathcal{O}(\as^2)$ higher-order terms, one will get results that are closer to each other and more in line with the fitting procedure. This approach, however, is not consistent with the perturbative expansion and for this reason we did not choose to do so in Articles~\cite{heavy_long,heavy_trans}.
It is left for future work to better understand the division between the LO and NLO terms.

\begin{figure}
	\centering
    \begin{subfigure}{0.49\textwidth}
        \centering
        \includegraphics[width=\textwidth]{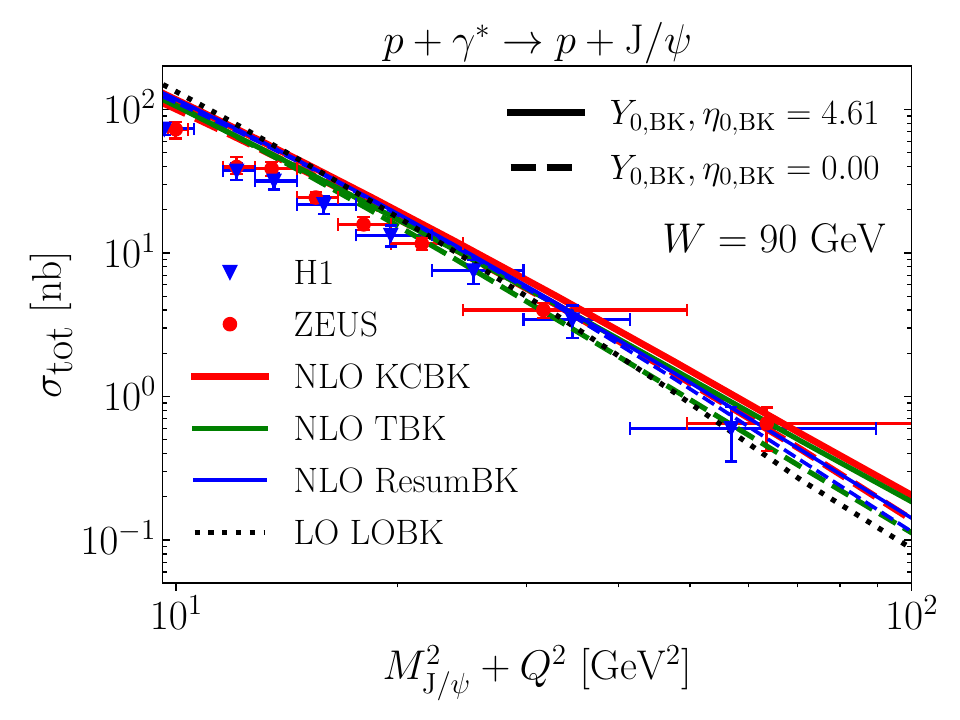}
        \caption{ Nonrelativistic limit. }
        \label{fig:jpsi_nonrel}
    \end{subfigure}
    \begin{subfigure}{0.49\textwidth}
        \centering
        \includegraphics[width=\textwidth]{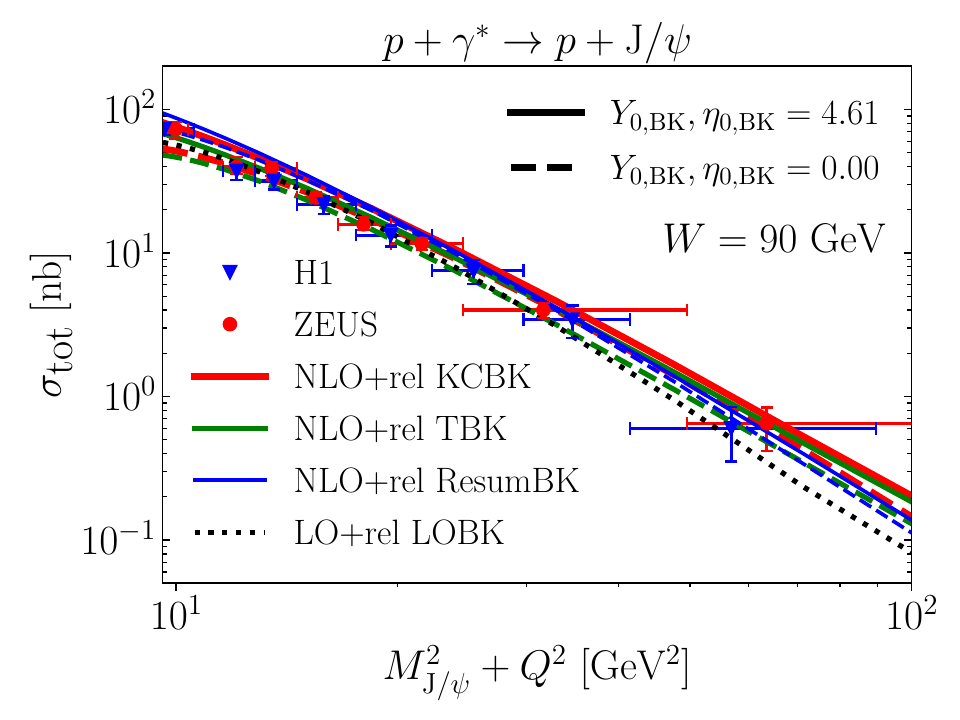}
        \caption{ With relativistic $\as^0 v^2$ corrections. }
        \label{fig:jpsi_rel}
    \end{subfigure}
	  \caption{ Total coherent $\jpsi$ production as a function of the photon virtuality $Q^2$ compared to the HERA data~\cite{ZEUS:2002wfj,ZEUS:2004yeh,H1:2005dtp,H1:2013okq}.
    Figures from Article~\cite{heavy_trans}, reproduced under the license CC BY 4.0.}
        \label{fig:jpsi_Q2_production}
\end{figure}

In addition to the NLO corrections in the nonrelativistic limit, we also included the relativistic corrections $\mathcal{O}(v^2)$ at  leading order. 
Their contribution is shown in Fig.~\ref{fig:jpsi_Q2_production} as a function of the photon virtuality, where it can be seen that the relativistic corrections are important even at NLO when considering $\jpsi$ production at small photon virtualities. In fact, in photoproduction the relativistic corrections seem to be even more important than the NLO corrections. 
For $\Upsilon$ production, the relativistic effects are estimated to be so small that they can be neglected for all photon virtualities, and the NLO effects are more important~\cite{heavy_trans}.

Finally, we would like to highlight that the main purpose of Articles~\cite{heavy_long,heavy_trans} was to calculate the NLO expression for heavy vector meson production in the nonrelativistic limit, and
thorough phenomenological analyses are left for future work. For example, varying the heavy quark mass $m$ within the uncertainties of the experimentally measured value has a large effect for small $Q^2$~\cite{Kowalski:2006hc}, which needs to be taken into account when comparing against the data.
Also, there exists data for $t$-differential $\jpsi$ production, which can be used to gain information about the impact parameter dependence of the dipole amplitude.
Another interesting prospect is to consider heavy vector meson production from heavy nuclei which is more sensitive to saturation. There is already a reasonable amount of data for $\jpsi$ production from Pb-Pb collisions measured in the ultra-peripheral collisions at the LHC~\cite{CMS:2016itn,CMS:2023snh,ALICE:2012yye,ALICE:2013wjo,ALICE:2019tqa,ALICE:2021gpt,LHCb:2021bfl}, with similar data for $\Upsilon$ expected to come soon.
Calculating this requires modeling the nuclear dipole amplitude in terms of the proton dipole amplitude using a relation such as Eq.~\eqref{eq:nuclear_dipole_amplitude}, which is yet to be done with the NLO dipole amplitude fits.

\subsection{Light vector meson production at large photon virtualities}
\label{sec:light_VM}

For light vector meson production to be perturbative one has to have a high enough photon virtuality. An additional simplification of the process can be done by assuming that the photon virtuality is higher than any momentum scale on the meson side, meaning that we can essentially neglect the meson mass and the transverse momenta of the quark and antiquark in the meson.
This can be given a precise mathematical formulation in the mixed space by noting that in the high-photon virtuality limit only dipoles of the size $1/Q^2$ contribute. We can then write the leading-order wave function as a Taylor series in the transverse dipole sizes,
\begin{equation}
    \label{eq:twist_expansion}
   \widetilde  \Psi^{q \bar q}(\rt, z) 
   = \widetilde \Psi^{q \bar q}(\mathbf{0}, z) + \mathcal{O}\bigl( \rt^2 \bigr), 
\end{equation}
and in the limit $Q^2 \to \infty$ only the first term contributes to light vector meson production. This $\rt$-independent term can be written in terms of the leading-twist distribution amplitude of the meson $\phi(z)$ which for longitudinal polarization has twist 2~\cite{Collins:1996fb,Chernyak:1983ej}. For transverse polarization, the leading term is proportional to $\rt \vdot \boldsymbol{\varepsilon}_\lambda$, corresponding to a twist-3 term~\cite{Chernyak:1983ej}.
The expansion~\eqref{eq:twist_expansion} can be considered as a twist expansion where higher-twist terms are suppressed by powers of $Q$, which also allows us to neglect transverse production in comparison to the longitudinal one as the transverse production is then suppressed by $\abs{\rt \vdot \boldsymbol{\varepsilon}_\lambda}^2 \sim 1/Q^2 $. Such an expansion also shows that the leading-order wave function for the Fock state $q \bar q g$ has twist 3, meaning that the nonperturbative diagram~\ref{fig:gluon_emission_nonperturbative} can be neglected~\cite{Ball:1998sk,Braun:1989iv}. 
Thus, the leading-twist contribution to light vector meson production at NLO can be written in terms of the twist-2 distribution amplitude for longitudinal polarization. 
This leads to a simplification of the calculation as the first term in the expansion~\eqref{eq:twist_expansion} corresponds to a delta function $\delta^2(\kt)$ in  momentum space. Also, all of the nonperturbative physics on the meson side is included in the twist-2 distribution amplitude $\phi(z)$.

The distribution amplitude $\phi(z)$ still needs to be renormalized in the calculation. This is done with the well-known Efremov--Radyushkin--Brodsky--Lepage (ERBL) equation~\cite{Efremov:1979qk,Lepage:1980fj} which corresponds to a resummation of gluon exchanges between the quark and antiquark with transverse momenta $\kt^2 < Q^2$. In general, the renormalized distribution amplitude $\phi(z,\mu_F)$ can be written in terms of the bare distribution amplitude $\phi_0(z)$ as
\begin{multline}
    \phi(z, \mu_F) = \phi_0(z)\\
    + \frac{\as C_F}{ 2\pi}
    \int_0^1 \dd{z'}  \phi_0(z') \left[ K(z, z') \left( \frac{2}{D-4} + \gamma_E -\ln(4\pi) + \ln(\frac{\mu_F^2}{\mu^2}) \right)  + f_\text{scheme}\right]
\end{multline}
where $K(z,z')$ is the ERBL kernel
\begin{equation}
    \begin{split}
    K(z,z') 
    =& \frac{z}{z'} \left(  1+ \frac{1}{z'-z}\right) \theta(z'-z-\alpha)
    + \frac{1-z}{1-z'} \left(  1+ \frac{1}{z-z'}\right) \theta(z-z'-\alpha)\\
    &+ \left( \frac{3}{2}+ \ln \frac{\alpha^2}{z(1-z)} \right) \delta(z'-z),
    \end{split}
\end{equation}
 $\mu$ is the scale from dimensional regularization and $f_\text{scheme}$ is a finite scheme-dependent term. This introduces a dependence on the factorization scale $\mu_F$ to the distribution amplitude,  which is given by the ERBL equation
\begin{equation}
    \frac{\partial}{\partial{\ln \mu_F^2}} \phi(z, \mu_F) = \frac{\as C_F}{2\pi}
    \int_0^1 \dd{z'} K(z,z')\phi(z', \mu_F).
\end{equation}
The ERBL equation can be solved exactly by
\begin{equation}
\label{eq:Gegenbauer_expansion}
    \phi(z,\mu_F) = \sum_{n=0}^\infty a_n(\mu_F) f_n(z)
\end{equation}
where the dependence on the factorization scale is contained in the coefficients $a_n$, and the eigenfunctions $f_n$ of the ERBL equation can be written in terms of the Gegenbauer polynomials $C_n^{\left( \frac{3}{2} \right)}$ as
\begin{equation}
    f_n(z) = 6z(1-z) C_n^{\left( \frac{3}{2} \right)} (2z-1).
\end{equation} 
It is interesting to note that in the limit $\mu_F \to \infty$ the solutions of the ERBL equation are driven towards the asymptotic limit 
$\phi(z, \mu_F = \infty) = 6z(1-z)a_0$.

In Article~\cite{light}, we choose $f_\text{scheme}= 0 $ in accordance with the $\msbar$ scheme. We note, however, that a regularization-scheme dependent choice might also be suitable. 
We do the NLO calculation in a formalism that allows one to use two different regularization schemes for gluon polarization vectors, the so-called ``conventional dimensional regularization'' and the ``four-dimensional helicity'' schemes~\cite{Bern:1991aq,Bern:2002zk}. 
With the $\msbar$ scheme, the resulting equations still contain a constant dependent on the regularization scheme which could be included in the renormalization of the distribution amplitude. 
This scheme dependence arises because the distribution amplitude is a nonperturbative quantity, and a similar scheme-dependent part should appear in other processes involving the distribution amplitude as well.
In the case of exclusive light vector meson production this scheme-dependent constant has an extremely small contribution~\cite{light}, and it vanishes if one takes the asymptotic form of the distribution amplitude.

The factorization scale $\mu_F$ should be chosen as the relevant scale of the process.
The NLO calculation suggests the choice $\mu_F^2 = 4 e^{-2 \gamma_E}/\rt^2$ which is also in line with the Fourier transform estimates for the running of the coupling constant~\cite{Kovchegov:2006vj}.
Another possible choice is $\mu_F^2 = Q^2$ which has the attractive property that the factorization scale is constant in the calculation.
The differences between these two choices are fairly small for reasonable forms of the distribution amplitude, less than $5\%$~\cite{light}.

The resulting expression for the longitudinal production amplitude is finite and can be numerically evaluated. It has been calculated before in a different framework in Ref.~\cite{Boussarie:2016bkq} where the resulting expressions are presented in momentum space. This makes comparisons between the two calculations difficult, as one has to perform complicated Fourier transforms from the momentum space to the coordinate space. So far, only parts of the calculation have been compared with agreeing results.

\begin{figure}
	\centering
    \begin{subfigure}{0.49\textwidth}
        \centering
        \includegraphics[width=\textwidth]{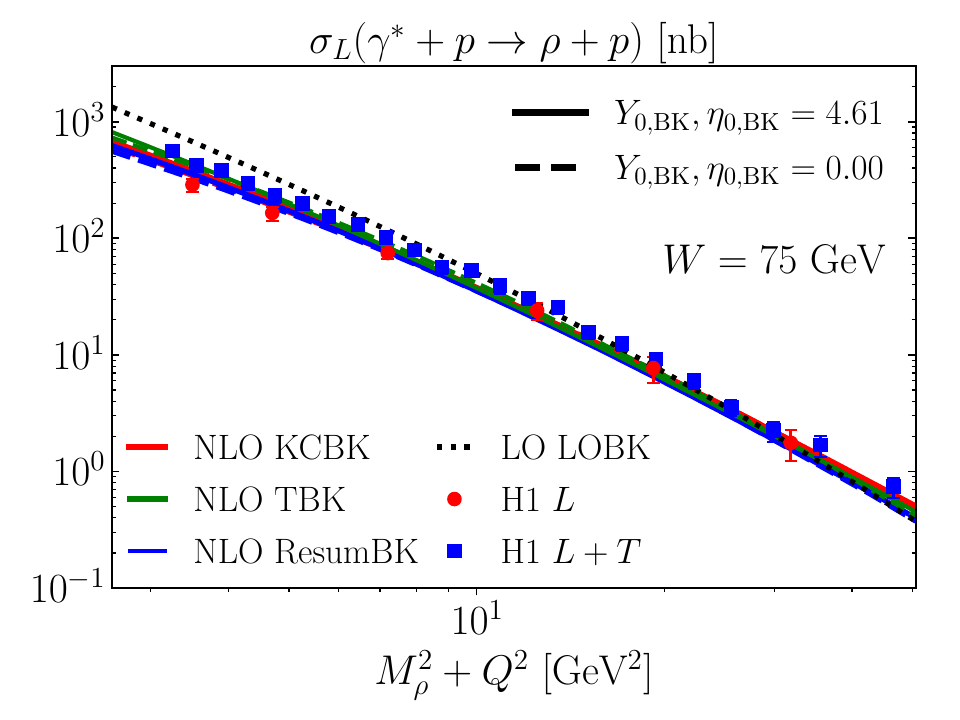}
        \caption{ Exclusive $\rho$ production. }
        \label{fig:rho_production}
    \end{subfigure}
    \begin{subfigure}{0.49\textwidth}
        \centering
        \includegraphics[width=\textwidth]{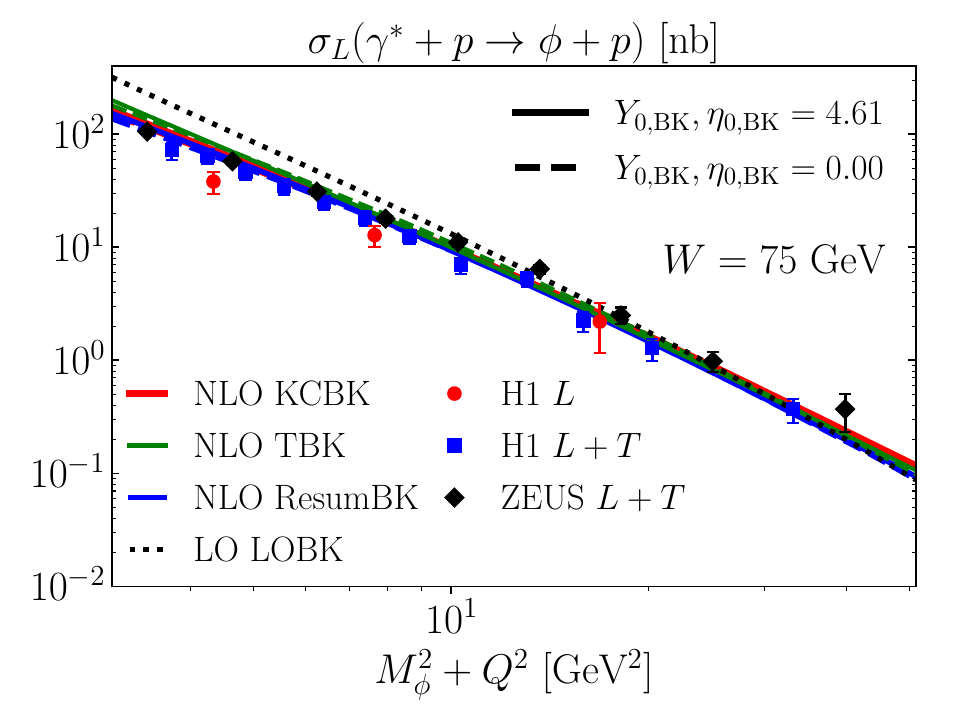}
        \caption{ Exclusive $\phi$ production. }
        \label{fig:phi_production}
    \end{subfigure}
	  \caption{Exclusive light vector meson at next-to-leading order as a function of the photon virtuality $Q^2$. Figures from Article~\cite{light}, reproduced under the license CC BY 4.0.}
        \label{fig:light_VM_NLO}
\end{figure}

The main numerical results for $\rho$ and $\phi$ meson production from Article~\cite{light} are shown in Fig.~\ref{fig:light_VM_NLO}. 
These calculations use a distribution amplitude that is close to the asymptotic form, including also the $n=2$ correction in Eq.~\eqref{eq:Gegenbauer_expansion} using numerical estimates from Refs.~\cite{Gao:2014bca,Polyakov:2020cnc}. The contributions from this correction to the asymptotic form are almost negligible in these figures, less than $10 \%$. In general, we find a very good agreement with the NLO results to the longitudinal production data. Note that the total production, longitudinal + transverse, should be dominated by the longitudinal production in the $Q^2 \to \infty$ limit, and thus it is consistent to compare to the total production data also. Only for smaller photon virtualities our results seem to overestimate the longitudinal production data, and it is expected that there the higher-twist effects should also contribute. Here the agreement between the different dipole amplitude fits is better than in the case of heavy vector meson production. This can be understood by the fact that in inclusive DIS, where the dipole amplitude fit is done, the same dipole sizes $1/ Q^2$ are probed as in light vector meson production. In contrast, for heavy vector meson production the probed dipole sizes are modified by the heavy meson mass as $1/(Q^2+ M_V^2)$.

\chapter{Inclusive diffraction}
\label{sec:DDIS}

Exclusive vector meson production is an example of an exclusive process where all of the produced particles are explicitly measured.
The opposite of this is an inclusive process where the final states are summed over. 
We have already briefly considered total inclusive particle production in Sec.~\ref{sec:inclusive_DIS} where the final state is not restricted.
In addition to this, we can consider inclusive diffraction where the interaction with the target is color neutral, resulting in final states that are in a color singlet.
This corresponds to the process in Fig.~\ref{fig:diffractive_process} with a sum over the final states $X$.

Inclusive diffraction has several advantages compared to exclusive vector meson production. First, we are not as restricted in our final state, leading to a higher cross section.
Second, the only nonperturbative part of the process is the interaction with the target, making inclusive diffraction a very clean probe of the target structure.
When compared to the total inclusive production, the advantage of inclusive diffraction is its higher sensitivity to the small-$x$ gluon distribution, enhancing the nonlinear effects.
Also, more differential quantities can be measured, such as the dependence on the momentum transfer $t$ and the invariant mass of the final state $M_X^2 = P_X^2$.
The momentum-transfer dependence gives us information about the geometry of the target in the transverse plane, similarly as in exclusive vector meson production in Ch.~\ref{sec:vector_meson_production}.
The dependence on the invariant mass is related to the relevant Fock states in the process as will be explained shortly.

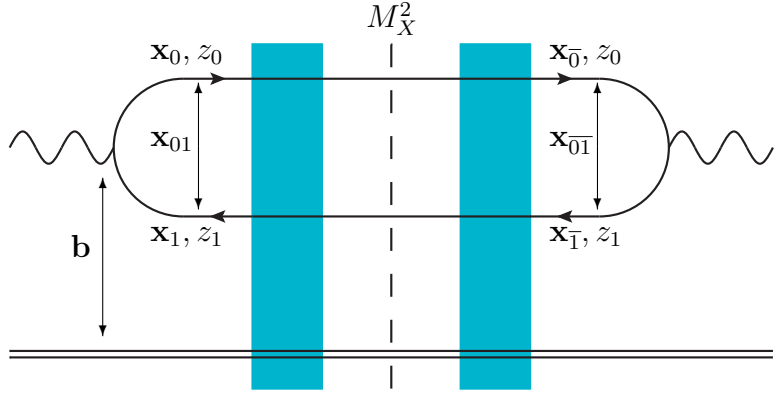
\begin{figure}[t]
    \centering
    \begin{overpic}[width=0.7\textwidth]{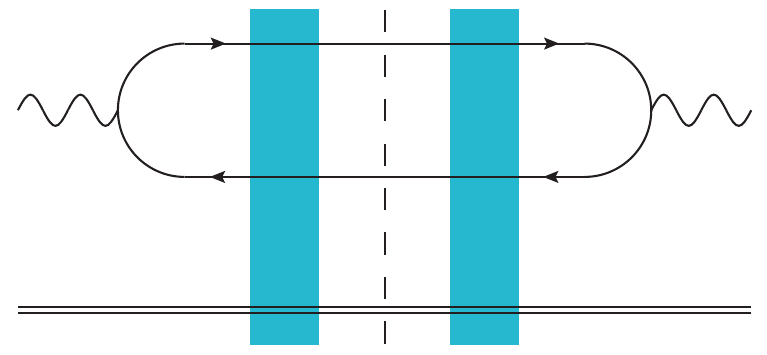}
     \put (47,47) {$M_X^2$}
     \put (10,18) {$\bt$}
     \put (14,18){\vector(0,1){10}}
     \put (14,18){\vector(0,-1){10}}
     \put (20,32) {$\xt_{01}$}
     \put (26,32){\vector(0,1){8}}
     \put (26,32){\vector(0,-1){8}}
     \put (20,43) {$\xt_0, z_0$}
     \put (20,20) {$\xt_1, z_1$}
     \put (70,32) {$\xt_{\ov{01}}$}
     \put (76,32){\vector(0,1){8}}
     \put (76,32){\vector(0,-1){8}}
     \put (70,43) {$\xt_{\ov 0}, z_{0}$}
     \put (70,20) {$\xt_{\ov 1}, z_{1}$}
    \end{overpic}
    \caption{
        Inclusive diffractive DIS at leading order in the dipole picture.
        The blue rectangle depicts the interaction with the target, and
        the dashed line corresponds to the final state which is set to have an invariant mass $M_X^2$.
    }
    \label{fig:DDIS_LO}
\end{figure}

Inclusive diffraction can be naturally calculated using Eqs.~\eqref{eq:general_amplitude} and \eqref{eq:general_cross_section} from Sec.~\ref{sec:eikonal_approximation}. This leads to the cross section
    \begin{equation}
        \label{eq:DDIS_cross_section}
        \frac{\dd[]{ \sigma^{\text{D}}_{\gamma^*_\lambda + A } }}{\dd{\abs{t}} \dd{M_X^2}} 
        =
        \sum_{\substack{\text{color-singlet} \\ \text{states } n}} 
        \int \dd{ [\ps]_n } 2q^+ (2\pi) \delta(q^+ - q_n^+)  \abs{\mathcal{M}_{\gamma_\lambda^* \to n}}^2
        \delta(M_X^2- M_n^2)
        \delta\left(\abs{t} - \Deltat^2\right)
    \end{equation}
where the delta functions specify the given momentum transfer $t$ and invariant mass $M_X^2$ of the final state.
At leading order, only the final state $X=q \bar q$ contributes,
and calculating the cross section corresponds to evaluating the Feynman diagram in Fig.~\ref{fig:DDIS_LO}.
The leading-order cross section evaluates to~\cite{Marquet:2007nf,Kowalski:2008sa}
\begin{equation}
    \label{eq:DDIS_LO}
    \begin{split}
         &\frac{\dd[]{ \sigma^{\text{D}}_{\gamma^*_{\lambda} + A } }}{\dd{\abs{t}} \dd{M_X^2}}  =
         2 \pi\aem  N_c Q^2\sum_f e_f^2
         \int \frac{
    \dd[2]{\xt_{0}}\dd[2]{\xt_{1}}
    \dd[2]{\xt_{\ov 0}}\dd[2]{\xt_{\ov 1}}
        }{(2\pi)^4} 
        \int_0^1  \dd[]{z_{0}} \dd[]{z_{1}} \delta(1-z_0 -z_1)   \\
         & \times \frac{1}{4\pi} 
         J_0\bigl(\abs{t} \abs{\bt -\ov \bt}\bigr)  
         J_0\bigl(\abs{\xt_{01}-\xt_{\ov 0 \ov 1}} M_X \sqrt{z_0 z_1}\bigr)
          \mathcal{F}_{\lambda}
          \biggl\langle 1-\dipole_{01} \biggr\rangle
          \biggl\langle1-\dipole_{\ov 0 \ov 1} \biggr\rangle^\ast
    \end{split}
    \end{equation}
    where
    \begin{equation}
    \mathcal{F}_{\lambda} = 
    \begin{cases}
    4 z_0^3 z_1^3 K_0(\abs{\xt_{01}} Q \sqrt{z_0 z_1} ) K_0(\abs{\xt_{\ov 0 \ov 1}} Q \sqrt{z_0 z_1} ) 
    & \lambda = L \\
     z_0^2 z_1^2 (z_0^2 +z_1^2 ) K_1(\abs{\xt_{01}} Q \sqrt{z_0 z_1} ) K_1(\abs{\xt_{\ov 0 \ov 1}} Q \sqrt{z_0 z_1} ) 
    \frac{\xt_{01} \vdot \xt_{\ov 0 \ov 1}}{\abs{\xt_{01}} \abs{\xt_{\ov 0 \ov 1}}}
      &\lambda = T
    \end{cases}
    \end{equation}
depends on the photon polarization $\lambda$.
It should be mentioned that this is the \textit{coherent} production cross section, in line with the notation used in Ch.~\ref{sec:vector_meson_production}. 
Total (i.e. coherent+incoherent) cross section corresponds to Eq.~\eqref{eq:DDIS_LO} with the average over the target fluctuations taken as 
$\biggl\langle \left( 1-\dipole_{01} \right) \left(1-\dipole_{\ov 0 \ov 1}\right)^\dag \biggr\rangle$ instead.

The inclusive diffractive cross section can be related to the diffractive structure functions, defined as
\begin{equation}
    \label{eq:DDIS_structure functions}
    \xpom F_\lambda^{\D(4)}(\beta, Q^2, \xpom, t) = \frac{Q^2}{(2\pi)^2 \aem} \frac{Q^2}{\beta} \frac{\dd{\sigma^\D_{\gamma_\lambda^* +A}}}{\dd{\abs{t}}\dd{M_X^2}}.
\end{equation}
The superscript indicates how many variables the structure function $\xpom F_\lambda^{\D(4)}$ depends on. 
The variables $\beta$ and $\xpom$ are defined for inclusive diffraction as
\begin{align}
    \beta &= \frac{Q^2}{2 q \vdot (P_n-P_n')} = \frac{Q^2}{Q^2 +M_X^2 - t} 
    \approx \frac{Q^2}{Q^2 +M_X^2 }, \\
    \xpom &= \frac{q \vdot (P_n -P_n')}{q \vdot P_n}
    = \frac{Q^2 +M_X^2 -t}{W^2 + Q^2 -m_n^2} 
    \approx \frac{Q^2 +M_X^2}{W^2 + Q^2 }.
\end{align}
Other interesting observables are the $t$-integrated diffractive structure functions
\begin{align}
     F_\lambda^{\D(3)}(\beta, Q^2, \xpom) &= \int_0^\infty \dd{\abs{t}}   F_\lambda^{\D(4)}(\beta, Q^2, \xpom, t),\\
     F_2^{\D(3)}(\beta, Q^2, \xpom) &=  F_L^{\D(3)}(\beta, Q^2, \xpom) + F_T^{\D(3)}(\beta, Q^2, \xpom),
\end{align}
and the most precise data is given in terms of the diffractive reduced cross section
\begin{equation}
    \label{eq:diffractive_reduced_cross_section}
    \sigma^{\D(3)}_r(\beta, Q^2, \xpom) = F_2^{\D(3)}(\beta, Q^2, \xpom) - \frac{y^2}{1+(1-y)^2} F_L^{\D(3)}(\beta, Q^2, \xpom)
\end{equation}
analogously to the inclusive case~\eqref{eq:reduced_cross_section}.

\begin{figure}[t]
    \centering
    \includegraphics[width=0.7\textwidth]{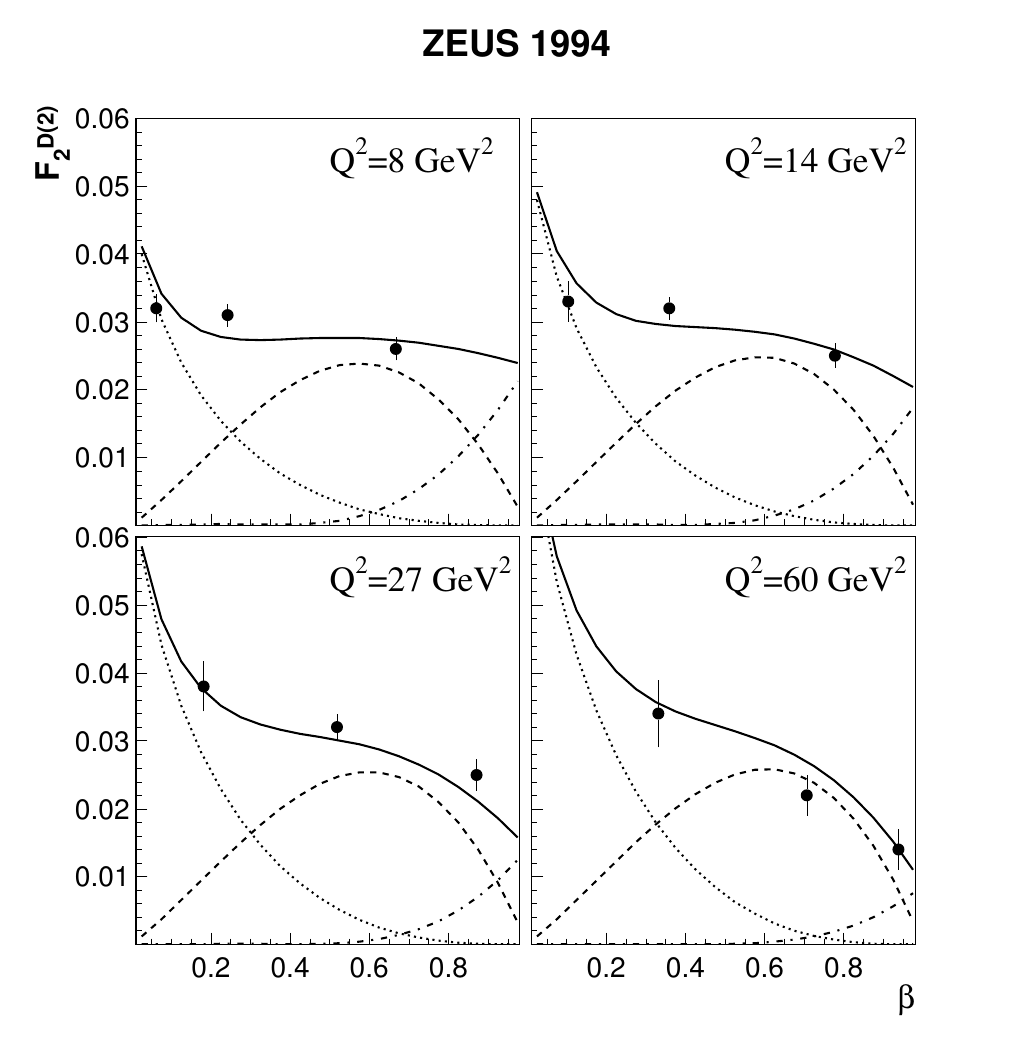}
    \caption{
        Theoretical predictions for the diffractive structure function $F_2^{\D(2)}(\beta, Q^2)$
        compared against the HERA data~\cite{ZEUS:1998rvb}.\\
        \textbf{Dashed line:} Leading-order $q \bar q$ contribution for transverse photons.\\
        \textbf{Dot-dashed line:} Leading-order $q \bar q$ contribution for longitudinal photons.\\
        \textbf{Dotted line:} Large-$Q^2$ contribution for the $q \bar q g$ state for transverse photons.\\
    Reprinted figure with permission from K. Golec-Biernat, and M. Wusthoff, Phys. Rev. D, 60, 114023, 1999. Copyright
(1999) by the American Physical Society.
    }
    \label{fig:DDIS_qqg}
\end{figure}

Experimental measurements of inclusive diffraction at DIS have been done at HERA~\cite{H1:1995cha,H1:1997bdi,H1:2006uea,H1:2006zyl,H1:2011jpo,H1:2012pbl,H1:2012xlc,ZEUS:1995sar,ZEUS:1996bqn,ZEUS:1997fox,ZEUS:1998rvb,ZEUS:2002cih,ZEUS:2005vwg,ZEUS:2008qxs} where it has been measured in $p+e^-$ collisions.
Comparisons of the data and theory then showed that the leading-order picture is not enough to describe diffractive DIS at low values of $\beta$~\cite{ZEUS:1998rvb}.
This is shown in Fig.~\ref{fig:DDIS_qqg} where it can be seen that the leading-order results~\eqref{eq:DDIS_LO} fall to zero when $\beta \to 0$, in contradiction with the measured data.
This started calculations of the process at  next-to-leading order, as it was found that gluon emission starts to dominate in the limit $\beta \to 0$, which was first calculated in Refs.~\cite{Wusthoff:1997fz,Wusthoff:1999cr,Bartels:1994jj,Golec-Biernat:1999qor} in the large-$Q^2$ limit.
Including these gluonic contributions to the cross section results in a very good agreement with the data as can be seen in Fig.~\ref{fig:DDIS_qqg}.
The more natural large-$M_X^2$ limit has later been calculated by several authors in Refs.~\cite{Kovchegov:1999ji,Bartels:1999tn,Kopeliovich:1999am,Kovchegov:2001ni,Munier:2003zb,Golec-Biernat:2005prq}, and it has been connected to the large-$Q^2$ result in Ref.~\cite{Marquet:2007nf}.

The reason for the importance of the gluonic contribution to the cross section stems from large logarithms $\ln 1/\beta$ that start to appear at NLO.
These logarithms are related to the rapidity interval $\Yshow = \ln 1/\beta$ between the electron and the parton shower $X$, and they can be resummed with the Kovchegov--Levin evolution equation~\cite{Kovchegov:1999ji} for the diffractive dipole amplitude.
In practice, the rapidity gap $\Ygap = \ln 1/\xpom$ between the parton shower $X$ and the target has to be large enough to be detectable, which means that
$ \ln 1/\beta = \ln \xpom/x$ is not too large
 in the HERA (or EIC) kinematics~\cite{Kowalski:2008sa}.
Thus, 
it is expected that
keeping only the first large logarithm $\ln 1/\beta$ of the gluonic contribution at NLO is enough for comparisons to the currently available data.

The above-mentioned calculations contained only a part of the full NLO calculation in certain limits. They were brought into a more systematic NLO framework in the dipole picture in Ref.~\cite{Beuf:2022kyp}
where it was shown they can be obtained from the part of the NLO calculation where a gluon emission happens before the shock wave in the amplitude and its complex conjugate\footnote{The large-$M_X^2$ limit in Ref.~\cite{Munier:2003zb} also requires an additional contribution from  gluon emission after the shock wave.}.
That article started the more systematic calculation of the full NLO equation for inclusive diffraction, which is being completed in an ongoing work of the author where the rest of the NLO diagrams are computed.
That work
will present the full NLO cross section for inclusive diffraction for the first time, and the purpose of this chapter is to give a rough outline of the full calculation.

\section{Diffractive DIS at  next-to-leading order}

\begin{figure}[t]
	\centering
    \begin{subfigure}{0.48\textwidth}
        \centering
        \includegraphics[width=\textwidth]{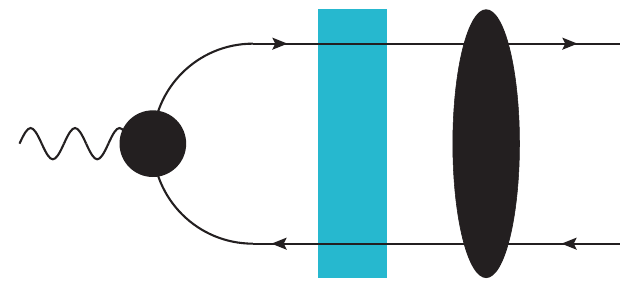}
        \caption{ }
        \label{fig:gamma_qq_qq}
    \end{subfigure}
    \begin{subfigure}{0.48\textwidth}
        \centering
        \includegraphics[width=\textwidth]{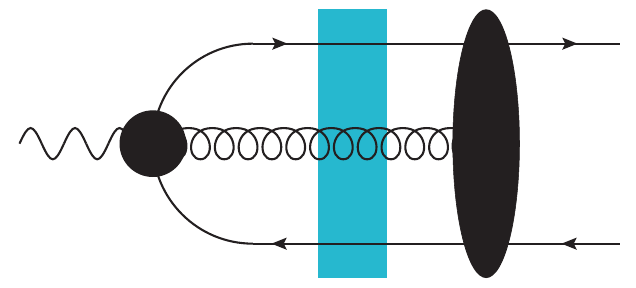}
        \caption{ }
        \label{fig:gamma_qqg_qq}
    \end{subfigure}
    \begin{subfigure}{0.48\textwidth}
        \centering
        \includegraphics[width=\textwidth]{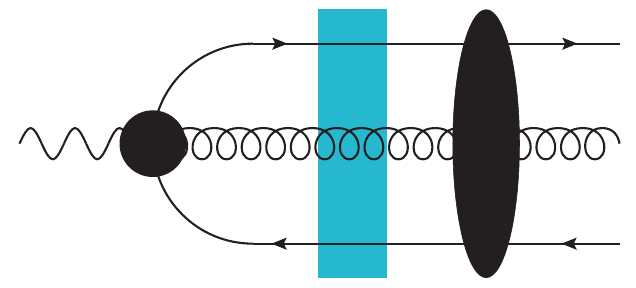}
        \caption{ }
        \label{fig:gamma_qqg_qqg}
    \end{subfigure}
    \begin{subfigure}{0.48\textwidth}
        \centering
        \includegraphics[width=\textwidth]{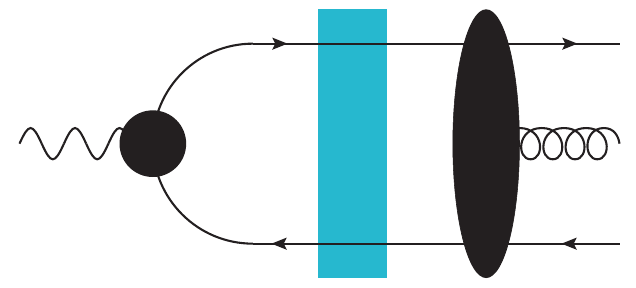}
        \caption{ }
        \label{fig:gamma_qq_qqg}
    \end{subfigure}
	  \caption{Contributions to the NLO invariant amplitude in diffractive DIS.
      The blue rectangle represents the interaction with the target, and the black ellipse effectively represents the wave function $\Psi^{n \to m}$ including the non-interacting case (see Eq.~\eqref{eq:out_wf_matrix_elements}). Note that the black blob for the $\gamma^* \to m$ vertex corresponds to the photon wave function $\Psi^{\gamma^* \to m}$ where also the NLO corrections are included.
      }
        \label{fig:DDIS_NLO}
\end{figure}

The full NLO calculation contains a lot of different diagrams that need to be accounted for. Perhaps the most simple classification of the calculation can be done by considering different Fock states at the shock wave and at the final state.
At NLO, this results in dividing the calculation into four terms shown in Fig.~\ref{fig:DDIS_NLO}. The full cross section can then be written as
\begin{equation}
    \label{eq:DDIS_cross_section_NLO}
    \frac{\dd[]{ \sigma^{\text{D}}_{\gamma^*_\lambda + A } }}{\dd{\abs{t}} \dd{M_X^2}} 
    = 
    \left[\frac{\dd[]{ \sigma^{\text{D}}_{\gamma^*_\lambda + A } }}{\dd{\abs{t}} \dd{M_X^2}} \right]_{q \bar q}
    +
    \left[
    \frac{\dd[]{ \sigma^{\text{D}}_{\gamma^*_\lambda + A } }}{\dd{\abs{t}} \dd{M_X^2}} 
    \right]_{q \bar qg}
\end{equation}
where
\begin{align}
    \left[\frac{\dd[]{ \sigma^{\text{D}}_{\gamma^*_\lambda + A } }}{\dd{\abs{t}} \dd{M_X^2}} \right]_{q \bar q}
    &=
     \int \dd{ [\ps]_{q \bar q} } 2q^+ (2\pi) \delta(q^+ - q_{q \bar q}^+)  \abs{\mathcal{M}_{ q \bar q}}^2
    \delta(M_X^2- M_{q \bar q}^2)
    \delta\left(\abs{t} - \Deltat^2\right), \\
    \left[\frac{\dd[]{ \sigma^{\text{D}}_{\gamma^*_\lambda + A } }}{\dd{\abs{t}} \dd{M_X^2}} \right]_{q \bar q g}
    &=
     \int \dd{ [\ps]_{q \bar q g} } 2q^+ (2\pi) \delta(q^+ - q_{q \bar q g}^+)  \abs{\mathcal{M}_{q \bar q g}}^2
    \delta(M_X^2- M_{q \bar q g}^2)
    \delta\left(\abs{t} - \Deltat^2\right),
\end{align}
and the corresponding invariant amplitudes are
\begin{align}
    \mathcal{M}_{q \bar q} =& \mathcal{M}_{q \bar q}^{\hyperref[fig:gamma_qq_qq]{\aamp}} + \mathcal{M}_{q \bar q}^{\hyperref[fig:gamma_qqg_qq]{\bamp}}, \\
    \mathcal{M}_{q \bar qg} =& \mathcal{M}_{q \bar qg}^{\hyperref[fig:gamma_qqg_qqg]{\camp}} + \mathcal{M}_{q \bar qg}^{\hyperref[fig:gamma_qq_qqg]{\damp}} .
\end{align}

Squaring the amplitudes, we end up with the following terms in the cross section:
\begin{equation}
    \label{eq:DDIS_cross_Section_terms}
\begin{split}
    \frac{\dd{\sigma^\textrm{D}_{\gamma^*_\lambda + A}}}{\dd[2]{\Deltat} \dd{M_X^2} } 
    =&
    \left[ \frac{\dd{\sigma^\textrm{D}_{\gamma^*_\lambda + A}}}{\dd[2]{\Deltat} \dd{M_X^2} } \right]_{\abs{\hyperref[fig:gamma_qq_qq]{\aamp}}^2}
    +
    \left[ \frac{\dd{\sigma^\textrm{D}_{\gamma^*_\lambda + A}}}{\dd[2]{\Deltat} \dd{M_X^2} } \right]_{\abs{\hyperref[fig:gamma_qqg_qq]{\bamp}}^2}\\
    +&
    \left[ \frac{\dd{\sigma^\textrm{D}_{\gamma^*_\lambda + A}}}{\dd[2]{\Deltat} \dd{M_X^2} } \right]_{\abs{\hyperref[fig:gamma_qqg_qqg]{\camp}}^2}
    +
    \left[ \frac{\dd{\sigma^\textrm{D}_{\gamma^*_\lambda + A}}}{\dd[2]{\Deltat} \dd{M_X^2} } \right]_{\abs{\hyperref[fig:gamma_qq_qqg]{\damp}}^2} \\
    +&2 \Re
    \left[ \frac{\dd{\sigma^\textrm{D}_{\gamma^*_\lambda + A}}}{\dd[2]{\Deltat} \dd{M_X^2} } \right]_{\hyperref[fig:gamma_qq_qq]{\aamp}\times \hyperref[fig:gamma_qqg_qq]{\bamp}^*}
    + 2 \Re
    \left[ \frac{\dd{\sigma^\textrm{D}_{\gamma^*_\lambda + A}}}{\dd[2]{\Deltat} \dd{M_X^2} } \right]_{\hyperref[fig:gamma_qqg_qqg]{\camp}\times \hyperref[fig:gamma_qq_qqg]{\damp}^*}
\end{split}
\end{equation}
where the subscript $i \times j^*$ refers to the product of invariant amplitudes $\mathcal{M}_i \times \mathcal{M}_j^*$ in the cross section.
The different contributions can be understood as follows:
\begin{itemize}
    \item $\abs{\hyperref[fig:gamma_qq_qq]{\aamp}}^2$:
    Contains the leading-order part, the NLO corrections to the photon wave function, the self-energy correction $\sqrt{Z_{q \bar q}} $ of the quark-antiquark pair, and a part of the final-state corrections corresponding to a gluon exchange between the quark-antiquark pair.
    \item $\abs{\hyperref[fig:gamma_qqg_qq]{\bamp}}^2$:
    A gluon is emitted and absorbed before the final state in both the amplitude and the complex conjugate.
    This contribution comes at the order $\mathcal{O}(\as^2)$ and can be neglected at NLO.
    \item $\abs{\hyperref[fig:gamma_qqg_qqg]{\camp}}^2$:
    A gluon is emitted before the shock wave in both the amplitude and its complex conjugate, and it is not absorbed in the final state. This is a finite contribution to the NLO cross section and has already been calculated in Ref.~\cite{Beuf:2022kyp}.
    \item $\abs{\hyperref[fig:gamma_qq_qqg]{\damp}}^2$:
    Contains the final-state corrections where a gluon is emitted after the shock wave in both the production amplitude and its complex conjugate.
    \item $\hyperref[fig:gamma_qq_qq]{\aamp}\times \hyperref[fig:gamma_qqg_qq]{\bamp}^*$:
    Contains the NLO corrections where a gluon crosses the shock wave and is absorbed in the final state.
    \item $\hyperref[fig:gamma_qqg_qqg]{\camp}\times \hyperref[fig:gamma_qq_qqg]{\damp}^*$:
    Contains the cross-terms where the gluon is emitted before the shock wave in the amplitude and after the shock wave in the complex conjugate, or vice versa. This contribution is finite.
\end{itemize}

To calculate these different terms, the wave functions $\Psi^{\gamma^* \to q \bar q}_\iin$ and $\Psi_\oout^{q \bar q \to q \bar q}$ are needed at the order $\mathcal{O}(\gs^2)$, and the wave functions $\Psi_\iin^{\gamma^* \to q \bar qg }$, $\Psi_\oout^{q \bar q \to q \bar qg}$ and $\Psi_\oout^{q \bar q g \to q \bar q}$ at the order $\mathcal{O}(\gs)$.
The wave functions with the photon have already been calculated in Refs.~\cite{Hanninen:2017ddy,Beuf:2016wdz,Beuf:2017bpd}, but the rest of the wave functions are new and have not been calculated before.
Feynman diagrams containing these wave functions, however, have already been calculated in Ref.~\cite{Caucal:2021ent}, so that the expressions for the wave functions can be compared to some extent to the results presented there.

With these wave functions, one can then calculate each term in Eq.~\eqref{eq:DDIS_cross_Section_terms} separately.
However,
contributions related to the final-state corrections in $\abs{\hyperref[fig:gamma_qq_qq]{\aamp}}^2$ and $\abs{\hyperref[fig:gamma_qq_qqg]{\damp}}^2$ turn out to be especially difficult to calculate,
as we need to Fourier transform the wave function $\Psi_\oout^{q \bar q \to q \bar q}$ to the mixed space used in the rest of the calculation.
It is not known how to do this Fourier transform analytically, and thus one is left with an additional integral which makes showing the cancellation of divergences in the full cross section highly nontrivial.
The purpose of the next section is to demonstrate how to deal with the final-state corrections in a more clever way.

\subsection{Final-state corrections}
\label{sec:final-state_corrections}

\begin{figure}
    \centering
    \begin{subfigure}{0.7\textwidth}
        \centering
        \begin{overpic}[width=\textwidth]{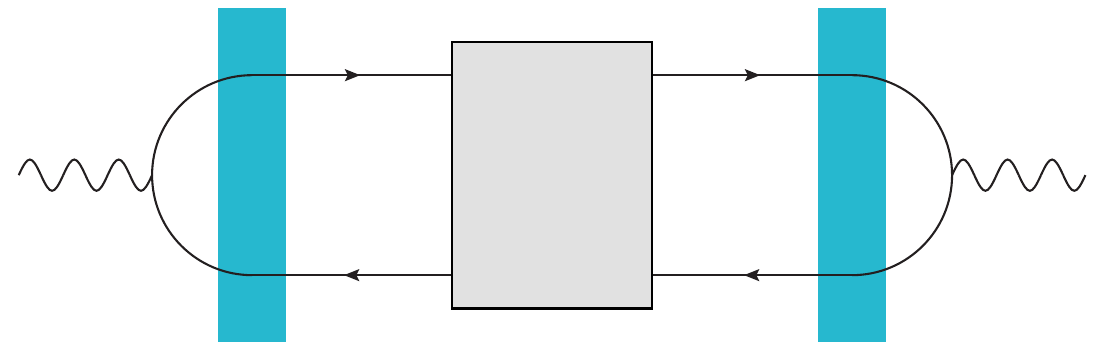}
            \put (48,16) {$F$}
            \put (38,16){\vector(0,1){8}}
            \put (38,16){\vector(0,-1){8}}
            \put (31,16) {$\Kt_{01}$}
            \put (62,16){\vector(0,1){8}}
            \put (62,16){\vector(0,-1){8}}
            \put (63,16) {$\Kt_{\ov{01}}$}
            \put (35,27) {$z_0$}
            \put (35,3)  {$z_1$}
            \put (61,27) {$z_{\ov 0}$}
            \put (61,3) {$z_{\ov 1}$}
        \end{overpic}
        \caption{ Factorization of the final-state corrections. }
        \label{fig:F}
    \vspace{1.0cm}
    \end{subfigure}
   \subfloat[Diagram $\azero$ \label{fig:A0}]{
        \centering
        \includegraphics[width=0.3\columnwidth]{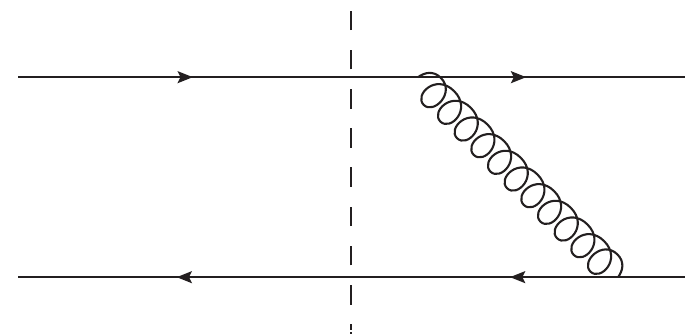}
        }
   \subfloat[Diagram $\aone$ \label{fig:A1}]{
        \centering
        \includegraphics[width=0.3\columnwidth]{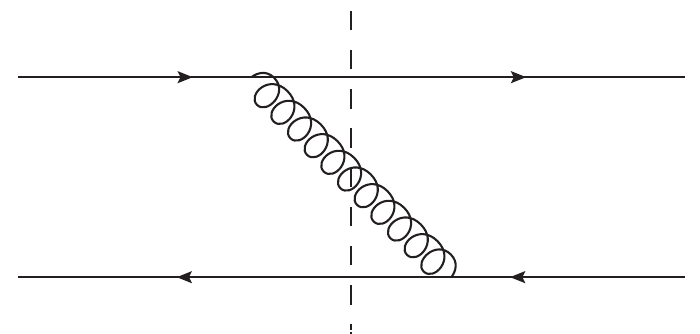}
        } 
   \subfloat[Diagram $\atwo$ \label{fig:A2}]{
        \centering
        \includegraphics[width=0.3\columnwidth]{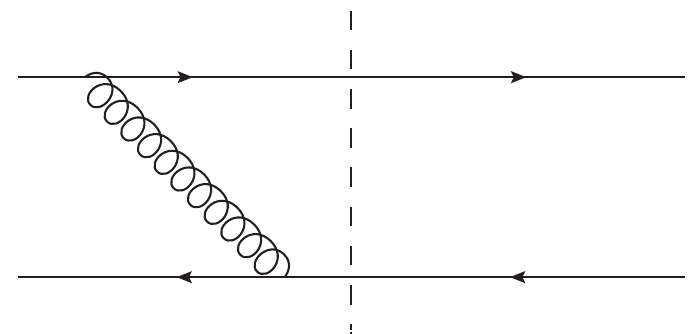}
        } \\
   \subfloat[Diagram $\bzero$ \label{fig:B0}]{
        \centering
        \includegraphics[width=0.3\columnwidth]{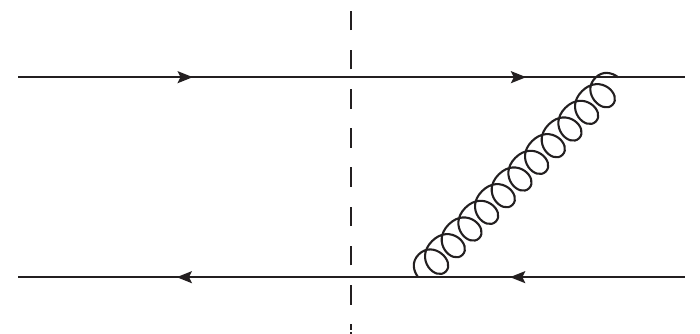}
        }
   \subfloat[Diagram $\bone$ \label{fig:B1}]{
        \centering
        \includegraphics[width=0.3\columnwidth]{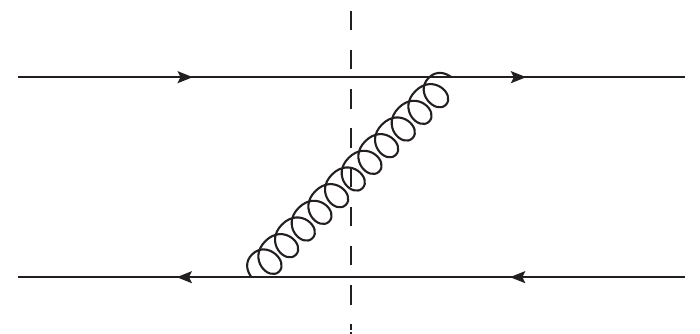}
        }
   \subfloat[Diagram $\btwo$ \label{fig:B2}]{
        \centering
        \includegraphics[width=0.3\columnwidth]{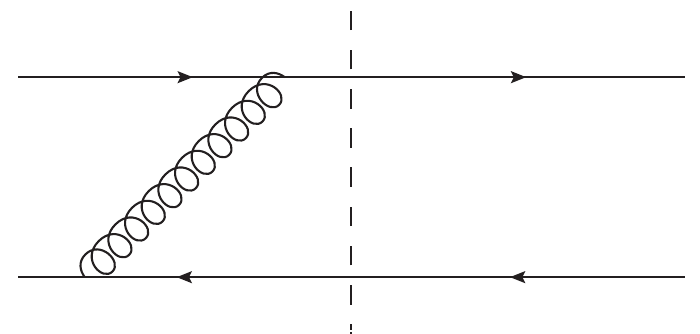}
        }\\
   \subfloat[Diagram $\czero$ \label{fig:C0}]{
        \centering
        \includegraphics[width=0.3\columnwidth]{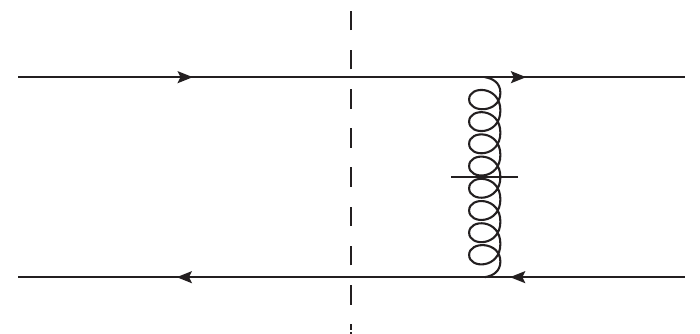}
        } 
   \hspace{0.31\textwidth
     }
   \subfloat[Diagram $\ctwo$ \label{fig:C2}]{
        \centering
        \includegraphics[width=0.3\columnwidth]{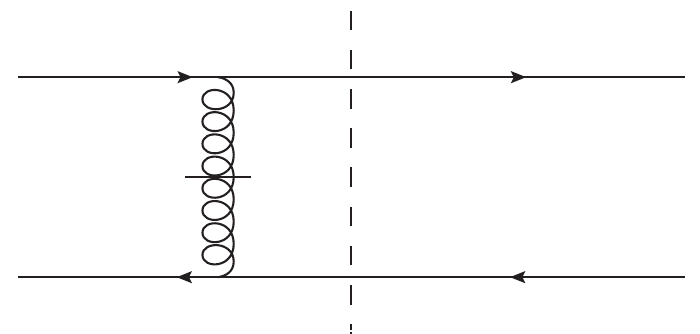}
        } \\
   \subfloat[Diagram $\done$ \label{fig:D1}]{
        \centering
        \includegraphics[width=0.3\columnwidth]{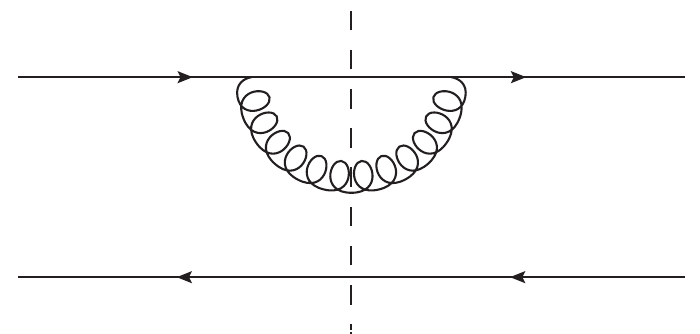}
        }
   \subfloat[Diagram $\eone$ \label{fig:E1}]{
        \centering
        \includegraphics[width=0.3\columnwidth]{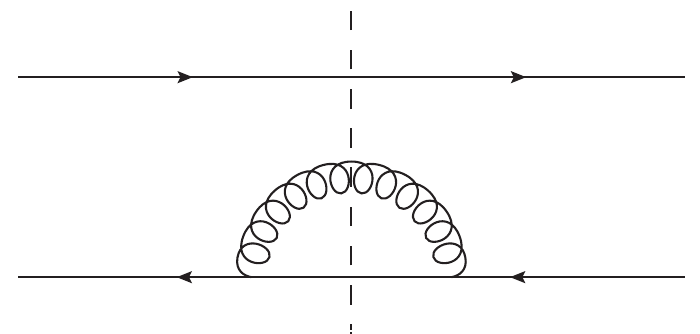}
        }
    \caption{Feynman diagrams contributing to the final-state corrections at NLO.
    }
\label{fig:final_state_NLO}
\end{figure}

The final-state corrections can be factorized out of the rest of the calculation. This is shown in Fig.~\ref{fig:F} where 
\begin{equation}
    F = \sum_{i \in \text{ diagrams}} F_i
\end{equation}
denotes the sum of the diagrams in Fig.~\ref{fig:final_state_NLO}.
As the final-state corrections, and thus their divergences, are deeply related to each other it is more natural to first sum them together in the momentum space and only then take the transverse Fourier transforms to the mixed space.
This is most convenient to do by grouping the different diagrams corresponding to the rows in Fig.~\ref{fig:final_state_NLO}.

Let us first consider the first row with the diagrams labeled as $\adiag i$. The corresponding contributions can be written as
\begin{equation}
    \begin{split}
    \F_{\adiag i} \propto &
    \int \frac{\dd[2]{\Kt_{01}} \dd[2]{\Kt_{\ov 0 \ov 1}}}{(2\pi)^{4}}e^{i \Kt_{\ov 0 \ov 1} \vdot \xt_{\overline 0 \overline 1}-i\Kt_{01} \vdot \xt_{0 1}}
    \frac{1}{(z_0 -z_{\ov 0})^3}(z_{\ov 0} \Kt_{01}-z_0 \Kt_{\ov 0 \ov 1}) \vdot (z_{\ov 1} \Kt_{01}- z_1 \Kt_{\ov 0 \ov 1}) \\
    &\times  \frac{\delta(M_X^2-M_{\adiag i}^2)}{[M_{\adiag i}^2-M_{\adiag j}^2 \pm i \delta] [M_{\adiag i}^2-M_{\adiag k}^2 \pm i \delta]}
    \end{split}
\end{equation}
where $M_{\adiag i}^2$ are the invariant masses at different parts of the Feynman diagrams, which come from the energy denominators in the light-cone perturbation theory.
The signs of the infinitesimals $\pm i \delta$ are determined by whether the energy denominator is in the amplitude or the complex conjugate.
The contributions $ \F_{\azero}$, $ \F_{\aone}$ and $ \F_{\atwo}$ differ only in the invariant masses and the signs of the infinitesimals, and thus they can be easily summed to obtain
\begin{equation}
    \label{eq:FA_sum}
    \begin{split}
    & \F_{\azero} + \F_{\aone} + \F_{\atwo}  \propto  
    \int \frac{\dd[2]{\Kt_{01}} \dd[2]{\Kt_{\ov 0 \ov 1}}}{(2\pi)^{4}}e^{i \Kt_{\ov 0 \ov 1} \vdot \xt_{\overline 0 \overline 1}-i\Kt_{01} \vdot \xt_{0 1}} 
    \frac{1}{(z_0 -z_{\ov 0})^3}\\
    &\times  
    (z_{\ov 0} \Kt_{01}-z_0 \Kt_{\ov 0 \ov 1}) \vdot (z_{\ov 1} \Kt_{01}- z_1 \Kt_{\ov 0 \ov 1}) 
    \biggl \{
        \frac{\delta(M_{\azero}^2-M_X^2)}{(M_{\azero}^2-M_{\aone}^2-i\delta)(M_{\azero}^2-M_{\atwo}^2-i\delta)} \\
        &+ 
        \frac{\delta(M_{\aone}^2-M_X^2)}{(M_{\aone}^2-M_{\azero}^2+i\delta)(M_{\aone}^2-M_{\atwo}^2-i\delta)}
        +
        \frac{\delta(M_{\atwo}^2-M_X^2)}{(M_{\atwo}^2-M_{\azero}^2+i\delta)(M_{\atwo}^2-M_{\aone}^2+i\delta)}
    \biggr \}.
    \end{split}
\end{equation}

We can get rid of the delta functions by writing
\begin{equation}
    \delta(x) = \frac{1}{2\pi i} \left[ \frac{1}{x-i\delta} - \frac{1}{x+i\delta}  \right].
\end{equation}
This is especially useful as there are cancellations between the three terms in Eq.~\eqref{eq:FA_sum}, resulting in 
\begin{equation}
    \begin{split}
        &\frac{\delta(M_{\azero}^2-M_X^2)}{(M_{\azero}^2-M_{\aone}^2-i\delta)(M_{\azero}^2-M_{\atwo}^2-i\delta)}  
        +
        \frac{\delta(M_{\aone}^2-M_X^2)}{(M_{\aone}^2-M_{\azero}^2+i\delta)(M_{\aone}^2-M_{\atwo}^2-i\delta)}\\
        &+
        \frac{\delta(M_{\atwo}^2-M_X^2)}{(M_{\atwo}^2-M_{\azero}^2+i\delta)(M_{\atwo}^2-M_{\aone}^2+i\delta)} \\
        =&\frac{1}{2\pi i} \left[\frac{1}{(M_X^2-M_{\azero}^2-i\delta)(M_X^2-M_{\aone}^2-i\delta)(M_X^2-M_{\atwo}^2-i\delta)}\right.\\
        &\left.-\frac{1}{(M_X^2-M_{\azero}^2+i\delta)(M_X^2-M_{\aone}^2+i\delta)(M_X^2-M_{\atwo}^2+i\delta)}\right].
    \end{split}
\end{equation}
Note that the correct signs for the infinitesimals $\pm i \delta$, discussed in Sec.~\ref{sec:lcpt}, are crucial for this trick.
While it is now easier to perform the Fourier transform without the delta function,
Fourier transforms with three different denominators are still in general quite complicated.
This can be simplified by noting that 
\begin{equation}
\begin{split}
    \left(z_{\ov 0} \Kt_{01}-z_{0}\Kt_{\ov 0 \ov 1} \right) &\vdot \left(z_{\ov 1}\Kt_{01}-z_{1}\Kt_{\ov 0 \ov 1} \right) \\
    = \frac{1}{2}(z_0 - z_{\ov 0}) \biggl[
        &
         z_0 z_{\ov 1} (M_X^2 - M_{\azero}^2) 
         +z_0 z_{\ov 1} (M_X^2 - M_{\atwo}^2) \\
        -& (z_{\ov 0} z_1 + z_0 z_{\ov 1}) (M_X^2 - M_{\aone}^2) 
        - (z_0 -z_{\ov 0}) M_X^2 \biggr]
\end{split}
\end{equation}
which allows us to cancel some of the denominators. This leads to
\begin{equation}
    \label{eq:FA_sum2}
    \begin{split}
    F_{\azero}+F_{\aone}+F_{\atwo}
     &\propto  \int \frac{\dd[2]{\Kt_{01}} \dd[2]{\Kt_{\ov 0 \ov 1}}}{(2\pi)^{4}}e^{i \Kt_{\ov 0 \ov 1} \vdot \xt_{\overline 0 \overline 1}-i\Kt_{01} \vdot \xt_{0 1}}
    \frac{1}{(z_0 -z_{\ov 0})^2} \\
        \times &  \biggl[
        z_0 z_{\ov 1} D_{\adiag 01} +z_0 z_{\ov 1} D_{\adiag 12} -(z_{\ov 0} z_1 + z_0 z_{\ov 1}) D_{\adiag 02} -(z_0 - z_{\ov 0}) M_X^2 D_{\adiag 012}
        \biggr]
    \end{split}
\end{equation}
where we have denoted
\begin{equation}
    \begin{split}
    D_{\adiag ij} 
    &= \frac{1}{\pi} \Im \left\{ \frac{1}{[M_X^2-M_{\adiag i}^2-i\delta][M_X^2-M_{\adiag j}^2-i\delta]} \right\}, \\
    D_{\adiag ijk} &= \frac{1}{\pi} \Im \left\{\frac{1}{[M_X^2-M_{\adiag i}^2-i\delta][M_X^2-M_{\adiag j}^2-i\delta][M_X^2-M_{\adiag k}^2-i\delta]}\right\}.
    \end{split}
\end{equation}
The crucial simplification here is that different terms in Eq.~\eqref{eq:FA_sum2} correspond to different divergences in terms of the gluon's plus-momentum cut-off $\alpha$.
The divergences come from the gluon's plus momentum going to zero, which corresponds to $z_2 = z_0 - z_{\ov 0} \to 0$, and we can see that Eq.~\eqref{eq:FA_sum2} is divergent in this limit.
Note that the energy denominators also depend on $z_0 - z_{\ov 0}$ so that we have roughly
\begin{equation}
    \frac{1}{M_X^2 - M_{\aone}^2 - i \delta} \sim (z_0 - z_{\ov 0}) \ln(z_0 - z_{\ov 0})
\end{equation}
after the Fourier integrals,
which alleviates the divergence for terms with this denominator. Other energy denominators do not have such a dependence on $z_0 -z_{\ov 0}$. This allows us to read the divergences of the different terms as follows:
\begin{enumerate}
    \item $D_{\adiag 01}$ and $D_{\adiag 12}$: a logarithmic divergence $\ln^2 \alpha$
    \item $D_{\adiag 02}$: a power divergence $1/\alpha$
    \item $D_{\adiag 012}$: no divergences
\end{enumerate}
Especially, the term $D_{\adiag 012}$ with three denominators is free of divergences.
The other terms with divergences contain only two denominators  and are thus simpler to study.

Similarly to Diagrams $\adiag i$, the second row in Fig.~\ref{fig:final_state_NLO} corresponding to Diagrams $\bdiag i$ can be shown to simplify in this way. 
For the third row with the instantaneous gluon exchange, Diagrams $\cdiag i$, we get a slightly different combination of denominators.
The contributions of Diagrams $\czero$ and $\ctwo$ read:
\begin{equation}
    \begin{split}
        F_{\czero} \propto & 
         \int \frac{\dd[2]{\Kt_{01}} \dd[2]{\Kt_{\ov 0 \ov 1}}}{(2\pi)^{4}} e^{i \Kt_{\ov 0 \ov 1} \vdot \xt_{\overline 0 \overline 1}-i\Kt_{01} \vdot \xt_{0 1}} \frac{1}{(z_0 -z_{\ov 0})^2}  \times \frac{\delta(M_X^2-M_{\atwo}^2)}{M_{\azero}^2-M_{\atwo}^2-i \delta },
    \end{split}
\end{equation} 
\begin{equation}
    \begin{split}
        F_{\ctwo} \propto & 
         \int \frac{\dd[2]{\Kt_{01}} \dd[2]{\Kt_{\ov 0 \ov 1}}}{(2\pi)^{4}} e^{i \Kt_{\ov 0 \ov 1} \vdot \xt_{\overline 0 \overline 1}-i\Kt_{01} \vdot \xt_{0 1}} \frac{1}{(z_0 -z_{\ov 0})^2}\times \frac{\delta(M_X^2-M_{\azero}^2) }{M_{\atwo}^2-M_{\azero}^2+i \delta }.
    \end{split}
\end{equation} 
These can be summed together using the identity
\begin{equation}
        \frac{\delta(M_X^2-M_{\atwo}^2)}{M_{\azero}^2-M_{\atwo}^2-i \delta}+\frac{\delta(M_X^2-M_{\azero}^2)}{M_{\atwo}^2-M_{\azero}^2+i \delta}
        = -D_{\cdiag 02}
    \end{equation}
which gives us
\begin{equation}
    \label{eq:FC_sum}
    \begin{split}
        F_{\czero}+ F_{\ctwo} \propto & 
         -\int \frac{\dd[2]{\Kt_{01}} \dd[2]{\Kt_{\ov 0 \ov 1}}}{(2\pi)^{4}} e^{i \Kt_{\ov 0 \ov 1} \vdot \xt_{\overline 0 \overline 1}-i\Kt_{01} \vdot \xt_{0 1}} \frac{1}{(z_0 -z_{\ov 0})^2}D_{\cdiag 02}.
    \end{split}
\end{equation} 
It turns out that we can combine the terms with the similar denominator structure $D_{\adiag 02}$, $D_{\bdiag 02}$ and $D_{\cdiag 02}$ as we have
\begin{equation}
    D_{\cdiag 02 } =
    \begin{cases}
        D_{\adiag 02} &\text{if } z_0 - z_{\ov 0 } >0 \\ 
        D_{\bdiag 02} &\text{if } z_0 - z_{\ov 0 } <0.
    \end{cases}
\end{equation}
While the coefficients in Eqs.~\eqref{eq:FA_sum2} and \eqref{eq:FC_sum} differ, one can show that the divergences $1/(z_0 - z_{\ov 0})^2$ and $1/(z_0 - z_{\ov 0})$ cancel in their sum so that the end result is finite.

This means that after summing together Diagrams $\adiag i$, $\bdiag i$ and $\cdiag i$ in the first three rows of Fig.~\ref{fig:final_state_NLO} the only divergences left come from terms with the energy denominator structure $D_{\adiag 01}$, $D_{\adiag 12}$, $D_{\bdiag 01}$ and $D_{\bdiag 12}$. The corresponding divergences are $\ln^2 \alpha$ in terms of the gluon's plus-momentum regulator $\alpha$.
The remaining divergences are then simpler to handle and can be shown to cancel with the rest of the calculation.
This concludes the handling of final-state corrections that have a form of a gluon exchange between the quark and the antiquark, i.e. Diagrams $\adiag i$, $\bdiag i$ and $\cdiag i$.

One still needs to evaluate Diagrams $\done$ and $\eone$ that are similar to the self-energy corrections of the quark and the antiquark.
For these diagrams, the Fourier integrals can be done analytically without any additional left-over integrals which makes treating the divergences simpler. Thus, no tricks are needed for calculating Diagrams $\done$ and $\eone$.
These diagrams contain both IR divergences in $\varepsilon$ and plus-momentum divergences of the form $\ln^2 \alpha$. It turns out that this $\ln^2 \alpha$ cancels exactly with the other final-state corrections so that the remaining plus-momentum divergences have only a  $\ln \alpha$ dependence on the cut-off.

\subsection{Cancellation of divergences}

It is important to highlight the nontrivial cancellation of different kinds of divergences appearing in the calculation. 
The appearing divergences can be classified into the following categories based on the Feynman diagrams of their origin:
\begin{enumerate}
    \item \label{enum:self-energy} $\abs{\hyperref[fig:gamma_qq_qq]{\aamp}}^2$:  Self-energy corrections to the final $q \bar q$ state. These contain both IR and UV divergences in terms of the $\varepsilon$ of dimensional regularization. 
    These divergences cancel each other when using the same $\varepsilon$ for both IR and UV divergences, and actually the self-energy corrections are identically zero in dimensional regularization.
    \item \label{enum:photon_wf} $\abs{\hyperref[fig:gamma_qq_qq]{\aamp}}^2$: NLO corrections to the $\gamma^* \to q \bar q$ wave function. These bring UV divergences in $\varepsilon$ and plus-momentum divergences of the form $\ln \alpha$.
    \item \label{enum:gluon_squared_shock-wave} 
    $\hyperref[fig:gamma_qqg_qq]{\bamp} \times \hyperref[fig:gamma_qq_qq]{\aamp}^*$: Gluons crossing the shock wave that are emitted and absorbed by the same particle.
    These are related to the self-energy corrections, except now the IR region is made finite by the different energy denominator structure on the photon side. Hence, these only contain UV divergences that do not cancel as in Point~\ref{enum:self-energy}. Also, plus-momentum divergences of the form $\ln \alpha$ are present.    
    \item 
    \label{enum:gluon_exchange_shock-wave}  
    $\hyperref[fig:gamma_qqg_qq]{\bamp} \times \hyperref[fig:gamma_qq_qq]{\aamp}^*$: Gluons crossing the shock wave that correspond to a gluon exchange between the quark and the antiquark. These only bring a plus-momentum divergence $\ln \alpha$.
    \item \label{enum:gluon_squared_final-state}  
    $\abs{\hyperref[fig:gamma_qq_qqg]{\damp}}^2$:
    Final-state corrections where the gluon is emitted by the same particle both in the amplitude and its complex conjugate, i.e. Diagrams $\done$ and $\eone$ in Fig.~\ref{fig:final_state_NLO}. 
    In terms of Feynman diagrams, these can be related to the self-energy corrections with a delta function setting the invariant mass of the  intermediate $q \bar q g$ state to be $M_X^2$. This delta function regulates the UV region, and thus these only contain IR divergences in $\varepsilon$. Also, plus-momentum divergences $\ln^2 \alpha$ are present.
    \item \label{enum:gluon_exchange_final-state}  
    $\abs{\hyperref[fig:gamma_qq_qq]{\aamp}}^2$ and $\abs{\hyperref[fig:gamma_qq_qqg]{\damp}}^2$: Final-state corrections that diagrammatically correspond to a gluon exchange between the quark and the antiquark, i.e. Diagrams $\adiag i$, $\bdiag i$ and $\cdiag i$ in Fig.~\ref{fig:final_state_NLO}. As described in Sec.~\ref{sec:final-state_corrections}, these contain $\ln^2 \alpha$ divergences when summed together.
\end{enumerate}

As the cancellation of these divergences is in general quite involved, we will not go through it here in its full complexity.
However, some general comments can be said about the cancellation. First of all, the final-state IR divergences in Point~\ref{enum:gluon_squared_final-state} are directly related to the IR divergences in the self-energy corrections.
The overall effect of the self-energy corrections can be thought of as converting the IR divergences to UV divergences which then cancel with the rest of the calculation.
With this in mind, the combined divergence structure of the final-state corrections in Points~\ref{enum:gluon_squared_final-state} and \ref{enum:gluon_exchange_final-state}  is the same as the photon wave function $\gamma^* \to q \bar q$ in Point~\ref{enum:photon_wf}. 
This can be understood by noting that the difference to the forward elastic scattering amplitude in inclusive DIS, Eq.~\ref{eq:inclusive_DIS_cross_section}, is in the final state and thus for both processes to be finite the divergence structure of the final state has to be the same.
Combining the final-state divergences with the rest of the calculation, the UV divergences cancel but there are still some logarithmic $\ln \alpha$ divergences left. 
These are related to the JIMWLK evolution of the dipole amplitudes in the production amplitude and its complex conjugate.
Absorbing this $\ln \alpha$ divergence to the evolution of the dipole amplitudes then cancels the final divergence, rendering the NLO cross section finite.

After canceling the divergences, the resulting expressions can be numerically evaluated for comparisons to the HERA data and predictions for the EIC~\cite{Accardi:2012qut,Aschenauer:2017jsk,AbdulKhalek:2021gbh}.
It will be especially interesting to compare diffractive structure functions for protons and heavy nuclei at the NLO accuracy to see the effects of saturation.
The numerical computations for these predictions are left for future work.

\chapter{Conclusions}

The focus of this thesis has been on diffractive processes in the high-energy limit where the nonlinear effects of QCD start to become relevant.
To probe this nonlinear region of QCD, it is crucial to get a better understanding of processes sensitive to saturation, which includes diffractive processes.
Calculations in the high-energy limit can be done in the dipole picture with the color-glass condensate effective field theory, outlined in Chs.~\ref{sec:dipole_picture} and \ref{sec:dipole-target_scattering_amplitude}.
The NLO calculations considered in this thesis are state-of-the-art in the dipole picture.

In Ch.~\ref{sec:vector_meson_production}, we discuss higher-order corrections to exclusive vector meson production. This includes relativistic corrections to heavy vector meson production, and NLO corrections to both heavy and light vector meson production.
In Article~\cite{relativistic}, it was shown that the relativistic corrections can be numerically important for small photon virtualities in heavy vector meson production.
The NLO corrections to heavy vector meson production in the nonrelativistic limit were calculated in Articles~\cite{heavy_long} and \cite{heavy_trans}, where it was found that the NLO corrections are numerically moderate when also using a dipole amplitude fitted at NLO. In general, a good agreement with the experimental data was found if also the relativistic corrections at leading order were included.
The NLO corrections to light vector meson production in the limit of large photon virtuality calculated in Article~\cite{light} were also considered.
Similarly to the heavy vector meson case, the NLO results were found to be comparable to the leading-order results, with an excellent agreement with the experimental data for values $Q^2 \gg M_V^2$ where the calculations are valid. 

In phenomenological applications, an important part of calculations in the dipole picture is the dipole amplitude discussed in Ch.~\ref{sec:dipole-target_scattering_amplitude}. 
Constraints for the dipole amplitude from massive quarks were considered in Article~\cite{structure} where heavy quark production was calculated numerically.
It was then shown that with properly parametrized initial conditions the dipole amplitude is universal in the sense that the same dipole amplitude can be used to calculate both total and charm quark production in DIS.
This was not possible at leading order when using a BK-evolved dipole amplitude, and thus the NLO corrections were found to be crucial for precision computations in the dipole picture.

Some discussion about an unpublished work on NLO corrections to inclusive diffraction in DIS was included in Ch.~\ref{sec:DDIS}.
The main difference to exclusive vector meson production is in the final state, where in inclusive diffraction the final state is defined to be any multiparticle state with a definite invariant mass $M_X^2$.
This means that calculating the NLO corrections to the final state is more complicated than in exclusive vector meson production and includes more Feynman diagrams, shown in Fig.~\ref{fig:final_state_NLO}.
The general strategy for dealing with these diagrams is discussed in Sec.~\ref{sec:final-state_corrections}.
Including all of the NLO corrections, along with the JIMWLK evolution of the dipole amplitude, one can then show that the end result is finite and suitable for numerical calculations.
While the NLO corrections have not been implemented numerically yet,
it will be interesting to see if they will modify the behavior of the diffractive structure functions.

This thesis is a part of a bigger picture of striving for precise predictions using the dipole picture. 
We note that other NLO calculations using the framework of this work also exist, some of which we have already addressed:
single inclusive hadron production~\cite{Chirilli:2011km,Chirilli:2012jd,Stasto:2014sea,Kang:2014lha,Xiao:2014uba,Altinoluk:2014eka,Watanabe:2015tja,Iancu:2016vyg,Ducloue:2016shw,Ducloue:2017dit,Ducloue:2017mpb,Liu:2019iml,Liu:2020mpy,Bergabo:2022zhe},
 total structure functions with massless~\cite{Hanninen:2017ddy,Beuf:2016wdz,Beuf:2017bpd} and massive quarks~\cite{Beuf:2021qqa,Beuf:2021srj,Beuf:2022ndu},
the dominating gluonic contribution to diffractive structure functions~\cite{Beuf:2022kyp},
exclusive production of light vector mesons~\cite{Boussarie:2016bkq},
diffractive jet production~\cite{Boussarie:2014lxa,Boussarie:2016ogo},
inclusive dijet production in DIS~\cite{Caucal:2021ent,Caucal:2022ulg,Caucal:2023nci},
diffractive dihadron production~\cite{Fucilla:2022wcg}, and
inclusive dihadron production in DIS~\cite{Bergabo:2022tcu,Bergabo:2023wed}.
Also, NLO corrections to the rapidity evolution of the dipole amplitude have been calculated, namely the NLO BK~\cite{Balitsky:2007feb} and NLO JIMWLK~\cite{Kovner:2013ona,Lublinsky:2016meo,Dai:2022imf} equations.
Analytical calculations in the high-energy limit have thus been largely promoted to the NLO precision era, with numerical implementations on the way.

Other interesting calculations in the dipole picture involve going beyond the eikonal approximation and including the so-called sub-eikonal effects~\cite{Altinoluk:2014oxa,Altinoluk:2015gia,Agostini:2019avp,Altinoluk:2020oyd,Altinoluk:2021lvu,Agostini:2022ctk,Altinoluk:2022jkk,Altinoluk:2023qfr,Chirilli:2018kkw,Chirilli:2021lif}.
While no numerical estimates of these sub-eikonal corrections exist at the moment of writing this thesis, it is expected that they will become important for lower energies. 
Computing these will give us guidelines for the accuracy of the eikonal approximation
and the validity of the high-energy factorization.

With these higher-order calculations in the dipole picture becoming available, it will be very interesting to implement them numerically and see if the results obtained at leading order are modified. It will also be very important to compare results from saturation models to calculations that do not involve saturation to see which observables can be used to search for saturation from the experimental data. 
Non-saturation calculations of these processes involve calculations in the BFKL framework and also the collinear framework: see for example Refs.~\cite{Eskola:2022vaf,Eskola:2022vpi,Eskola:2023oos} for NLO exclusive heavy vector meson production in  collinear factorization.
Numerical comparisons to these results will be crucial for distinguishing between saturation and non-saturation effects.

\tailmatter

\cleardoublepage
\phantomsection
\addcontentsline{toc}{chapter}{References}
\bibliography{refs}

\backmatter

\phantomsection
\includedarticles

\begin{article}{relativistic}
\arttitle{Relativistic corrections to the vector meson light front wave function}
\artauthor{Tuomas Lappi, Heikki Mäntysaari, Jani Penttala}
\artpublish{Physical Review D}
\artpublishmore{102, 054020}
\artyear{(2020)}
\artpages{22}
\arthide
\end{article}

\begin{article}{heavy_long}
\arttitle{Exclusive heavy vector meson production at next-to-leading order in the dipole
picture}
\artauthor{Heikki Mäntysaari, Jani Penttala}
\artpublish{Physics Letters B}
\artpublishmore{823, 136723}
\artyear{(2021)}
\artpages{14}
\arthide
\end{article}

\begin{article}{light}
\arttitle{Exclusive production of light vector mesons at next-to-leading order in the dipole picture}
\artauthor{Heikki Mäntysaari, Jani Penttala}
\artpublish{Physical Review D}
\artpublishmore{105, 114038}
\artyear{(2022)}
\artpages{25}
\arthide
\end{article}

\begin{article}{heavy_trans}
\arttitle{Complete calculation of exclusive heavy vector meson production at next-to-leading order in the dipole picture}
\artauthor{Heikki Mäntysaari, Jani Penttala}
\artpublish{Journal of High Energy Physics}
\artpublishmore{08, 247}
\artyear{(2022)}
\artpages{45}
\arthide
\end{article}

\begin{article}{structure}
\arttitle{Proton structure functions at NLO in the dipole picture with massive quarks}
\artauthor{Henri Hänninen, Heikki Mäntysaari, Risto Paatelainen, Jani Penttala}
\artpublish{Physical Review Letters}
\artpublishmore{130, 192301}
\artyear{(2023)}
\artpages{7}
\arthide
\end{article}

\printindex

\end{document}